\documentclass[11pt,a4paper]{article}

\usepackage{setspace}

\usepackage{cite}

\usepackage[left=2cm,right=2cm,top=2.5cm,bottom=3.5cm]{geometry}

\usepackage{amssymb,amsmath,mathrsfs,bm}
\usepackage{comment}
\usepackage{booktabs}
\usepackage{mdframed}
\usepackage{graphicx}
\usepackage{setspace}
\usepackage{tcolorbox}
\usepackage{cite}
\usepackage{hyperref}
\usepackage{enumitem}
\usepackage{titlesec}
\usepackage{empheq}
\usepackage{xcolor}
\usepackage{makecell}
\usepackage{dsfont}
\usepackage{diagbox}

\usepackage[scr=esstix]{mathalpha}

\usepackage{tikz}

\usepackage{hyperref}
\hypersetup{
  linktocpage,
  colorlinks  = true, 
  urlcolor    = blue, 
  linkcolor   = blue, 
  citecolor   = red 
}

\makeatletter 
\@addtoreset{equation}{section}
\makeatother

\def \o{{O}}
\newcommand \widebar [1] {\overline{#1}}

\def\xb{\bar{x}}
\def\yb{\bar{y}}
\def\dbar#1{\widebar{D}_{#1}}

\def\dfour{\Delta^{(4)}}

\def\S{{\bf s}}

\def\beq 	{\begin{equation}}
\def\eeq	{\end{equation}}
\def\bea 	{\begin{eqnarray*}}
\def\eea	{\end{eqnarray*}}
\def \be  	{\begin{equation}}
\def \ee  	{\end{equation}}
\def \ba  	{\begin{eqnarray*}}
\def \ea  	{\end{eqnarray*}}

\def\B{\ensuremath{\mathcal{B}}}
\def\H{\ensuremath{\mathcal{H}}}
\def\Ht{\ensuremath{{\mathcal{L}}}}

\def\F{\ensuremath{\mathcal{F}}}
\def\G{\ensuremath{\mathcal{G}}}
\def\M{\ensuremath{\mathcal{M}}}
\def\P{\ensuremath{\mathcal{P}}}
\def\Q{\ensuremath{\mathcal{Q}}}
\def\a{\ensuremath{\mathbf{a}}}

\def\S{\ensuremath{\mathbb{S}}}
\def\A{\ensuremath{\mathbb{A}}}
\def\I{\mathds{1}}

\def\hb{\ensuremath{\bar{h}}}
\def\jb{\ensuremath{\bar{j}}}

\def\Li{\text{Li}}

\def\bs{  {\bold s} }
\def\bt{  {\bold t} }
\def\bu{ {\bold u} }

\def\twoladder{\begin{tikzpicture}[scale=0.1]
\def\shift{0.3}
\draw[line width=.5pt] (-.6-\shift,-0.5)--(-.9-\shift,-1);
\draw[line width=.5pt] (-.6-\shift,1)--(-.9-\shift,1.5);
\draw[line width=.5pt] (.6+\shift,-0.5)--(.9+\shift,-1);
\draw[line width=.5pt]  (.6+\shift,1)--(.9+\shift,1.5);
\draw[line width=.5pt] (-.6-\shift,-0.5)--(.6+\shift,-0.5);
\draw[line width=.5pt]  (-.6-\shift,1.)--(.6+\shift,1.);
\draw[line width=.5pt] (-.6-\shift,-0.5)--(-.6-\shift,1.);
\draw[line width=.5pt]  (.6+\shift,-0.5)--(.6+\shift ,1.);
\draw[line width=.5pt]  (0,-0.5)--(0 ,1.);
\end{tikzpicture}
}

\def\oneladder{\begin{tikzpicture}[scale=0.1]
\def\shift{0.0}
\draw[line width=.5pt] (-.6-\shift,-0.5)--(-.9-\shift,-1);
\draw[line width=.5pt] (-.6-\shift,1)--(-.9-\shift,1.5);
\draw[line width=.5pt] (.6+\shift,-0.5)--(.9+\shift,-1);
\draw[line width=.5pt]  (.6+\shift,1)--(.9+\shift,1.5);
\draw[line width=.5pt] (-.6-\shift,-0.5)--(.6+\shift,-0.5);
\draw[line width=.5pt]  (-.6-\shift,1.)--(.6+\shift,1.);
\draw[line width=.5pt] (-.6-\shift,-0.5)--(-.6-\shift,1.);
\draw[line width=.5pt]  (.6+\shift,-0.5)--(.6+\shift ,1.);
\end{tikzpicture}
}

\begin{document}
\thispagestyle{empty}

~\\[-2.25cm]

\begin{flushright}
UUITP-02/26
\end{flushright}

\begin{center}
\vskip 0.5truecm 
{\Large\bf
	{\LARGE Quantum Gravity on AdS$_3\times$S$^3$ from CFT: \\\vskip.1in
	Bootstrapping $\emph{n=21}$ 
		}
}\\
\vskip 1.25truecm
	{\bf Francesco Aprile$^{1}$, Hynek Paul$^{2}$, Michele Santagata$^{3}$ \\
	}
		\vskip 0.4truecm
	{\it
		$^{{1})}$ Departamento de F\'{i}sica Te\'{o}rica \& IPARCOS, Facultad de Ciencias F\'{i}sicas,\\
		Universidad Complutense, 28040 Madrid\\\vskip .2truecm
		$^{{2})}$ Instituut voor Theoretische Fysica, KU Leuven, Celestijnenlaan 200D,\\
		B-3001 Leuven, Belgium\\\vskip .2truecm
		$^{{3})}$  Department of Physics and Astronomy, Uppsala University, Box 516, 75120 Uppsala, Sweden \\\vskip .2truecm	
		\vskip .2truecm   }
	\vskip .5truecm

\textit{E-mail:} ~\href{mailto:faprile@ucm.es}{{\tt faprile@ucm.es}},~\href{mailto:hynek.paul@kuleuven.be}{{\tt hynek.paul@kuleuven.be}},~\href{mailto:michele.santagata@physics.uu.se}{{\tt michele.santagata@physics.uu.se}}
\end{center}

\vskip 1.25truecm

\centerline{\bf Abstract}
\vskip .4truecm

\noindent 
We consider the simplest four-point scattering amplitude of $SO(n)$ tensor multiplets in six-dimensional (2,0) supergravity on AdS$_3\times$S$^3$. Using crossing symmetry and the consistency of the operator product expansion in the dual CFT, we explicitly construct the one-loop contribution to the correlator at order $1/c^2$, both in position space and in Mellin space. We show that a strong form of the bootstrap equations imposes constraints on the value of $n$. Remarkably, we find that our bootstrap approach uniquely determines $n=21$, which corresponds to the spectrum of IIB string theory compactified on K3. This stands in sharp contrast to the tree-level correlator for which $n$ is unconstrained. We also analyse the spectrum of unprotected double-trace operators and solve the mixing problem in the first case that involves both tensor and graviton correlators. When $n=21$, the anomalous dimensions rationalise and one of them vanishes. Lastly, we study the flat-space limit of the correlator and find perfect agreement with the one-loop amplitude recently obtained in \cite{Huang:2025nyr}.

{	
\hypersetup{linkcolor=blue}
\newpage
\thispagestyle{empty}
\setcounter{tocdepth}{2}
\tableofcontents
}

		\baselineskip=17pt
		\parskip=5pt

\newpage
\setcounter{page}{1}\setcounter{footnote}{0}
\section{Introduction and summary of results}\label{sec:intro}

In holography, a weakly coupled theory of gravity in AdS 
with light particles of spin up to two is dual to a CFT with a large 
central charge and a sparse low-energy spectrum. The condition on 
the spectrum translates into the existence of a parametrically large gap between the 
low lying single-particle operators and the dimension of the lightest 
stringy states, which in turn implies that 
the CFT must be strongly coupled. 
In this context one would like to leverage properties of the CFT to 
understand if all weakly coupled theories of gravity in AdS arise from 
string theory, or if there are counterexamples \cite{Hartman:2022zik}.
In particular, solution techniques based on the conformal bootstrap 
might offer another perspective on the swampland program 
\cite{Vafa:2005ui} and the string lamppost principle (SLP), i.e.~that all consistent 
quantum theories belong to the string landscape, see e.g.~\cite{Agmon:2022thq} for a review.

This work connects to the SLP in a particular setup, namely $(2,0)$ supergravity 
in six dimensions coupled to $n$ tensor multiplets and compactified on AdS$_3\times$S$^3$ with RR flux. This theory is well known from type IIB string theory and brane engineering: 
it is the low-energy theory of the D1-D5 system with transverse space  
K3 \cite{Horowitz:1996ay}, which provides one of the paradigmatic 
examples of the AdS/CFT correspondence \cite{Maldacena:1997re}. 
The string theory construction determines $n=21$,  however, the classical field
equations for this theory make sense with arbitrary $n$ \cite{Romans:1986er,Riccioni:1997np}.  
Rephrasing the question from above, one would like to use 
CFT principles to understand how the value $n=21$ 
emerges in the low-energy supergravity theory. 

From the low-energy point of view, it should be obvious that a constraint on $n$ can only arise due to quantum effects. 
A way to arrive at $n=21$ is by considering the gravitational anomaly \cite{Alvarez-Gaume:1983ihn} 
and observe that there is only one way in which the massless 
representations of the $(2,0)$ supersymmetry in six dimension, 
i.e.~the graviton multiplet and the tensor multiplets, 
combine together to give a (unitary) anomaly free theory \cite{Townsend:1983xt}. 
Here we will follow a different approach, which is based on the physics of scattering amplitudes in AdS combined with analytic bootstrap methods in the dual CFT. The main result of this paper is the construction of the 
one-loop contribution to the four-point correlator  
\begin{equation}\label{intro_4pt1111}
\langle s_1^{i_1}  s_1^{i_2} s_1^{i_3} s_1^{i_4}\rangle\,,
\end{equation}
where the $s_1^i$ denote the half-BPS chiral primary operators dual to the lightest
Kaluza-Klein mode of the tensor multiplets, labelled by their flavour index $i=1,\ldots n$. 
Our construction will show how the value of $n=21$ 
emerges from the consistency of the Operator Product Expansion (OPE) 
in the dual CFT and crossing symmetry.

Our bootstrap approach stems from the basic observation that the Kaluza-Klein modes 
of the supergravity fields are dual to the single-particle operators in the 
CFT and the low-energy spectrum of the theory is built out of the single-particle 
operators and the multi-particle operators obtained by taking the OPE of the latter.
Then, since the leading order OPE of two single-particle operators closes 
on double-particle operators, their leading order CFT data, i.e.~anomalous 
dimensions and three-point couplings, can be reconstructed from the study 
of tree-level four-point amplitudes and disconnected contributions. 
With this data, the OPE predicts the maximal logarithmic 
discontinuities of the correlator to all orders in the large central charge expansion.

The one-loop bootstrap programme outlined above, which we will refer to as the double-particle bootstrap, has been successfully 
applied to a variety of setups that can be studied with the unifying formalism of Grassmannian field theories \cite{Doobary:2015gia}. 
This includes the four-point scattering of supergravitons in AdS$_5\times$S$^5$ \cite{Aprile:2017bgs,Aprile:2017qoy,Aprile:2019rep,Alday:2019nin,Aprile:2020luw}, 
supergluons in AdS$_5\times$S$^3$ \cite{Alday:2021ajh,Huang:2024rxr}, hypermultiplets in AdS$_2\times$S$^2$ \cite{Heslop:2023gzr}, and
the current state-of-the-art are the two-loop results in AdS$_5\times$S$^5$ \cite{Huang:2021xws,Drummond:2022dxw} and AdS$_5\times$S$^3$ \cite{Huang:2023oxf}.

Here we demonstrate, for the first time, that the double-particle bootstrap generalises to supergravity on an AdS$_3\times$S$^3$ background.
However, it turns out that this is a non-trivial generalisation because the theory
has many new features. It is precisely this richness that will have interesting consequences
on the quantum gravity theory, such as the value of $n=21$ for instance.

In order to apply AdS unitary method to our problem, we have to study the 
$s_1^{i}\times s_1^{\,j}$ OPE. To leading order in the large central charge expansion it takes the schematic form
\begin{equation}\label{OPEIntro}
s_1^{i} (z_1)\times s_1^{\,j}(z_2)  \sim \ {\bf 1} +
\sum_k P_{\I}^{ij} [{\cal O}{\cal O}]^{\phantom{ij}}_{\I,k} + \sum_k
	P_\mathbb{S}^{ij} [{\cal O}{\cal O}]^{\phantom{ij}}_{\mathbb{S},k}+ 
	\sum_k P_\mathbb{A}^{ij} [{\cal O}{\cal O}]^{\phantom{ij}}_{\mathbb{A},k}\, + O\!\left(1/\sqrt{c}\right),
\end{equation}  
where ${\bf 1}$ denotes the identity operator, and $[{\cal O}{\cal O}]_{\a,k}$ stands for 
the various kinds of double-particle operators.
Here we have split the $SO(n)$ indices into the three irreducible representations $\a$ which 
appear in the tensor product of the fundamental representation with itself: the 
singlet $\I$, the symmetric traceless $\mathbb{S}$, and the
anti-symmetric $\mathbb{A}$.

The set of all double-particle operators, $[{\cal O}{\cal O}]_{\a,k}$, contains many 
different composite operators. There are the operators $:\!s_{p}^{i}\partial^n\bar{\partial}^m s_{p}^{ j}\!:\,$, where $s_p$ 
denotes a KK-mode of the six-dimensional tensor multiplets on S$^3$ 
at level $p$, but  there are \emph{also} the double-particle operators constructed 
from the gravity multiplet. The half-BPS primary operators in this case are both 
the scalar fields $\sigma_p$  and the vector fields $V_p^{\pm}$.  
In general, the double-particle operators are degenerate at leading order, i.e.~there 
is more than one operator with the same leading-order dimension, and they all enter the OPE 
in \eqref{OPEIntro}. Even so, we can distinguish the two flavoured reps from the singlet,
because only the double-particle operators $:s_{p}^{i}\partial^n\bar{\partial}^m s_{p}^{j}:\,$ 
participate in $\mathbb{S}$ and $\mathbb{A}$, while it is only in the singlet of $SO(n)$ 
that double-particle operators built out of the tensor multiplets mix with double-particle 
operators built out of the graviton multiplet.

At each order in the large central charge expansion, 
the OPE predicts the maximal logarithmic divergences of the four-point correlator 
$\langle s_1^{i_1}  s_1^{i_2} s_1^{i_3} s_1^{i_4}\rangle$ in the short distance limit, say $z_1\rightarrow z_2$ in~\eqref{OPEIntro}. 
Specifically, at order $1/c^l$ and in the basis of (super)conformal blocks, 
the maximal logarithmic divergence is of order $l$ and is determined by the coefficients
\begin{equation}\label{pred_OPE_intro}
\sum_{k} C^{(0)\phantom{n+1}}_{s_1s_1; [{\cal O}{\cal O}]_{\a,k}}\left( \gamma^{(1)}_{[{\cal O}{\cal O}]_{\a,k}}\right)^{l} C^{(0)\phantom{n+1}}_{s_1s_1; [{\cal O}{\cal O}]_{\a,k}}\,,
\end{equation}
where the sum is over the different degenerate double-particle operators labelled by $k$.
For each of them, $\gamma^{(1)}_{[{\cal O}{\cal O}]}$ denotes the tree-level anomalous 
dimension and $C^{(0)}_{s_1s_1;  [{\cal O}{\cal O}]}$ is the three-point coupling with the external operators. In order to compute the one-loop divergence, $l=2$, in all $SO(n)$ representations~$\a$, directly from \eqref{pred_OPE_intro}, 
one needs $\gamma^{(1)}_{[{\cal O}{\cal O}]_{\a,k}}$ and $C^{(0)}_{s_1s_1;[{\cal O}{\cal O}]_{\a,k}}$ for all degenerate operators $[{\cal O}{\cal O}]_{\a,k}$. As will be explained later on, an alternative and more straightforward approach to obtain the one-loop divergence which avoids the explicit resolution of the mixing problem is to glue together (infinite) families of correlators which contain the same OPE data. In any case, the unmixing is simpler in the flavoured reps because it can be brought to a conventional 
single-correlator problem thanks to the (hidden) 6d conformal symmetry of the tree-level 
correlators \cite{Rastelli:2019gtj,Giusto:2020neo}. 
On the contrary, the task in the singlet representation is more complicated. Only in the 
special case of $h=\hb=1$, i.e.~minimum-twist and spin $0$, there is one operator (given by $:s_{1}^{i}s_{1}^{j}:\,$) and hence no mixing. In this paper we 
will make progress for $h=\hb=2$, i.e.~the next-to-minimum twist at spin $0$, 
in which case there are a total of three operators which mix. However, 
more work needs to be done in order to have complete control on the 
spectrum in the singlet channel (even algorithmically).

At this point we observe that in our correlator, $\langle s_1^{i_1}  s_1^{i_2} s_1^{i_3} s_1^{i_4}\rangle$,
some of the information that we would get from the spectrum can be traded 
in favor of crossing symmetry. To see this, let us observe that the flavour 
structure of our correlator is fixed to be of the form
\begin{align}
\langle s_1^{i_1}  s_1^{i_2} s_1^{i_3} s_1^{i_4}\rangle\ \propto\ 
\delta^{i_1i_2}\delta^{i_3 i_4} {\cal H}_s + \delta^{i_1i_4}\delta^{i_2 i_3} {\cal H}_t + \delta^{i_1i_3}\delta^{i_2 i_4} {\cal H}_u\,.
\end{align}
As a consequence of crossing symmetry, out of the three functions ${\cal H}_s$, ${\cal H}_t$ and ${\cal H}_u$, there is only one that is independent, which we will take to be ${\cal H}_t$. Then, 
if one succeeds in bootstrapping ${\cal H}_t$ at one loop, the other two orientation, 
${\cal H}_{u}$ and ${\cal H}_s$, are given for free. Let us note that a bonus of ${\cal H}_t$ 
is  that its maximal logarithmic discontinuity  from \eqref{pred_OPE_intro} 
is the one coming from  the flavoured reps, $\mathbb{S}$ and $\mathbb{A}$, which, 
as mentioned above, are special since the mixing problem has an underlying 
6d conformal symmetry.

Elaborating on the relation between the OPE and crossing, there is an additional 
observation that we can make. For illustration, consider the anomalous dimensions 
of the minimum twist operators at spin zero, $h=\hb=1$.
There is one operator in the singlet $\I$ and one operator in the symmetric traceless 
rep, $\mathbb{S}$, and their tree-level anomalous dimensions are
\begin{align}\label{anom_intro_11}
	\gamma^{(1)}_{\,:s^{\phantom{(i}}_1\!\!s^{\phantom{j)}}_1\!\!\!:}=-2+\frac{n}{3}\,,\qquad\qquad \gamma^{(1)}_{\,:s^{(i}_1s^{j)}_1:}=-2\,.
\end{align}
Importantly, the anomalous dimension in the singlet depends on the value of $n$. The OPE predictions in \eqref{pred_OPE_intro} then imply
the following  equation:
\begin{align}\label{equation_crossing}
{\cal H}^{(2)}_{\I}\Big|_{\log^2(U)} = 
\bigg[ {\cal H}^{(2)}_{t}+ {\cal H}^{(2)}_{u}+ n\, {\cal H}^{(2)}_{s}\bigg] _{\log^2(U)} = \frac{1}{8}\left(-2+\frac{n}{3}\right)^{\!2} B_{1,1}  + 
\sum_{h,\bar{h}} \big[\ldots\big]_{h,\bar{h}} B_{h,\bar{h}}\,,
\end{align}
where $B_{h,\bar{h}}$ are conformal blocks and we are highlighting the 
exchange of $:s_1 s_1\!:$ because we quoted the anomalous dimension 
in \eqref{anom_intro_11} -- but otherwise all operators in the sum over $h,\hb$
are important.
Our observation is that the LHS of \eqref{equation_crossing} only depends 
on ${\cal H}^{(2)}_{t}$ and its images under crossings, therefore if the latter 
happens to be $n$-independent, the LHS is linear in $n$, but the RHS is quadratic in $n$  
by construction. Under these circumstances, which we refer to as the strong version of the bootstrap equations, 
there is a constraint on the value 
of $n$ that follows from \eqref{equation_crossing}. An intriguing question comes 
up at this point: what is this function? And what is the corresponding value of $n$?

Even without giving further details, there are two basic reasons to expect that a 
one-loop function solution of the strong version of the bootstrap equations 
exists.  First, the $t$- or $u$-channel $\log^2$ discontinuities do 
not depend on $n$, i.e.~the anomalous dimension of double-particle operator in 
$\mathbb{S},\mathbb{A}$ have this property. Second, the space of one-loop functions in AdS 
can be restricted to diagrams whose seeds are the three crossing orientations 
of the two-loop ladder function $\phi^{(2)}$, whose functional form is given in terms of classical polylogarithms \cite{Usyukina:1992jd,Isaev:2003tk}.
These two facts will provide the basis of our bootstrap approach to constructing ${\cal H}^{(2)}_t$.
We will then show that such a one-loop function exists, and moreover 
it takes the very simple form
\begin{equation}\label{intro_Ft}
	{\cal H}^{(2)}_t = {\cal F}_t +  \sum_{i=1}^{4} \beta_i\,{B}_{i}\,,
\end{equation}
where ${\cal F}_t$ is given in terms of the two-loop ladder and its lower-weight completion. 
We shall see that ${\cal F}_t$ can be written as a differential operator acting on the two-loop ladder 
and hence admits a concise differential representation advocated in \cite{Huang:2024dck}.
On the other hand, the $B_i$ are tree-like ambiguities given in terms of $\dbar{}$-functions, and their coefficients cannot be fixed by the double-particle bootstrap alone. We will later show that two of them can be fixed by considering the flat-space limit, so that only two genuine ambiguities remain. However, the important point here is that the ambiguities do not contribute to \eqref{equation_crossing}, so by plugging our function into \eqref{equation_crossing} we find that there is only one value of $n$ that solves the constraint and this is $n=21$.

Of course, the way to leave $n$ unconstrained is to allow the one-loop function $\H_t^{(2)}$ to explicitly depend on $n$, thus making both RHS and LHS \eqref{equation_crossing} 
polynomials in $n$ of the same degree. Then, we
conclude that the most general parametrisation of the one-loop function is
\begin{equation}
{\cal H}^{(2)}_t = {\cal F}_t +  \sum_{i=1}^{4} \beta_i\,{B}_{i} + (n-21)\,{\cal R}_t\,,
\end{equation}
where the remainder function ${\cal R}_t$ crucially does not contribute to the $\log^2$ discontinuities in the flavoured reps $\mathbb{S}$ and $\mathbb{A}$, since these are already taken into account by ${\cal F}_t$. This requirement on ${\cal R}_t$ is again very constraining 
and as a result we shall see that necessarily it comes with a new ingredient beyond single-valued polylogarithms: 
a weight-three function  with an extra letter compared to the alphabet that describes 
single-valued polylogarithms. It is not accidental that precisely the function that 
multiplies $(n-21)$ has this feature, and in fact there is a clear physical picture for this which is manifest in Mellin space.

The Mellin transform of our one-loop correlator ${\cal H}^{(2)}_t$ defines a Mellin amplitude, 
denoted by $\M^{(2)}(s,t)$, which is a function of the two Mellin variables $s$ and $t$. In the flat-space 
limit \cite{Penedones:2010ue}, these Mellin variables become Mandelstam invariants of a $2\rightarrow2$ 
scattering process and our CFT result must be consistent with scattering massless tensor multiplets in 
six dimensions. The higher dimensionality can be seen already 
from \eqref{pred_OPE_intro} together with the fact that this quantity is constructed by 
summing over double-particle operators which crucially involve operators from all KK-levels of the S$^3$. 
In particular, an explicit computation shows that
\begin{equation}
\lim_{\rm flat\,space} {\cal F}_t\ \propto\  - s^2 {\cal B}^{\oneladder}_{\text{6d}}(s,t) + (s\leftrightarrow u)\,,
\end{equation}
where ${\cal B}^{\oneladder}_{\text{6d}}(s,t)$ is the 6d box diagram, and
\begin{equation}\label{singlelogintro}
\lim_{\rm flat\,space}{\cal R}_t\ \propto \ t \log\Big({-}\frac{t}{\mu^2}\Big)\,.
\end{equation}
Our result here can be checked against the flat-space amplitude recently 
constructed in \cite{Huang:2025nyr}. Remarkably,  
we find a perfect match for ${\cal B}^{\oneladder}_{\text{6d}}(s,t)$, 
including the overall factors that we are omitting above. 
In addition, this comparison with the known flat-space amplitude fixes two out of the four tree-like ambiguities in \eqref{intro_Ft}.
Finally, we also match the structure of the remainder when $n\neq 21$.
In particular,  the coefficient of the single-log term, i.e. the one in \eqref{singlelogintro} 
that depends on the renormalisation scale~$\mu$, vanishes for $n=21$ in which case the flat-space amplitude is ultraviolet (UV) finite, and descends from string theory. \\[-.4cm]

\noindent{\bf Outline of the paper:} 

In Section \ref{sec:basics}, we review basic 
facts about the $(2,0)$ supergravity on AdS$_3\times$S$^3$ coupled to $n$ tensor multiplets 
and its CFT$_2$ dual, with particular emphasis on the kinematics of the four-point correlation function of tensor multiplets. 
In Section \ref{sec:leading_log}, we compute the leading logarithmic discontinuity of our one-loop correlator 
using the double-particle bootstrap. To this end we first study the mixing of double-particle operators, 
reviewing the known anomalous dimensions of the flavoured double-particle operators \cite{Aprile:2021mvq}, 
and also computing new data in the singlet. As mentioned previously, we will compute the 
mixing matrix for the next-to-minimum twist double-particle operators at $h=\hb=2$. This is the 
first case in which  the mixing involves double-particle operators from the graviton multiplet, 
and the matrix is of size $3\times 3$. The resulting anomalous dimensions depend on $n$, but, 
surprisingly, they become rational numbers for $n=21$ and moreover one of the three anomalous dimensions vanishes.

In Section \ref{sec:one-loop} we collect all the available CFT data, in particular resumming the maximal 
logarithmic discontinuities in the flavoured reps, and we set up the strong version of the bootstrap 
problem for the one-loop function. We will explain that this version of the bootstrap translates 
into the assumption that our ansatz is built out of single-valued polylogarithms. Then, by solving 
the bootstrap equations we shall find the remarkable result that the value of $n$ is determined to 
be $n=21$. In Section \ref{sec_Mellin_amp}, we calculate the Mellin amplitude of the obtained 
one-loop function and show that it can be written as a differential representation with tree-like ambiguities. 
We also compute the flat-space limit and fix two of the four ambiguities. 
In Section \ref{sec:one-loop_extended} we come back to the one-loop function when $n\neq 21$, 
and repeat our bootstrap approach for the remainder function. We conclude in Section \ref{sec:discussion} 
with a summary of results, and a 
discussion, pointing out some interesting avenues for future work.

Lastly, Appendix \ref{APPA} contains details about the block expansion and the mixing problem, and in Appendix \ref{app:bpl} we discuss the bulk-point limit as an alternative formulation of the flat-space limit in Mellin space.

\section{Four-point function generalities}\label{sec:basics}
We shall start by reviewing some basic facts about the spectrum of six-dimensional $\mathcal{N}=(2,0)$ 
supergravity on AdS$_3\times$S$^3$ coupled to $n$ tensor multiplets, focusing on the half-BPS superconformal 
primary operators that are dual to the single-particle fields in the bulk \cite{Deger:1998nm,deBoer:1998kjm}.
These fields are the Kaluza-Klein modes on $S^3$ of the $(2,0)$ multiplets in six-dimensions, i.e. the graviton and the $n$ tensor multiplets.  We will collectively denote them by
\begin{equation}
{\cal O}_p\in\{ s_p^i, \sigma_p, V^+_p, V^-_p\}\,.
\end{equation} 
Each superconformal primary is characterised by its quantum numbers under 
$SL(2;\mathbb{R})_L\times SU(2)_L \times SL(2;\mathbb{R})_R\times SU(2)_R$, i.e.~the 
maximal bosonic subgroup of the superconformal group.
We denote these quantum numbers with 
$(h,j,\bar{h},\bar{j})$, see Table~\ref{table:qnumbers}. 
Note that $s_p^{i}$ and $\sigma_p$ are scalar primaries, whereas $V_p^\pm$ are of spin $\ell=1$, 
where the 2d spin is defined as $\ell = |h - \bar{h}|$.

The $s_p^{i}$ are the scalar primaries 
from  the tensor multiplets. The label 
$p\ge 1$ specifies the KK-level, and an index $i=1,\ldots,n$ that is a fundamental 
index of the $SO(n)$ flavour group. The primaries from the graviton 
multiplet are the scalars $\sigma_p$ and the vectors $V_p^+$ and $V_p^-$, where again the label $p$ specifies the KK-level. 
The  tower of $\sigma_p$ starts at $p=2$. The towers of $V_p^\pm$ start 
from $p=1$, however, the lowest modes $V_1^\pm$ are non dynamical.\footnote{
The $V_1^\pm$ are purely holomorphic or anti-holomorphic,  
since either $\bar{h}=\bar{j}=0$ or ${h}={j}=0$, thus their correlators are completely fixed by Ward identities and do not contain dynamical information.} 
We shall then group together $\sigma_{p}$ and $V^{\pm}_{p}$ for $p\ge 2$ as the dynamical
modes coming from graviton multiplet. 

\begin{table}
\begin{center}\def\arraystretch{1.5}
\begin{tabular}{|c||c|c|c|c||c|c|}\hline
 ${\cal O}_p$ & $h$ &  $j$ &  $\bar{h}$ & $\bar{j}$ & spin $\ell$ & $SO(n)$\\\hline
$s_p^i$ & $\frac{p}{2}$ &  $\frac{p}{2}$ &  $\frac{p}{2}$ &$\frac{p}{2}$ & 0 & $\mathbf{n}$ \\\hline
$\sigma_p$ & $\frac{p}{2}$ &  $\frac{p}{2}$ &  $\frac{p}{2}$ &$\frac{p}{2}$ & 0 & $\mathbf{1}$ \\\hline
$ V_p^+$ & $\frac{p+1}{2}$ &  $\frac{p+1}{2}$ &  $\frac{p-1}{2}$ &$\frac{p-1}{2}$ & 1 & $\mathbf{1}$ \\\hline
$ V_p^-$ & $\frac{p-1}{2}$ &  $\frac{p-1}{2}$ &  $\frac{p+1}{2}$ &$\frac{p+1}{2}$ & 1 & $\mathbf{1}$ \\\hline
\end{tabular}
\caption{Quantum numbers of the superconformal primary operators dual to the single-particle spectrum of 6d (2,0) supergravity on AdS$_3\times$S$^3$ coupled to $n$ tensor multiplets. The 2d spin $\ell$ is defined as $\ell = |h-\bar{h}|$.}
\label{table:qnumbers}
\end{center}
\end{table}

For each super-primary it will be convenient to contract the $SU(2)_L\times SU(2)_R$ R-symmetry 
indices with auxiliary polarisation vectors $w_\alpha$, $\bar{w}_{\dot{\alpha}}$. 
These are spin-$\frac{1}{2}$ representations of $SU(2)_L$, $SU(2)_R$, respectively.
In this way we define the operators,
\begin{align}\label{listofssigmaV}
\begin{split}
s_p^{i}({z,\bar{z},w,\bar{w}}) &= w_{\alpha_1}\cdots  w_{\alpha_p}\,\bar{w}_{\dot{\alpha}_1}\ldots \bar{w}_{\dot{\alpha}_p}
\,s_p^{{i};\alpha_1 \ldots \alpha_p \dot{\alpha}_1\ldots \dot{\alpha}_p}({z,\bar{z}})\,, \\[2pt]
\sigma_p({z,\bar{z},w,\bar{w}}) &= w_{\alpha_1}\cdots  w_{\alpha_p}\,\bar{w}_{\dot{\alpha}_1}\ldots \bar{w}_{\dot{\alpha}_p}\, \sigma_p^{\alpha_1 \ldots \alpha_p\dot{\alpha}_1\ldots \dot{\alpha}_p}({z,\bar{z}})\,, \\[2pt]
V_p^+({z,\bar{z},w,\bar{w}}) &= w_{\alpha}w_{\beta}\, w_{\alpha_1}\cdots  w_{\alpha_{p-1}}\,\bar{w}_{\dot{\alpha}_1}\ldots \bar{w}_{\dot{\alpha}_{p-1}}\,V_p^{+;\alpha \beta \alpha_1 \ldots \alpha_{p-1} \dot{\alpha}_1\ldots \dot{\alpha}_{p-1}}({z,\bar{z}})\,, \\[2pt]
V_p^-({z,\bar{z},w,\bar{w}}) &= \bar{w}_{\dot{\alpha}} \bar{w}_{\dot{\beta}} \, w_{\alpha_1}\cdots  w_{\alpha_{p-1}} \, \bar{w}_{\dot{\alpha}_1}\ldots \bar{w}_{\dot{\alpha}_{p-1}}\,V_p^{-;\alpha_1 \ldots \alpha_{p-1} \dot{\alpha}\dot{\beta }\dot{\alpha}_1\ldots \dot{\alpha}_{p-1}}({z,\bar{z}})\,.
\end{split}
\end{align}
By construction, the polarisation vectors are null, i.e.~$\epsilon^{\alpha\beta}\,w_\alpha w_\beta=\epsilon^{\dot{\alpha}\dot{\beta}}\,\bar{w}_{\dot{\alpha}}\bar{w}_{\dot{\beta}}=0$.
The null condition is preserved by rescalings, therefore we will make the following choice without loss of generality,
\begin{align}
	w_\alpha=\begin{pmatrix}1\\w\end{pmatrix},\qquad\bar{w}_{\dot{\alpha}}=\begin{pmatrix}1\\\bar{w}\end{pmatrix}.
\end{align}
Note that on the LHS of \eqref{listofssigmaV} we anticipated the use of $w,\bar{w}$ given above.

Contraction between different polarisation vectors gives, 
$w_{ij}\equiv \epsilon_{\alpha\beta}\,w_i^\alpha w_j^\beta=w_i-w_j\,,$ and similarly for $\bar{w}_{ij}=\bar{w}_i-\bar{w}_j$. Then $|w_{ij}|^2=w_{ij}\bar{w}_{ij}$. In the same way, 
the two-dimensional distance between spacetime coordinates is simply $|z_{ij}|^2=z_{ij}\bar{z}_{ij}$ where $z_{ij}\equiv z_i-z_j$ and $\bar{z}_{ij}\equiv \bar{z}_i-\bar{z}_j$. In the following, we will sometimes omit the dependence on the anti-holomorphic variables. 

The KK spectrum of six-dimensional supergravities that arise from (AdS$_3\times$S$^3)\times{\cal M}_4$ compactifications can be interpreted in terms of some two dimensional conformal field theory whose target space is the symmetric product $S^N({\cal M}_4)$ \cite{deBoer:1998kjm}. In our setup this conformal field theory is a SCFT$_2$ with $(4,4)$ superconformal symmetry, which is usually understood as a marginal deformation of the free orbifold theory $S^N({\cal M}_4)$. This theory has central charge $c=6N$, and from now on we will rewrite the large central charge expansion as the large-$N$ expansion.

\subsection{Four-point function of tensor multiplets}
The main focus of this work is the correlator $\langle s^{i_1}_1s^{i_2}_1s^{i_3}_1s^{i_4}_1\rangle$
of four tensor multiplets with lowest KK-level, $p=1$. By superconformal symmetry, we can extract 
a kinematical factor of propagators such that the correlator depends only on conformal and R-symmetry cross-ratios:
\begin{align}\label{sec1_1111}
	\langle s^{i_1}_1(z_1,w_1)s^{i_2}_1(z_2,w_2)s^{i_3}_1(z_3,w_3)s^{i_4}_1(z_4,w_4)\rangle 	=\frac{|w_{12}|^2|w_{34}|^2}{|z_{12}|^2|z_{34}|^2}\times G^{i_1i_2i_3i_4}(x,\bar{x},y,\bar{y})\,.
\end{align}
The relevant cross ratios are $x,\bar{x}$ and $y,\bar{y}$ defined as 
\begin{equation}\label{holo_cross}
\frac{z_{12} z_{34}}{z_{13} z_{24}}=x,\qquad\frac{w_{12} w_{34}}{w_{13} w_{24} }= y\,,
\end{equation}
and similarly for $\bar{x},\bar{y}$.
The variables $x,y$, as well as the variables $\bar{x},\bar{y}$ may be interpreted 
as the eigenvalues of a matrix-valued cross-ratio in analytic superspace \cite{Aprile:2021pwd}.  
The more standard conformal cross ratios, $(U,V)$, and the  R-symmetry 
cross ratios, $(\widetilde{U},\widetilde{V})$, are defined as
\begin{align}\label{eq:cross-ratios}
	U=\frac{|z_{12}|^2 |z_{34}|^2}{|z_{13}|^2 |z_{24}|^2} \,,\qquad V=\frac{|z_{14}|^2 |z_{23}|^2}{|z_{13}|^2 |z_{24}|^2}\,,\qquad	\widetilde{U}
	=\frac{|w_{12}|^2 |w_{34}|^2}{|w_{13} |^2 |w_{24}|^2 }\,,\qquad \widetilde{V}=  \frac{|w_{14} |^2 |w_{23}|^2 }{|w_{13} |^2 |w_{24}|^2 }\,.
\end{align}
The relation between \eqref{holo_cross} and \eqref{eq:cross-ratios} is
\begin{equation}
U=x \bar{x},\qquad V=(1-x)(1-\bar{x}),\qquad \widetilde{U}=y \bar{y},\qquad \widetilde{V}=(1-y)(1-\bar{y})\,.
\end{equation}
Note however that, unlike higher-dimensional CFTs, only the simultaneous exchange 
of $x\leftrightarrow \xb$ and $y\leftrightarrow \bar{y}$ is a symmetry of the correlator, i.e.~$G^{i_1i_2i_3i_4}(x,\bar{x},y,\bar{y})$.

We will consider the large-$N$ expansion of the correlator \eqref{sec1_1111} given by
\begin{align}\label{eq:largeNexp}
\begin{split}
	G^{i_1i_2i_3i_4} = G^{(0),i_1i_2i_3i_4} + \frac{1}{N}\,G^{(1),i_1i_2i_3i_4} + \frac{1}{N^2}\,G^{(2),i_1i_2i_3i_4} + \ldots\,.
\end{split}
\end{align}
The leading-order term is a disconnected contribution given by the product of two-point functions, which for unit normalised external operators reads
\begin{align}\label{eq:G0}
	G^{(0),i_1i_2i_3i_4}  = \delta^{i_1i_2}\delta^{i_3i_4}+\frac{U}{\widetilde{U}}\,\delta^{i_1i_3}\delta^{i_2i_4}+\frac{U\widetilde{V}}{\widetilde{U}V}\,\delta^{i_1i_4}\delta^{i_2i_3}\,.
\end{align}
In principle, all the other terms in the large-$N$ expansion can be derived from the AdS$_3$ 
effective action through an expansion in Witten diagrams. However, this AdS$_3$ effective action is only known up to cubic couplings \cite{Mihailescu:1999cj,Arutyunov:2000by,Behan:2024srh}, 
and moreover its derivation becomes increasingly more complicated. For example, the contact 
interactions relevant to compute four-point contact diagrams are not known. 
An alternative solution to this problem  was put forward in  \cite{Rastelli:2019gtj} by exploiting the fact that the diagrammatic expansion in AdS$_3$ does not manifest the superconformal Ward identities, 
\begin{equation}\label{eq:SCWI}
\Big[ (\partial_x +\partial_y)G^{i_1i_2i_3i_4}(x,\bar{x},y,\bar{y}) \Big]_{x=y} =0\,,\qquad\qquad \Big[ (\partial_{\xb} +\partial_{\yb})G^{i_1i_2i_3i_4}(x,\bar{x},y,\bar{y}) \Big]_{\xb=\yb} =0\,,
\end{equation}
which therefore can be used as a constraint. 

Following \cite{Rastelli:2019gtj}, 
the tree-level supergravity contribution is determined by imposing the superconformal Ward identity
on an ansatz built out of the relevant exchange and contact Witten diagram in AdS$_3$. 
The result can be written as 
\begin{align}\label{eq:G1}
	G^{(1),i_1i_2i_3i_4} &= -\delta^{i_1i_2}\delta^{i_3i_4}-\frac{U}{\widetilde{U}}\,\delta^{i_1i_3}\delta^{i_2i_4}-\frac{U\widetilde{V}}{\widetilde{U}V}\,\delta^{i_1i_4}\delta^{i_2i_3}
	+  (x-y)(\xb-\yb)\,\frac{U}{\widetilde{U}}\,\mathcal{H}^{(1),i_1i_2i_3i_4}(x,\xb)\,,
\end{align}
where
\begin{align}\label{eq:H1}	
	\mathcal{H}^{(1),i_1i_2i_3i_4}(x,\xb) &= \delta^{i_1 i_2}\delta^{i_3 i_4}\,\dbar{2211}+\delta^{i_1 i_4}\delta^{i_2 i_3}\,\dbar{1221}+\delta^{i_1 i_3}\delta^{i_2 i_4}\,\dbar{1212}\,,
\end{align}
with\footnote{We refer to \cite[Appendix D]{Arutyunov:2002fh} for the general definition of the $\dbar{}$-functions.}
\begin{align}\label{d1221scritta}
	\dbar{1221} = - \partial_V \Bigg[ \frac{\log(x\xb) (\log(1-x)-\log(1-\xb) )+ 2 \Li_2(x)-2\Li_2(\xb) }{ x-\xb}\Bigg],
\end{align}
and the other two $\dbar{}$-function in \eqref{eq:H1} are related by crossing transformations. An independent derivation of the above tree-level result has been obtained through a supergravity approach in \cite{Giusto:2018ovt,Giusto:2019pxc}.

We note that (up to a sign) the free-like contribution to 
$G^{(1),i_1i_2i_3i_4}$ in \eqref{eq:G1} is the same as the disconnected
contribution $G^{(0),i_1i_2i_3i_4}$.  This is somewhat surprising,
but recall that $G^{(1),i_1i_2i_3i_4}$ is simply the result of 
rewriting the relevant Witten diagrams by making manifest the superconformal 
Ward identities \eqref{eq:SCWI}, and therefore  there is no straightforward interpretation in terms of Wick contractions.
Alternatively, one can infer the free-like contribution in $G^{(1),i_1i_2i_3i_4}$ 
by matching the $O(1/N)$ contribution to the large-$N$ expansion of the 
identity super-Virasoro block.\footnote{
However, this matching at $O(1/N)$ only requires the product of the identity Virasoro block and the $U(1)$ affine block \cite{Fitzpatrick:2015zha,Hijano:2015qja,Giusto:2018ovt}.
It would be interesting to construct the full super-Virasoro block to all orders in $1/N$.}

Beyond tree level, we shall parametrise the one-loop contribution ${G}^{(2),i_1i_2i_3i_4}$ in the form
\begin{align}\label{generalG2}
G^{(2),i_1i_2i_3i_4} &= \G^{(2),i_1i_2i_3i_4}(x,\xb;y,\yb) +  (x-y)(\xb-\yb)\,\frac{U}{\widetilde{U}}\,\mathcal{H}^{(2),i_1i_2i_3i_4}(x,\xb)\,,
\end{align}
where the function $\G^{(2),i_1i_2i_3i_4}(x,\xb;y,\yb)$ is constrained by the Ward identity \eqref{eq:SCWI} and by the chiral algebra twist \cite{Rastelli:2019gtj}. 
Following the chiral algebra argument \cite{Beem:2013sza}, this function is expected to be a rational function.
It will be the purpose of this paper to compute the reduced correlator $\mathcal{H}^{(2),i_1i_2i_3i_4}(x,\xb)$.

\subsection{The superconformal block decomposition}

A key tool in our analysis  is the decomposition into superconformal blocks. For external scalar chiral primary operators there 
are three relevant superconformal blocks: half-BPS, semi-short and long.\footnote{In this context, semi-short is the same as $\frac{1}{4}$-BPS.} 
These have been constructed  in \cite{Aprile:2021mvq,Behan:2021pzk,Behan:2024srh}, and from their explicit expressions
one finds that $G^{}(x,\xb,y,\yb)$ in \eqref{sec1_1111}  (with $SO(n)$ indices omitted to simplify the notation)
can be written in the following form,
\begin{align}\label{eq:superconf_decomposition}
	G^{}(x,\xb,y,\yb) = \mathcal{C} + \big[(x-y)\,\mathcal{S}(x,y) + (\xb-\yb)\,\mathcal{S}(\xb,\yb)\big] + (x-y)(\xb-\yb)\,\frac{U}{\widetilde{U}}\,\Ht(x,\xb)\,,
\end{align}
where ${\cal C}$ is a constant, ${\cal S}(x,y)$ is a holomorphic function, while $\Ht(x,\xb)$ is a two-variable function (with no dependence on $y,\yb$, and the reason for this is that we are considering the correlator of the lowest KK-modes $s_1^i$).
Each term in \eqref{eq:superconf_decomposition} has a large-$N$ expansion that follows from \eqref{eq:largeNexp}. Moreover, the constant ${\cal C}$ and the function $\mathcal{S}(x,y)$ are special since they are determined by  
protected data.
With respect to \eqref{eq:superconf_decomposition}, a half-BPS block has coefficient functions in $\mathcal{C}$, $\mathcal{S}(x,y)$ and $\Ht(x,y)$. 
A semi-short block has coefficient functions in $\mathcal{S}(x,y)$ and $\Ht(x,\bar{x})$. 
Finally, a long block has  a coefficient function which contributes only to $\Ht(x,\bar{x})$. In each case, the  
functions involved depend non-trivially on the quantum numbers of the exchanged superconformal primary  operator.
The precise form of these superblocks can be found in \cite{Aprile:2021mvq,Behan:2024srh}, but it will not be needed here. For our purposes it will be enough to use an auxiliary block decomposition in terms of standard 2d bosonic blocks.

Shortly we will be interested in extracting the CFT data of long exchanged operators that are contained in ${\cal L}^{i_1i_2i_3i_4}(x,\xb)$.
Analogously to the correlator itself, we define the large-$N$ expansion
\begin{align}\label{eq:H_large_N}
	\Ht^{i_1i_2i_3i_4}(x,\xb) = \Ht^{(0),i_1i_2i_3i_4}(x,\xb) + \frac{1}{N}\,\Ht^{(1),i_1i_2i_3i_4}(x,\xb) + \frac{1}{N^2}\,\Ht^{(2),i_1i_2i_3i_4}(x,\xb) + \ldots\,.
\end{align} 
The leading contribution is obtained by rewriting $G^{(0)}$ from \eqref{eq:G0} in the form \eqref{eq:superconf_decomposition}, which yields
\begin{align}\label{eq:H0}
	\Ht^{(0),i_1i_2i_3i_4}(x,\xb) = \frac{1}{U}\,\delta^{i_1i_3}\delta^{i_2i_4}+ \frac{1}{UV}\,\delta^{i_1i_4}\delta^{i_2i_3}\,.
\end{align}
The tree-level contribution obtained from $G^{(1)}$ given in \eqref{eq:G1} reads
\begin{align}\label{eq:L1}
	\Ht^{(1),i_1i_2i_3i_4}(x,\xb) = -\frac{1}{U}\,\delta^{i_1i_3}\delta^{i_2i_4}- \frac{1}{UV}\,\delta^{i_1i_4}\delta^{i_2i_3}+\H^{(1),i_1i_2i_3i_4}(x,\xb)\,,
\end{align}
and it contains the $\dbar{}$ functions which are present in $\H^{(1),i_1i_2i_3i_4}(x,\xb)$, see \eqref{eq:H1}.

Before giving more details about the long operators and their 
contributions to ${\cal L}^{i_1i_2i_3i_4}$ ,  
we want to decompose the $SO(n)$ structure in irreducible representations. 
Let us recall that $s_1^{i}$ transforms 
in the vector representation $\mathbf{n}$ of $SO(n)$. Then, the 
OPE $s_1^{i_1}\times s_1^{i_2}$ splits into 
irreducible representations of $SO(n)$ which appear in the 
tensor product of the vector representation with itself,
\begin{align}
	\mathbf{n}\otimes\mathbf{n}=\mathds{1}\oplus\mathbb{S}\oplus\mathbb{A}\,.
\end{align}
In order, we find the singlet $\I$, the symmetric traceless $\mathbb{S}$, and the antisymmetric representation $\mathbb{A}$, with dimensions given by $\big\{1,\frac{n(n+1)}{2}-1,\frac{n(n-1)}{2}\big\}$, respectively. These irreps will often be referred to as the three different flavour channels.

At the level of the 4pt-function, it is convenient to introduce projectors onto these irreps: 
\begin{align}\label{eq:projectors_irreps}
\begin{split}
	P_{\I}^{i_1i_2i_3i_4} &= \frac{1}{n}\,\delta^{i_1 i_2}\delta^{i_3 i_4}\,,\\
	P_\mathbb{S}^{i_1i_2i_3i_4} &=\frac{1}{2}\big(\delta^{i_1 i_4}\delta^{i_2 i_3}+\delta^{i_1 i_3}\delta^{i_2 i_4}\big) - \frac{1}{n}\,\delta^{i_1 i_2}\delta^{i_3 i_4}\,,\\
	P_\mathbb{A}^{i_1i_2i_3i_4} &= \frac{1}{2}\big(\delta^{i_1 i_4}\delta^{i_2 i_3}-\delta^{i_1 i_3}\delta^{i_2 i_4}\big)\,.
\end{split} 
\end{align}
The projectors $P_\mathbf{a}^{i_1i_2i_3i_4}$, with $\mathbf{a}\in\big\{\mathds{1},\mathbb{S},\mathbb{A}\big\}$, are properly normalised and as such they satisfy the relations
\begin{align}
	P_\mathbf{a}^{i_1i_2i_3i_4}P_\mathbf{b}^{{i_4i_3}i_5i_6}=\delta_{\mathbf{a},\mathbf{b}}\,P_\mathbf{b}^{i_1i_2i_5i_6}\,,\qquad\text{Tr}(P_\mathbf{a})\equiv\delta_{i_1i_4}\delta_{i_2i_3}P_\mathbf{a}^{i_1i_2i_3i_4}=\text{dim}(\mathbf{a})\,.
\end{align}
By making use of these projectors we decompose $\Ht^{i_1i_2i_3i_4}(x,\xb)$ into the three flavour channels,
\begin{equation}\label{eq:H_projector_decomp}
	\Ht^{i_1i_2i_3i_4}(x,\xb)=\Ht_{\I}(x,\xb)\,P_{\I}^{i_1i_2i_3i_4}+\Ht_\mathbb{S}(x,\xb)\,P_\mathbb{S}^{i_1i_2i_3i_4}+\Ht_\mathbb{A}(x,\xb)\,P_\mathbb{A}^{i_1i_2i_3i_4}\,.
\end{equation}
Due to the orthogonality of projectors, the OPE decomposition can be treated separately in each invariant subspace, i.e.~in each flavour channel.  

The function $\Ht_{\mathbf{a}}(x,\xb)$ admits an expansion that descends 
from the decomposition of the correlator into the different superconformal blocks, i.e. involving components of the half-BPS, semi-short and long blocks. In all cases we can rewrite it as an auxiliary decomposition in terms of 2d bosonic blocks,
\begin{align}\label{eq:block_decomposition}
	\Ht_{\mathbf{a}}(x,\xb) = \sum_{h,\bar{h}} A_{\mathbf{a},h,\bar{h}} \,{B}_{h,\bar{h}}(x,\xb)\,,
\end{align}
with the following two ingredients:
\begin{itemize}
\item[1)] The coefficients $A_{\mathbf{a},h,\bar{h}}$ are given by the square of the OPE coefficients between the two 
external tensor multiplets and an exchanged long operator $\o_{\mathbf{a},h,\hb}$. Schematically,
\begin{equation}
{\mathbf{a},h,\bar{h}}\sim
 P_\mathbf{a}^{i_1i_2i_3i_4}\,C_{s^{i_1}_1s_1^{i_2}\o_{\mathbf{a},h,\hb}}C_{s_1^{i_3}s_1^{i_4}\o_{\mathbf{a},h,\hb}}\,.
\end{equation}
\item[2)]
The bosonic block ${B}_{h,\bar{h}}(x,\xb)$ reads\footnote{The 
factor $\frac{1}{U}$ is introduced to match our conventions in \eqref{eq:superconf_decomposition} where we defined ${\cal L}$ through 
$G\sim  U{\cal L}$. The combination $g_h (x)g_{\bar{h}}(\bar{x})+ g_{\bar{h}} (x)g_{h}(\bar{x})$ is then a conventional 2d conformal block with shifted weights $h,\hb$ \cite{Osborn:2012vt} .
}
\begin{equation}\label{eq:long_block}
	{B}_{h,\bar{h}}(x,\xb) = \frac{1}{x \bar{x}}\Big[ g_h (x)g_{\bar{h}}(\bar{x})+ g_{\bar{h}} (x)g_{h}(\bar{x})\Big]\,,\quad g_h (x) = x^{h} {}_2 F_1 (h+1,h+1,2h+2;x)\,.
\end{equation}
\end{itemize}
Let us note that since all long operators exchanged in $\langle s^{i_1}_1s^{i_2}_1s^{i_3}_1s^{i_4}_1\rangle\,$ are R-symmetry singlets
(their superprimaries have $j=\jb=0$) the corresponding long blocks only depend on $(h,\hb)$, i.e. the dimensions of the exchanged primary.

For later purposes, we note that the blocks $B_{h,\bar{h}}(x,\xb)$ in \eqref{eq:long_block} satisfy an eigenvalue equation which reads
\begin{align}\label{eq:eigenvalue_long_block}
	\dfour B_{h,\hb}(x,\xb) = h(h+1)\hb(\hb+1) B_{h,\hb}(x,\xb)\,.
\end{align}
The differential operator $\dfour$ is related to the Casimir of the conformal group acting on points 1 and 2, and is given by
\begin{align}\label{eq:dfour}
	\dfour (x,\xb)= \partial_x\partial_{\xb}(1-x)(1-\xb)\partial_x\partial_{\xb}(x\xb)^2\,.
\end{align}
Note that  $\dfour$ inherits the  $1 \leftrightarrow 2$ symmetry from the conformal blocks, namely it satisfies
\begin{equation}
\dfour (x', \xb')=  (1-x)^2(1-\xb)^2 	\dfour(x, \xb) \frac{1}{(1-x)^2(1-\xb)^2 }\,,
\end{equation}
where $x'\equiv x/(x-1)$, $\xb'\equiv \xb'/(\xb'-1)$.
Moreover, as we will explain later on, the eigenvalue of $\dfour$ on the RHS of 
\eqref{eq:eigenvalue_long_block} appears in the numerator of certain tree-level anomalous 
dimensions. We will then see that we can considerably simplify our results by making use of this differential operator.

\subsection{The OPE decomposition at large $N$}\label{subsec:large-N}
At leading order in the large-$N$ expansion, 
the operators $\o_{\mathbf{a},h,\hb}$ appearing in the $s_1^{i_1}\times s_1^{i_2}$ OPE
are the identity operator ${\bf 1}$ and an infinite tower of double-particle operators $[{\cal O}{\cal O}]_{\mathbf{a},h,\hb}$ in each flavour channel:
\begin{equation}
s_1^{i} (z_1)\times s_1^{\,j}(z_2) 
\sim \ {\bf 1} +
\sum_k P_{\I}^{ij} [{\cal O}{\cal O}]^{\phantom{ij}}_{\I,k} + \sum_k
	P_\mathbb{S}^{ij} [{\cal O}{\cal O}]^{\phantom{ij}}_{\mathbb{S},k}+ 
	\sum_k P_\mathbb{A}^{ij} [{\cal O}{\cal O}]^{\phantom{ij}}_{\mathbb{A},k}\, + O\!\left( \frac{1}{\sqrt{N}}\right).
\end{equation}
The latter are long operators with classical conformal weights $(h,\hb)$ which receive $1/N$ corrections:
\begin{align}\label{eq:OPE_data_large_N}
\begin{split}
	h'_{\a}&=h+\frac{\gamma^{(1)}_{\a} }{N}+\frac{\gamma^{(2)}_{\a}}{N^2}+\ldots\,,\qquad \hb'_{\a}=\hb+\frac{\gamma^{(1)}_{\a}}{N}+\frac{\gamma^{(2)}_{\a}}{N^2}+\ldots\,,
\end{split}
\end{align}
and their anomalous dimensions $\gamma^{(n)}_\a$ will depend on $h$ and $\hb$ as well as the flavour channel $\a$.
A similar large-$N$ expansion holds for the OPE coefficients in \eqref{eq:block_decomposition},
\begin{align}\label{eq:A_large_N}
A_{\a,h,\bar{h}} &= A^{(0)}_{\a,h,\bar{h}} + \frac{1}{N}\,A^{(1)}_{\a,h,\bar{h}} + \frac{1}{N^2}\,A^{(2)}_{\a,h,\bar{h}} + \ldots\,.
\end{align}
We will specify below the precise definition of $A^{(0)}_{\a,h,\bar{h}}$ in terms of three-point couplings.

In preparation for our one-loop bootstrap calculation, we recall that the superconformal 
block decomposition in \eqref{eq:block_decomposition} is a non-perturbative statement. 
It follows that the \emph{perturbative} large-$N$ expansion
inherits a rigid structure, and in particular, by plugging \eqref{eq:OPE_data_large_N}-\eqref{eq:A_large_N} back 
into \eqref{eq:block_decomposition}, one finds logarithmic contributions in the cross-ratio $U=x\xb$. 
Focusing on the maximal logarithmic contribution (leading log) at each order, we find
\begin{align}\label{eq:logu_stratification}
\begin{split}
	\Ht^{(0)}_\a(x,\xb)&=\sum_{h,\hb}\,\big[A^{(0)}_{\a}\big]_{h,\bar{h}}\,B_{h,\hb}(x,\xb)\,,\\[3pt]
	\Ht^{(1)}_\a(x,\xb)&=\log(U)\sum_{h,\hb}\,\big[A^{(0)}_{\a}\gamma^{(1)}_{\a}\big]_{h,\bar{h}}\,B_{h,\hb}(x,\xb)+\ldots\,,\\[3pt]
	\Ht^{(2)}_\a(x,\xb)&=\frac{1}{2}\log^2(U)\sum_{h,\hb}\,\big[A_\a^{(0)}(\gamma^{(1)}_\a)^2\big]_{h,\bar{h}}\,B_{h,\hb}(x,\xb)+\ldots\,,
\end{split}
\end{align}
where the lower logs are not important at this stage. A few additional comments are in order:

First, the spectrum of double-trace operators $[{\cal O}{\cal O}]_{\a,h,\hb}$ is actually \textit{degenerate}, meaning that 
generically there is more than one double-particle operator of given conformal weights $(h,\hb)$ 
and therefore these operators can mix. A precise counting of this degeneracy, denoted by $d_{\a}(h,\hb)$, 
will be provided in the next section. For now, we simply introduce the additional label $k=1,\ldots,d_{\a}(h,\hb)$ 
that runs over the degenerate operators, 
and define more carefully the quantities appearing in \eqref{eq:logu_stratification}:
\begin{align}
	\label{eq:average}
	\big[A^{(0)}_{\a}\big]_{h,\bar{h}}&=  \sum_{k=1}^{d_{\a}(h,\hb)}C_{s_1s_1[{\cal O}{\cal O}]_{\a,h,\hb,k}}^{(0)} C_{s_1s_1[{\cal O}{\cal O}]_{\a,h,\hb,k}}^{(0)}\,,\\
	\label{eq:average1}
	\big[A^{(0)}_{\a}\gamma^{(1)}_{\a}\big]_{h,\bar{h}}&=  \sum_{k=1}^{d_{\a}(h,\hb)}C_{s_1s_1[{\cal O}{\cal O}]_{\a,h,\hb,k}}^{(0)} \gamma^{(1)}_{\a,h,\bar{h},k} C_{s_1s_1[{\cal O}{\cal O}]_{\a,h,\hb,k}}^{(0)}\,,
\end{align}
where we reinstated all quantum numbers on the RHS.

Second, the log$^2(U)$ discontinuity at one loop is entirely determined by solving the mixing 
problem defined by \eqref{eq:average}-\eqref{eq:average1}. In fact, from the  knowledge of the individual leading-order OPE coefficients 
together with the tree-level anomalous dimensions in \eqref{eq:average}-\eqref{eq:average1},
$$C_{s_1s_1[{\cal O}{\cal O}]_{\a,h,\hb,k}}^{(0)}\,,\qquad \gamma^{(1)}_{\a,h,\bar{h},k}\,,$$
it is immediate to construct the quantity
\begin{align}\label{exactlythatequation}
\big[A^{(0)}_{\a}(\gamma^{(1)})^2\big]_{h,\bar{h}} =  \sum_{k=1}^{d_{\a}(h,\hb)}C_{s_1s_1[{\cal O}{\cal O}]_{\a,h,\hb,k}}^{(0)} \big(\gamma^{(1)}_{\a,h,\bar{h},k}\big)^2C_{s_1s_1[{\cal O}{\cal O}]_{\a,h,\hb,k}}^{(0)}\,.
\end{align}

Finally, since the $\log^2(U)$ discontinuity of $\H^{(2)}$ and that of $\Ht^{(2)}$ are identical, we find that
\begin{align}\label{eq:leading_log}
	\H^{(2)}_{\mathbf{a}}(x,\xb)\Big|_{\log^2(U)} =\frac{1}{2}\sum_{h,\bar{h}}\big[ A^{(0)}_{\a}(\gamma_{\a}^{(1)})^2\big]_{h,\bar{h}}\,B_{h,\hb}(x,\xb)\,.
\end{align}
Computing \eqref{eq:leading_log} is at the heart of our one-loop bootstrap approach, 
as it allows us to obtain a part of the one-loop correlator purely from lower-order OPE data.

\section{Unmixing supergravity and leading logs}\label{sec:leading_log}
In this section we discuss the exchanged double-particle operators 
in each of the three reps of $SO(n)$, namely $\mathds{1}$, $\mathbb{S}$, and $\mathbb{A}$. 
Then we will discuss the mixing problem at tree level. As stated earlier, resolving the mixing will allow us to compute the OPE coefficients in \eqref{exactlythatequation}
and determine the maximal log discontinuities of the one-loop correlator 
${\cal H}^{(2)}_{\a}(x,\xb)$ via \eqref{eq:leading_log}.\footnote{The solution of the mixing 
problem determines the maximal log discontinuities to all orders in the large-$N$ expansion, 
but for our purposes we will focus only on the one-loop contributions.}

\subsection{The spectrum of double-particle operators}\label{subsec:double-trace_spectrum}
To understand the exchanged double-particle operators $[{\cal O}{\cal O}]_{\a,h,\hb}$ 
and their degeneracy, recall that the spectrum of single-particle superprimaries ${\cal O}_p$
consists of four infinite towers of operators: $s_p^i$, $\sigma_p$, $V_p^+$, and $V_p^-$. 
These are the basic building blocks for describing multi-particle states in the supergravity 
low-energy theory, in particular the double-particle operators.

The relevant long operators exchanged in  the $\langle s^{i_1}_1s^{i_2}_1s^{i_3}_1s^{i_4}_1\rangle$ correlator 
are singlets under the R-symmetry, i.e.~they have $j=\jb=0$. 
The selection rules for the $SU(2)_L\times SU(2)_R$ subgroup of the superconformal group 
dictate that the constituent operators need to have the same values of $j$ and $\jb$. A quick 
look at the quantum numbers of the single-particle 
operators (c.f. Table \ref{table:qnumbers}) shows that these conditions leave us with 
four possible families of double-particle operators:
\begin{align}\label{eq:dt_list}
	:\!s_{p}^{i_1}\partial^n\bar{\partial}^ms_{p}^{i_2}\!:\,,\quad:\!\sigma_p\partial^n\bar{\partial}^m\sigma_p\!:\,,\quad:\!V_p^+\partial^n\bar{\partial}^mV_p^+\!:\,,\quad:\!V_p^-\partial^n\bar{\partial}^mV_p^-\!¨:\,.
\end{align}
The quantum numbers $(h,\hb)$ of these operators depend on $p$ and the number 
of holomorphic and antiholomorphic derivatives $n$ and $m$. 
In each case there is a space of degenerate operators,
and the next few paragraphs are dedicated to count how many operators there are for 
given quantum numbers $(h,\hb)$ in each representation $\a=\I,\S,\A$. 
Along the way we discuss what is known about the mixing problem.

\subsubsection*{Symmetric and antisymmetric channels}\vspace{-0.2cm}
In these two representations, the only relevant operators are built out of the tensor multiplets. 
The other double-particle operators are automatically singlets of $SO(n)$.
For given quantum numbers $(h,\hb)$ we find the following operators, 
\begin{align}\label{eq:dt_ops_SA}
\begin{matrix}
	\S: &\qquad P_{\S}^{i_1 i_2}\, :\!s_k^{i_1}\partial^{h-k}\bar{\partial}^{\bar{h}-k}s_k^{i_2}\!:\vert_{(0,0)}\,,\\[5pt] 
	\A: &\qquad P_{\A}^{i_1 i_2}\, :\!s_k^{i_1}\partial^{h-k}\bar{\partial}^{\bar{h}-k}s_k^{i_2}\!:\vert_{(0,0)}\,,
\end{matrix}\qquad k=1,\ldots,\frac{\tau}{2}\,,
\end{align}
where the twist $\tau$ is defined by $\tau\equiv\Delta-\ell=2\min(h,\hb)$. 
In \eqref{eq:dt_ops_SA}, the $SO(n)$ indices have been projected onto the desired rep, 
and the subscript $(0,0)$ indicates the projection onto the R-symmetry singlet $j=\jb=0$.

The total degeneracy for given $(h,\hb)$, denoted hereafter with $d_{\a}(h,\hb)$, is
\begin{align}\label{eq:degeneracy_SA}
	d_{\S,\A}(h,\hb)=\min(h,\hb)=\frac{\tau}{2}\,.
\end{align}
This is analogous to the counting of double-trace degeneracies in higher-dimensional 
holographic setups, see e.g. \cite{Aprile:2017xsp} for the analogous computation in AdS$_5\times$S$^5$ 
supergravity.\footnote{However, in contrast to CFTs in higher dimensions, note that specifying 
the quantum numbers $(\tau,\ell)$ of an operator in  $d=2$ is not equivalent to 
specifying the conformal weights $(h,\hb)$ due to the absolute value in the definition of the spin $\ell=|h-\hb|$. 
Nevertheless, the notion of twist and spin is still useful since the supergravity spectrum is symmetric 
under simultaneous exchange of $h\leftrightarrow\hb$ and $j\leftrightarrow\jb$. 
}

The double-particle operators \eqref{eq:dt_ops_SA} are expected to 
acquire an anomalous dimension at order $1/N$ induced by the tree-level supergravity 
contribution $\H^{(1)}$. This is indeed what happens, as discussed in \cite{Aprile:2021mvq}. 
Let us review the results of \cite{Aprile:2021mvq} in the current notation. First, 
the operator mixing can be resolved 
by considering the family of correlators $\langle s^{i_1}_ps^{i_2}_ps^{i_3}_qs^{i_4}_q\rangle$, $p,q\geq1$.
Then, the unmixed tree-level anomalous dimensions can be written as
\begin{align}\label{eq:anom_dims_S,A}
\begin{split}
	\gamma^{(1)}_{\S,k} &= -\frac{h(h+1)\hb(\hb+1)}{(\ell+2k-1)_2}\,,\qquad\ell=0,2,4,\ldots\,,\\
	\gamma^{(1)}_{\A,k} &= -\frac{h(h+1)\hb(\hb+1)}{(\ell+2k-1)_2}\,,\qquad\ell=1,3,5,\ldots\,.
\end{split}
\end{align}
These expressions are symmetric in $h\leftrightarrow\hb$, meaning that operators with $h$ and $\hb$ exchanged acquire the same anomalous dimension. One also notes that the formula is the same for both channels, the only difference is that the spin $\ell$ takes even or odd values depending on the symmetry of the representation.

We point out that the numerator of these
anomalous dimensions is given by a familiar combination of $h$ and $\hb$. It is precisely 
the eigenvalue of the differential operator $\dfour$ on the long blocks, c.f. \eqref{eq:eigenvalue_long_block}. 
As will be discussed in the next section, this fact will be useful for the computation 
of the leading log in the symmetric and antisymmetric channel.

Another fact that is useful to observe is the spin dependence in \eqref{eq:anom_dims_S,A}. In particular, 
there is a peculiarity for the lowest quantum numbers, $h=1$ or $\hb=1$, corresponding to twist $\tau=2$ and arbitrary spin $\ell$ (since $\tau=2\min(h,\hb)$ and $\ell=|h-\hb|$). 
In this case there is only one operator, given by $:\!s_1^{i_1}s_1^{i_2}\!:\,$, and its anomalous dimension actually does not depend on the spin because numerator and denominator simplify, 
yielding the result
\begin{equation}
\gamma^{(1)}_{\S,\A}\Big|_{\tau=2}= -2\,.
\end{equation}
It is only for twists $\tau\geq4$, i.e. for $\min(h,\hb)\ge2$, that one finds non-trivial spin dependence. 
For instance, at twist $\tau=4$ there are two operators and their anomalous 
dimensions are given by 
\begin{align}
\gamma^{(1)}_{\S,\A}\Big|_{\tau=4}= \left\{ \frac{6(\ell+2)}{\ell+4},\frac{6(\ell+3)}{\ell+1}\right\}.
\end{align}
More generally, one can check that the large-spin limit of the anomalous dimensions $\gamma^{(1)}_{\a,k}$ in  \eqref{eq:anom_dims_S,A} 
asymptotes to a constant, but otherwise the spin dependence is non-trivial.

We also quote the result for the unmixed three-point couplings that can be extracted from \cite{Aprile:2021mvq},
\begin{align}\label{3pointc11}
\begin{split}
\big(C_{s_1s_1[{\cal O}{\cal O}]_{k}}\big)^2 =\ &
\frac{\Gamma(h+1)^2\Gamma(\bar{h}+1)^2 }{
	\Gamma(2h+1) \Gamma(2\bar{h}+1)}\,\times \\
	&
\Bigg[\frac{4\Gamma(k+\frac{1}{2})}{\pi\Gamma(k)}\frac{(2\ell+4k-1 )}{h(h+1)\hb(\hb+1)}\frac{\Gamma(\ell+k+\frac{1}{2})}{\Gamma(\ell+k)} \frac{\Gamma(\frac{\tau}{2}-k+\frac{3}{2})\Gamma(\frac{\tau}{2}+\ell+k+1)}{\Gamma(\frac{\tau}{2}-k+1)\Gamma(\frac{\tau}{2}+\ell+k+\frac{1}{2})}\,  \Bigg].
\end{split}
\end{align}
Before moving on, it is worth emphasising 
that the CFT data computed above, namely the anomalous dimension in \eqref{eq:anom_dims_S,A} 
and the three-point couplings in \eqref{3pointc11}, are both $n$-independent. 
This will not be the case for the singlet channel, that we study next.

\subsubsection*{Singlet channel}\vspace{-0.2cm}
The situation in the singlet channel is considerably more involved, as all 
four types of operators listed in \eqref{eq:dt_list} can contribute. 
First, there are the  operators built out of the tensor multiples and projected onto the singlet: 
\begin{align}\label{eq:dt_ops_0}
	P_{\I}^{i_1 i_2}\, :\!s_k^{i_1}\partial^{h-k}\bar{\partial}^{\bar{h}-k}s_k^{i_2}\!:\vert_{(0,0)}\,,&\qquad k=1,\ldots,\frac{\tau}{2}\,.
\end{align}
Then, a careful analysis shows that the following additional operators all have the same quantum numbers:\footnote{We 
would like to thank Connor Behan and Rodrigo Pitombo for illuminating discussions on this point.}
\begin{align}\label{eq:dt_ops_I}
\begin{split}
	:\!\sigma_k\partial^{h-k}\bar{\partial}^{\bar{h}-k}\sigma_k\!:\vert_{(0,0)}\,,&\qquad k=2,\ldots,\frac{\tau}{2}\,,\\
	:\!V_k^{+}\partial^{h-k-1}\bar{\partial}^{\bar{h}-k+1}V_k^{+}\!:\vert_{(0,0)}\,,&\qquad k=2,\ldots,\min(h-1,\bar{h}+1)\,,\\
	:\!V_k^{-}\partial^{h-k+1}\bar{\partial}^{\bar{h}-k-1}V_k^{-}\!:\vert_{(0,0)}\,,&\qquad k=2,\ldots,\min(h+1,\bar{h}-1)\,.
\end{split}
\end{align}
For a given value of $(h,\hb)$, the total number of degenerate double-particle operators in the singlet channel is therefore given by summing the various degeneracies. For $h,\hb\geq1$, this leads to the total degeneracy
\begin{equation}\label{eq:degeneracy_I}
	d_{\I}(h,\hb)=\tau-1 +\max(\min(h-1,\bar{h}+1)-1,0)+ \max(\min (h+1,\bar{h}-1)-1,0)\,,
\end{equation}
where again, $\tau=2\min(h,\hb)$.
Compared to the degeneracy in the symmetric and antisymmetric channels, given above in \eqref{eq:degeneracy_SA}, one observes that the number of operators in the singlet grows much faster, see Table \ref{tab:degeneracies} for an explicit evaluation of \eqref{eq:degeneracy_I} for the first few values of $(h,\hb)$.
\definecolor{forestgreen(web)}{rgb}{0.13, 0.55, 0.13}
\begin{table}
\begin{center}
\begin{tabular}{|c|c|c|c|c|c|c|c|c}\hline
\diagbox{$h$}{$\hb$} \rule{0pt}{.7cm}& 0 & 1 & 2 & 3 & 4 & 5 & 6 & $\cdots$\\\hline
0 & $\times$ & $\times$ & $\times$ & $\times$ & $\times$ & $\times$ & $\times$ & $\cdots$\\\hline
1 & $\times$ & {\color{forestgreen(web)}1} & {\color{forestgreen(web)}1} & 2 & 2 & 2 & 2 & $\cdots$\\\hline
2 & $\times$ & {\color{forestgreen(web)}1} & {\color{forestgreen(web)}3} & 4 & 5 & 5 & 5 & $\cdots$\\\hline
3 & $\times$ & 2 & 4 & 7 & 8 & 9 & 9 & $\cdots$\\\hline
4 & $\times$ & 2 & 5 & 8 & 11 & 12 & 13 & $\cdots$\\\hline
5 & $\times$ & 2 & 5 & 9 & 12 & 15 & 16 & $\cdots$\\\hline
6 & $\times$ & 2 & 5 & 9 & 13 & 16 & 19 & $\cdots$\\\hline
$\vdots$ & $\vdots$ & $\vdots$ & $\vdots$ & $\vdots$ &$\vdots$  & $\vdots$ & $\vdots$ & $\ddots$
\end{tabular}
\caption{Number of degenerate double-trace operators in the singlet channel. The entries highlighted 
in {\color{forestgreen(web)}green} mark values of $(h,\hb)$ for which operators built from the primaries $V_p^{\pm}$ are absent.}
\label{tab:degeneracies}
\end{center}
\end{table}

Concerning the leading order anomalous dimensions $\gamma^{(1)}_{\I,k}$ of the 
operators \eqref{eq:dt_ops_0}-\eqref{eq:dt_ops_I}, the corresponding mixing problem has
not been explored yet. So far, there is a technical challenge due 
to the large amount of degeneracy together with the fact that the required correlators, still  
of the form $\langle ppqq\rangle$, involve all possible combinations 
of pairs built out of  
\begin{equation}
s_p^{i_1}s_p^{i_2}\,,\quad \sigma_p\sigma_p\,,\quad V_p^+V_p^+\,,\quad  V_p^-V_p^-\,,
\end{equation}
and, at present, the correlators involving the spin-1 superprimaries $V_p^\pm$ have not been 
calculated.\footnote{A six-dimensional generating function for these correlators was conjectured in \cite{Giusto:2020neo}. However, a more detailed analysis is required to explicitly 
extract the needed correlators and to corroborate their proposal.}

In this paper we shall make progress for the values of $(h,\hb)$ for which the double-particle 
operators involving $V_p^\pm$ are absent. These are the entries highlighted in 
green in Table \ref{tab:degeneracies}, i.e. the minimum twist operator with $(h,\hb)=(1,1)$ and the next-to-minimum twist operator with $(h,\hb)=(2,2)$.\footnote{
The operators with $(h,\hb)=(1,2)$ and $(2,1)$ have odd spin $\ell=1$ and are not exchanged in the R-symmetry singlet channel.}
In those two cases, all necessary ingredients to resolve the mixing are already available. 
The details of this calculation are given in Appendix \ref{app:unmixing}, where we use the known tensors-tensors \cite{Rastelli:2019gtj,Giusto:2020neo} and 
tensors-gravitons \cite{Giusto:2020neo} correlators and also 
the gravitons-gravitons correlators in \cite{Behan:2024srh}. Here we only quote the results for the anomalous dimensions:
\begin{itemize}
\item $(h,\hb)=(1,1)$: There is only one operator, $:\!s_1^{}s_1^{}\!:\,$, and its anomalous dimension is given by
\begin{align}\label{eq:anom_dim_I1}
	\gamma^{(1)}_{\I,k=1}=-2+\frac{n}{3}\,.
\end{align}
This agrees with the result obtained in \cite{Ceplak:2021wzz}.

\item $(h,\hb)=(2,2)$: Here we find three degenerate operators, $k=1,2,3$, since we count
\begin{equation}
:\!s_1^{}\partial\bar{\partial}s_1^{}\!:\,,\qquad :\!s_2^{}s_2^{}\!:\,,\qquad :\!\sigma_2\sigma_2\!:.
\end{equation} Their anomalous dimensions read 
\begin{align}\label{eq:anom_dim_I2}
	\vec{\gamma}^{(1)}_{\I}=\Big\{3(n-6),\,-\tfrac{3}{5}\big(13+\sqrt{13^2+5(n-21)}\big),\,-\tfrac{3}{5}\big(13-\sqrt{13^2+5(n-21)}\big)\Big\}\,.\,
\end{align}
This is a new result. Remarkably, it has the striking feature that 
for $n=21$ the anomalous dimensions become \textit{rational} and, 
in particular, one of them \textit{vanishes}:
\begin{align}\label{eq:anom_dim_I3}
	\vec{\gamma}^{(1)}_{\I}\Big|_{n=21}=\Big\{45,\,-\tfrac{78}{5},\ 0\Big\}\,.\,
\end{align}
\end{itemize}
This is the first time we see that the value $n=21$ is special. To have a better understanding of what is going on, let us write the unmixing problem in matrix form,
\begin{equation}
\bigg( \big[A^{(0)}_{\I}\big]_{2,2} \bigg)_{ppqq}= \mathbb{C}.\mathbb{C}^T\,,\qquad
	\bigg( \big[A^{(0)}_{\I}\gamma^{(1)}_{\I}\big]_{2,2}\bigg)_{ppqq}= \mathbb{C}.{\rm diagonal}\big[\vec{\gamma}^{(1)}_{\I}\big].\mathbb{C}^T\,,
\end{equation}
where the anomalous dimension are given in \eqref{eq:anom_dim_I2}. The matrix of three couplings, for $n=21$, reads
\begin{equation}\label{matrix3ptcou}
\mathbb{C}\Big|_{n=21}=\left[\begin{array}{ccc} 
C^{(0)}_{s_1s_1 [k=1] }  & C^{(0)}_{s_1s_1 [2]} &  C^{(0)}_{s_1s_1 [3]}  \\[.2cm]  
C^{(0)}_{s_2s_2 [1] }  & C^{(0)}_{s_2s_2 [2]} &  C^{(0)}_{s_2s_2 [3]}  \\[.2cm]
C^{(0)}_{\sigma_2\sigma_2 [1] }  & C^{(0)}_{\sigma_2\sigma_2 [2]} &  C^{(0)}_{\sigma_2\sigma_2 [3]}  
\end{array}\right] = 
\left[\begin{array}{ccc} 
\frac{-1}{6\sqrt{5}} &  \frac{-1}{3\sqrt{26}}  & \frac{-\sqrt{7}}{\sqrt{390}}  \\
\frac{1}{3\sqrt{5}} &  \frac{-1}{6\sqrt{26}}  & \frac{-\sqrt{7}}{2\sqrt{390}}  \\
0 &  \frac{-\sqrt{21}}{2\sqrt{26}}  & \frac{\sqrt{5}}{2\sqrt{26}}  \end{array}\right].
\end{equation}\\[0cm]
It is clear from the numbers in \eqref{matrix3ptcou} that there are indeed three exchanged operators. However, note that 
$\big[A^{(0)}_{\I}\gamma^{(1)}_{\I}\big]_{2,2}$ only sees two of them because one operator has a vanishing anomalous dimension.
As a result, the explicit form of $\big[A^{(0)}_{\I}\gamma^{(1)}_{\I}\big]_{2,2}$ depends only on the $2\times3$ submatrix of $\mathbb{C}$ 
(i.e. with the last column containing $C_{{\cal O}_p{\cal O}_p[3]}$ removed).
Since anomalous dimensions do typically not vanish by accident, our result strongly calls for a greater explanation.

In the flat-space theory, the value of $n=21$ 
has long been understood as a consequence of the cancellation of 
the gravitational anomaly \cite{Erler:1993zy}. More recently \cite{Huang:2025nyr} has shown 
that $n=21$ also renders one-loop amplitudes UV finite.\footnote{It has been proposed in \cite{Kallosh:2026roi} that the UV finiteness may be understood as a consequence of the absence of the superconformal anomaly in the associated 6D (2,0) \emph{conformal supergravity} theory. See also \cite{Kallosh:2025pmu} for a possible explanation involving the $SO(5,21)$ symmetry of the theory.}
However it is not immediately obviously how to relate these facts to our computation in \eqref{eq:anom_dim_I3}.

We will come back to this point in our final discussion in Section \ref{sec:discussion}.

\subsection{Predictions for the leading log via the OPE}\label{subsec:leading_logs}
Having understood the degeneracy in the spectrum of long double-particle 
operators $[{\cal O}{\cal O}]_{\a,h,\hb,k}$,  we can proceed to explicitly calculate 
the leading-log contributions discussed in Section \ref{subsec:large-N}  
at one-loop order. In each flavour channel $\a=\I,\S,\A$, these predictions read
\begin{align}\label{eq:leading_log2}
	\H^{(2)}_{\mathbf{a}}(x,\xb)\Big|_{\log^2(U)} =\frac{1}{2}\sum_{h,\bar{h}}\big[ A^{(0)}_{\a}(\gamma_{\a}^{(1)})^2\big]_{h,\bar{h}}\,B_{h,\hb}(x,\xb)\,.
\end{align}
To compute this quantity  we can either use the explicit results from the unmixing, 
when they are available, 
or we can streamline the computation as we now describe. 
Despite the intrinsic interest in unmixing the  double-particle spectrum, 
the streamlined computation is applicable even without a detailed understanding 
of the individual three-point couplings and anomalous dimensions. 
This is of course redundant for the symmetric and antisymmetric channels where we have explicit data, see~\eqref{eq:anom_dims_S,A}-\eqref{3pointc11},
but it will turn out to be very useful in discussing the singlet.

For illustration, let us consider the example of a single exchanged operator, as is 
the case for $h=\hb=1$ for instance. Obvious manipulations in that case yield
\begin{equation}
	\big[A^{(0)}_{\a}(\gamma^{(1)}_{\a})^2\big]_{h=1,\bar{h}=1}
	=
	\bigg( C^{(0)}_{11;\a}(\gamma^{(1)}_{\a})C^{(0)}_{pp;\a} \bigg) 
	\Bigg(\frac{1}{  C^{(0)}_{pp;\a} C^{(0)}_{pp;\a}}\Bigg)
	\bigg( C^{(0)}_{pp;\a} (\gamma^{(1)}_{\a}) C^{(0)}_{11;\a}\bigg)\quad{\rm with}\ p=1\,.
\end{equation}
On the RHS we recognise two factors of $[A^{(0)}_{\a}\gamma^{(1)}_{\a}]_{h=1,\hb=1}$ 
and the inverse factor  $[A^{(0)}_{\a}]_{h=1,\hb=1}$, both extracted from $\langle s_1s_1s_1s_1\rangle $, 
the first at tree-level and the second from the disconnected contribution.  This shows that we can compute the 
quantity we are interested in, $[A^{(0)}_{\a}(\gamma^{(1)}_{\a})^2]_{h=1,\bar{h}=1}$, 
directly from the coefficients of a block decomposition. 
This same idea can be exploited in the general case.  In fact, the formula that takes into 
account the exchange of degenerate operators is a natural generalisation of the one above:\footnote{More details about the derivation of this formula can be found in e.g. \cite[Section 3.1]{Aprile:2019rep}.}
\begin{equation}\label{eq:leading_log4}
	\bigg(\big[A^{(0)}_{\a}(\gamma^{(1)}_{\a})^2\big]_{h,\bar{h}}\bigg)_{1111}= \sum_{p=1}^{d_{\bf a}(h,\hb)}
	\bigg(\big[A^{(0)}_{\a}\gamma^{(1)}_{\a}\big]_{h,\hb} \bigg)_{\!11pp}
	\bigg(\big[A^{(0)}_{\a}\big]_{h,\hb} \bigg)^{-1}_{\!pppp}
	\bigg(\big[A^{(0)}_{\a}\gamma^{(1)}_{\a}\big]_{h,\hb} \Bigg)_{\!pp11}\,. 
\end{equation}
As in the previous example, in order to compute \eqref{eq:leading_log4}  we only need to know the block decomposition of two quantities: the tree-level correlators
\begin{equation}
\langle s_1^{i_1}s_1^{i_2} {\cal O}_p {\cal O}_p\rangle^{(1)}\,,
\end{equation}
where ${\cal O}_p\in\{ s_p^i, \sigma_p, V^+_p, V^-_p\}$ with $p$ spanning the degenerate 
space of operators for given quantum numbers  $(h,\hb)$, and the disconnected contributions from
\begin{equation} \label{eq:leading_log4444}
\langle {\cal O}_p {\cal O}_p {\cal O}_p {\cal O}_p\rangle^{(0)}\,. 
\end{equation}
Therefore, formula \eqref{eq:leading_log4} is a shortcut which requires only minimal input 
compared to the full unmixing problem, since in that case we would also need to consider other correlators of the form 
$\langle {\cal O}_p {\cal O}_p {\cal O}_q {\cal O}_q\rangle^{(1)}$.

We will now put \eqref{eq:leading_log4} to use and compute \eqref{eq:leading_log2} for each 
of the three flavour channels.

\subsubsection*{Symmetric and antisymmetric channels}\vspace{-0.2cm}
The double-particle operators in $\S,\A$ 
are built only from the tensor multiplet operators. 
Therefore, the needed tree-level correlators entering \eqref{eq:leading_log4} are
\begin{equation}
\langle s_1^{i_1}s_1^{i_2}s_p^{i_3}s_p^{i_4}\rangle^{(1)}\,,
\end{equation}
whose explicit expressions are given in Appendix \ref{app:KK_correlators}. Using the blocks from \eqref{eq:long_block}, we find that 
\begin{align}\label{eq:A0g1_11pp}
	\bigg(\big[A^{(0)}_{\a}\gamma^{(1)}_{\a}\big]_{h,\hb} \bigg)_{\!11pp}
	= -\frac{\Gamma(h+1)^2\Gamma(\bar{h}+1)^2}{\Gamma(p+1)^2 \Gamma(2h+1)\Gamma(2\bar{h}+1)} \frac{\Gamma(\frac{\tau}{2}+p+1)}{\Gamma(\frac{\tau}{2}-p+1)}\,.
\end{align}
In addition, we need the block decomposition of 
$\langle s_p^{i_1}s_p^{i_2}s_p^{i_3}s_p^{i_4}\rangle^{(0)}$. The order $O(c^0)$ is just 
given by disconnected two-point functions, 
from which we find
\begin{align}\label{eq:A0_pppp}
\bigg(\big[A^{(0)}_{\a}\big]_{h,\hb} \bigg)^{}_{\!\!pppp}
	= \frac{1}{p!^4}\frac{\Gamma(h)\Gamma(h+1)\Gamma(h+p+1)\Gamma(\hb)\Gamma(\bar{h}+1)\Gamma(\bar{h}+p+1) }{
	(h+1)\Gamma(2h+1)\Gamma(h+1-p) (\hb+1)\Gamma(2\bar{h}+1)\Gamma(\bar{h}+1-p)}\,.
\end{align}
Note that the above expressions are the same for both channels $\a=\S$,$\A\,$, but only 
even (odd) spins $\ell=|h-\hb|$ are exchanged in the symmetric (antisymmetric) channel.

With the OPE data from \eqref{eq:A0g1_11pp}-\eqref{eq:A0_pppp} one can assemble
\begin{equation}\label{H22sym_and_anty}
	\H^{(2)}_{\mathbf{a}}(x,\xb)\Big|_{\log^2(U)} = \frac{1}{2}\sum_{h,\bar{h}} \bigg(
	\sum_{k=1}^{d_\a(h,\hb)} A^{(0)}_{\a,h,\hb,k}(\gamma^{(1)}_{\a,k})^2
	\bigg)\, B_{h,\hb}(x,\xb)\,,
\end{equation}
and perform the summation over blocks.
Instead of giving the resummations right away, 
we note that numerator of the anomalous dimensions $\gamma^{(1)}_{\a,k}$  in \eqref{eq:anom_dims_S,A}
contains a factor of the eigenvalue of $\dfour$, therefore we expect that pulling out a power of $\dfour$ 
leads to simpler expressions for \eqref{H22sym_and_anty}.
In fact, in the symmetric channel, we find
\begin{align}\label{eq:leading_log_res_S}
	\mathcal{H}^{(2)}_{\mathbb{S}}\Big|_{\log^2(U)} = \dfour \bigg[\frac{(1-U-V)(\log^2(1-x)-\log^2(1-\xb))-4U(\text{Li}_2(x)-\text{Li}_2(\xb))}{4(x-\xb)^3}\bigg]\,,
\end{align}
while for the antisymmetric channel we find
\begin{align}\label{eq:leading_log_res_A}
\begin{split}
	\mathcal{H}^{(2)}_{\mathbb{A}}\Big|_{\log^2(U)} &= \dfour\bigg[\frac{\log(1-x)+\log(1-\xb)}{(x-\xb)^2}\\
	&\qquad\quad~+\frac{(1-U-V)(\log^2(1-x)-\log^2(1-\xb))-4(V-1)(\text{Li}_2(x)-\text{Li}_2(\xb))}{4(x-\xb)^3}\bigg]\,.
\end{split}
\end{align}
It should be noted that after the action of $\dfour$ the power of the denominator increases, 
and moreover, an additional factor of $1/V$ is generated. The final denominator then reads 
$(x-\xb)^7V$. Note also that despite the apparent  poles at $x=\xb$ these functions have a 
regular expansion at $x=\xb$, as expected for a Euclidean correlator. 
Furthermore, let us point out a peculiarity in the twist 2 contributions, namely one finds that 
they contain no $\log^2(1-x)$ terms.\footnote{One can see this explicitly from the small $x,\bar{x}$ expansion of the leading log coefficient functions in \eqref{eq:leading_log_res_S}
and \eqref{eq:leading_log_res_A}. For the first one we find
\begin{align}\notag
	\mathcal{H}^{(2)}_{\mathbb{S}}\Big|_{\log^2(U)}=\frac{1}{1-x}+\bar{x} \bigg[ \frac{6}{x} + \frac{4}{1-x}+\frac{18\log(1-x)}{x^2} - \frac{6(-2+x)\log^2(1-x)}{x^3}\bigg]+\ldots\,,
\end{align}
and for the other
\begin{align}\notag
\begin{split}
	\mathcal{H}^{(2)}_{\mathbb{A}}\Big|_{\log^2(U)}=&\ \bigg[ \frac{2}{x}+\frac{1}{1-x}+\frac{2\log(1-x)}{x^2} \bigg]\ +\\
	&\ \ \bar{x}\bigg[ \frac{2}{x} + \frac{4}{1-x}+\frac{48}{x^2}+\frac{(96-46x)\log(1-x)}{x^3} - \frac{6(-2+x)(\log^2(1-x)+4 {\rm Li}_2(x))}{x^3}\bigg]  + \ldots\,.
\end{split}
\end{align}
The contributions at order $O(\xb^0)$ correspond to the exchange of twist $2$ double-particle operators and do not contain $\log^2(1-x)$ or ${\rm Li}_2(x)$ terms, 
which instead appear for the first time at linear order in $\xb$, corresponding to twist $4$.
}
Physically, this is a reflection of the fact that the anomalous dimensions of the twist $2$ double-particle operators are exactly constant as function of spin. 
The fact that \eqref{H22sym_and_anty} involves a sum over degenerate operators with spin-dependent anomalous dimensions is therefore crucial to obtaining a $\log^2(1-x)$ contribution, as well as the other weight-two contribution proportional to ${\Li}_2(x)$.

The results in \eqref{eq:leading_log_res_S} and  \eqref{eq:leading_log_res_A} constitute predictions for the one-loop correlator in the symmetric and antisymmetric channels.

\subsubsection*{Singlet channel}\vspace{-0.2cm}
The double-particle operators that contribute to the singlet channel are built from all four types of single-particle superprimaries, thus we have to consider the tensor fields, $s^i_p$, and the graviton fields, $\sigma_p$, $V_p^{\pm}$, see \eqref{eq:dt_ops_0}-\eqref{eq:dt_ops_I}. As already mentioned, differently from the symmetric and anti-symmetric reps, in the singlet channel we cannot unmix the double-particle operators for all values of $(h,\hb)$ because the tree-level correlators with external vector primaries $V_p^\pm$ are not available.

What we can do instead is to compute the predictions for the $\log^2(U)$ coefficients for the two values $(h,\hb)=(1,1)$ and $(2,2)$, since in these cases we only need input from tensor-tensor and tensor-graviton correlators. We find,
\begin{align}\label{eq:leading_log_res_singlet}
	\big[A^{(0)}_{{\I}}(\gamma_{{\I}}^{(1)})^2\big]_{h,\hb} = \begin{cases}
	~\frac{1}{4}\big({-}2+\frac{n}{3}\big)^2\,, &\quad (h,\hb)=(1,1)\,,\\[5pt]
	~\frac{1}{100}\big(5 n^2-56 n+200\big)\, &\quad (h,\hb)=(2,2)\,.\\
\end{cases}
\end{align}
For convenience, we have recorded the necessary lower-order OPE data to arrive at these expressions in Appendix \ref{app:unmixing}.

A distinct feature of the OPE predictions \eqref{eq:leading_log_res_singlet}
is that they depend \textit{quadratically} on $n$, whereas the $n$-dependence introduced by the projectors is only \textit{linear}. This fact will be of crucial importance later on. We shall find that, even though we only have partial information about the $\log^2(U)$ part in the singlet channel, we will still be able to provide strong enough constraints on the one-loop correlator.

Even if we cannot solve the unmixing for generic values of $h,\hb$,  
we can still say something important about the  $n$-dependence. 
The streamlined version of the predictions discussed in \eqref{eq:leading_log4}-\eqref{eq:leading_log4444} is now very useful.  
Taking into account the degeneracy of double-particle operators, we have to consider the following family of tree-level correlators, 
\begin{equation}\label{again11ppsinglet}
\langle s_1^{i_1}s_1^{i_2} {\cal O}_p {\cal O}_p\rangle^{(1)}\,,
\end{equation}
labelled by $p=1,\ldots,d_{\I}(h,\bar{h})$. It is convenient to split  the collective field ${\cal O}_p$ into tensor fields and graviton fields. 
Schematically the prediction in \eqref{eq:leading_log4} will be 
given by summing over three types of sums, 
\begin{align}\label{explainspin_bigger0}
	\bigg(
	\big[A^{(0)}_{\I}(\gamma^{(1)}_{\I})^2\big]_{h,\bar{h}}\bigg)_{1111}  
	= &\  \sum_{\textrm{tensors}\, s_p\phantom{~^{\pm}}}^{}\!\!\!\!\!\!\bigg(\ldots \bigg)_{} \ 
	+ \!\! \sum_{\textrm{gravitons}\,\sigma_p\phantom{~^{\pm}}}^{}\!\!\!\!\!\!\!\bigg(\ldots \bigg)_{}\ +\!\!
	\sum_{\textrm{gravitons}\,V^{\pm}_p}^{} \!\!\bigg(\ldots \bigg)_{},
\end{align}
where each parenthesis on the RHS is build by `squaring' the tree-level correlator 
and `dividing' by disconnected free theory, as in \eqref{eq:leading_log4}. One can check that free theory brings no $n$ dependence. 
Therefore, all the $n$ dependence comes from the tree-level correlators.  
For the first two terms in \eqref{explainspin_bigger0}, we will find the following structure
\begin{align}
\label{11pptensorsinglet1}
\bigg(\big[A^{(0)}_{\I}\gamma^{(1)}_{\I}\big]_{h,\hb} \bigg)_{\!11s_ps_p} &=\ \,n\ c_p(h)\,\delta_{h,\hb} + \bigg(\big[A^{(0)}_{\S}\gamma^{(1)}_{\S}\big]_{h,\hb} \bigg)_{\!11s_ps_p}, \\
\label{11pptensorsinglet2}
\bigg(\big[A^{(0)}_{\I}\gamma^{(1)}_{\I}\big]_{h,\hb} \bigg)_{\!11\sigma_p\sigma_p}  &= \sqrt{n}\,(p^2-1)\,{c}_p(h)\,\delta_{h,\hb}\,.
\end{align}
The case of $\langle 11\sigma_p\sigma_p \rangle^{(1)}$ is special because it gives a contribution 
that is proportional to that of the tensor-tensor correlators at $h=\hb$ (spin-zero), 
as pointed out in \cite{Behan:2024srh}. This is why we used the same $c_p(h)$ in both \eqref{11pptensorsinglet1} and \eqref{11pptensorsinglet2}. 
Finally, the last term in  \eqref{explainspin_bigger0}  involves $V^{\pm}$ and it has the structure
\begin{equation}
\label{11pptensorsinglet3}
\bigg(\big[A^{(0)}_{\I}\gamma^{(1)}_{\I}\big]_{h,\hb} \bigg)_{\!11V^{\pm}_pV^{\pm}_p}  = \sqrt{n}\ {c}^{\pm}_p(h,\hb)\,.
\end{equation}
In particular, the sum over $V^{\pm}$ correlators in \eqref{explainspin_bigger0}
contributes only linearly in $n$.

The conclusion of the above discussion is that we can further refine the $n$-dependence of 
the OPE predictions at one-loop order by distinguishing the case $h=\hb$ from $h\neq\hb$:
\begin{align}
	\label{refine_singletpred1}
	\bigg(\big[A^{(0)}_{\I}(\gamma^{(1)}_{\I})^2\big]_{h= \bar{h}}\bigg)_{1111}  = &\  \sum_{\textrm{tensors}\, s}^{}\!\!
	\bigg(\ldots \bigg)_{} \ + \!\! \sum_{\textrm{gravitons}\,\sigma, V^{\pm}}^{}\!\!\bigg(\ldots \bigg)_{} = \textrm{polynomial in \emph{n} of deg}\,\leq 2 \,,\\[.2cm]
	\label{refine_singletpred2}
	\bigg(\big[A^{(0)}_{\I}(\gamma^{(1)}_{\I})^2\big]_{h\neq \bar{h}}\bigg)_{1111}  = &\  \sum_{\textrm{tensors}\, s}^{}\!\!
	\bigg(\ldots \bigg)_{} \ + \!\! 
	\sum_{\textrm{gravitons}\,V^{\pm}\phantom{\,\sigma,}}^{} \!\!\bigg(\ldots \bigg)_{} = \textrm{polynomial in \emph{n} of deg}\leq 1\,.
\end{align}
This structure, and in particular the dependence on $n$, 
should match the structure that one obtains by projecting the correlator onto the singlet channel, 
thus
\begin{equation}\label{againequaconstraintn}
{\cal H}^{(2)}_{\I}\Big|_{\log^2(U)} = 
\bigg[{\cal H}^{(2)}_{t}+ {\cal H}^{(2)}_{u}+ n {\cal H}^{(2)}_{s}\bigg]_{\log^2(U)} =\frac{1}{2}\sum_{h,\bar{h}}\big[ A^{(0)}_{\I}(\gamma_{\I}^{(1)})^2\big]_{h,\bar{h}}\,B_{h,\hb}(x,\xb)\,.
\end{equation}
In the Introduction, see \eqref{equation_crossing}, we wrote the same equation as above, 
and for simplicity we highlighted only the first contribution at $h=\hb=1$. Now we see that the latter is the first 
of a tower of predictions at $h=\hb$ for $h\ge 1$, whose $n$ dependence is at most quadratic. 
For $h\neq h$ we also have predictions that are $n$ dependent, 
but these depend at most linearly in $n$. Crucially, crossing symmetry of our correlator implies that out of the three 
functions that parametrise the reduced correlator, 
\begin{equation}
	\mathcal{H}^{i_1i_2i_3i_4}(x,\xb) = \delta^{i_1i_2}\delta^{i_3i_4}\,{\cal H}_s+ \delta^{i_1i_4}\delta^{i_2i_3}\,{\cal H}_t + \delta^{i_1i_3}\delta^{i_2i_4}\,{\cal H}_u \,,
\end{equation}
only one is independent. Therefore, the claim is the following: if an $n$-independent one-loop function exists, 
the  $h=\hb$ sector of the OPE equations in \eqref{againequaconstraintn}, 
through  \eqref{refine_singletpred1}, put a constraint on the value of $n$, which implies that $n$ cannot be arbitrary. 
In the next section we will answer the questions: What is this one-loop function, and what is this value of $n$?

\section{The one-loop bootstrap}\label{sec:one-loop}
In this section we set up a bootstrap problem to compute the reduced one-loop correlator for $\langle s_1s_1s_1s_1\rangle$.
Our approach will be to make a list of all constraints from crossing symmetry and the OPE, as the $\log^2(U)$ 
discontinuity computed in the previous section as well as other predictions from the unmixing, 
and then construct a function that solves all these constraints in a non-trivial way.

Given our one-loop function, we shall find that the consistency of the OPE in the singlet channel determines the value of $n$ to be $n=21$. In this section we will present our result in position space taking advantage of a simplification due to the differential operator $\dfour$. Finally, we will compute the one-loop anomalous dimensions  of the lightest double-particle operators, providing new data for the spectrum of the D1-D5 CFT beyond the planar limit.

\subsection{Constructing an ansatz}\label{subsec:ansatz}

We start by making manifest crossing symmetry, using a basis 
of delta-functions for the flavour indices, and parametrising the correlator 
in terms of a \textit{single} function $\F(x,\xb)$ as follows:
\begin{align}\label{eq:H2_F}
	\mathcal{H}^{i_1i_2i_3i_4}(x,\xb) = \delta^{i_1i_2}\delta^{i_3i_4}\,\F(1-x,1-\xb) + \delta^{i_1i_4}\delta^{i_2i_3}\,\F(x,\xb) + \delta^{i_1i_3}\delta^{i_2i_4}\,\frac{\F(x',\xb')}{V^2}\,,
\end{align}
where we recall that $x'\equiv x/(x-1)$, $\xb'\equiv \xb'/(\xb'-1)$, and 
moreover we have that $\F(x,\xb)$ obeys 
\begin{align}\label{eq:crossing_F}
	\F(x,\xb)=\frac{1}{U^2}\,\F\left(\frac{1}{x},\frac{1}{\xb}\right)\,.
\end{align}
One can easily change basis in flavour space and go back to the basis 
of projectors given in \eqref{eq:projectors_irreps}. The change of basis reads
\begin{align}\label{eq:irrep_basis}
\begin{split}
	\H^{}_{\I} &= \F(x,\xb)+\frac{\F(x',\xb')}{V^2}+n\,\F(1-x,1-\xb)\,,\\
	\H^{}_{\S} &= \F(x,\xb)+\frac{\F(x',\xb')}{V^2}\,,\\
	\H^{}_{\A} &= \F(x,\xb)-\frac{\F(x',\xb')}{V^2}\,.
\end{split}
\end{align}
We thus see that $\F(x,\xb)$ is simply related to the sum of the symmetric and antisymmetric channels:
\begin{align}\label{eq:F_as_sum}
	\F(x,\xb) = \frac{1}{2}\Big(\H^{}_{\S}+\H^{}_{\A}\Big)\,.
\end{align}
Our next step is to write an ansatz for the one-loop contribution $\F^{(2)}(x,\xb)$.

At one-loop, the relation \eqref{eq:F_as_sum} shows that the  $\log^2(U)$ discontinuity of $\F^{(2)}(x,\xb)$
 is given by the $\log^2(U)$ discontinuities of $\H^{(2)}_{\S}$ and $\H^{(2)}_{\A}$, see 
\eqref{eq:leading_log_res_S}-\eqref{eq:leading_log_res_A}. 
The latter contain transcendental functions with the following two properties:
First, their transcendental weight is at most 2, and the relevant polylogs have arguments in the 
alphabet $\{x,\xb,1-x,1-\xb\}$. Second, their top weight-2 parts are antisymmetric under the exchange of $x\leftrightarrow\xb$.

Considering the above properties, we will construct an ansatz for $\F^{(2)}(x,\xb)$ 
consisting of weight-4 antisymmetric transcendental functions, together with 
their lower-weight completions, that are within the class of single-valued 
harmonic polylogarithms (SVHPLs) built out of the alphabet $\{x,\xb,1-x,1-\xb\}$. At this point, 
this choice of alphabet is a rephrasing of what we referred to as the  strong version of the bootstrap constraints in the introduction.
The reason is that we are considering the minimal set of functions that can uplift the $n$-independent $\log^2(U)$ discontinuities 
of $\H^{(2)}_{\S}$ and $\H^{(2)}_{\A}$ in \eqref{eq:leading_log_res_S}-\eqref{eq:leading_log_res_A} 
to a full one-loop function.
As our calculations will show, it turns out that this minimal ansatz is sufficient to solve all the bootstrap constraints.

Let us now spell out the details about the basis of functions.
The top-weight function in our basis of SVHPLs, which will provide the building block for the correlator,  
is the well known weight-4 two-loop ladder integral (in four 
dimensions with massive external legs \cite{Usyukina:1992jd}\footnote{One should not be confused by the dimensionality: 
we are just rewriting elements from the basis of SVHPLs, and the latter has no dimensionality.}):
\begin{align}
\begin{split}
	\phi^{(2)}(x,\xb) &= \frac{1}{2}\log^2(U)\big[\Li_2(x)-\Li_2(\xb)\big]-3\log(U)\big[\Li_3(x)-\Li_3(\xb)\big]+6\big[\Li_4(x)-\Li_4(\xb)\big]\,.
\end{split}
\end{align}
Apart from products of simple logs, the lower-weight completion can be constructed by 
taking derivatives of the two-loop ladder. First, we introduce the following derivative,
\begin{align}
	\Psi(x,\xb) \equiv \big[x\partial_x-\xb\partial_{\xb}\big]\phi^{(2)}(x,\xb)\,,
\end{align}
which defines a pure transcendental function of weight 3 that is symmetric under $x\leftrightarrow\xb$. 
Similarly, we introduce a pure weight-3 antisymmetric function by taking the following derivative 
\begin{equation}\label{weight3derivphi2}
\big[x\partial_x +\xb \partial_{\xb} \big] \phi^{(2)}(x,\xb)= -\frac{1}{2}\log(x\xb) \phi^{(1)}(x,\xb)\,,
\end{equation} 
where $\phi^{(1)}$ is the one-loop box function (in four dimensions):
 \begin{align}\label{oneloopscalar}
\phi^{(1)}(x,\xb) &= \log(x\xb)\big[\log(1-x)-\log(1-\xb)\big]+2\big[\Li_2(x)-\Li_2(\xb)\big]\,.
\end{align}
It should be noted that also $\phi^{(1)}$ can be written as derivatives acting on $\phi^{(2)}$, 
so in practice all functions that we need descend from the two-loop ladder upon taking derivatives. 
Finally, in order to have a complete basis we consider the orbits generated by acting with crossing 
transformations. For example, one has that
\begin{align}
\begin{split}
	\phi^{(2)}(x,\xb)&=-\phi^{(2)}(1/x,1/\xb)\,,\\
	\phi^{(1)}(x,\xb)&=-\phi^{(1)}(1/x,1/\xb)=-\phi^{(1)}(1-x,1-\xb)\,,\\
\end{split}
\end{align}
which means $\phi^{(2)}$ has three distinct crossing orientations, while $\phi^{(1)}$ has only one. 
Repeating this for all transcendental elements we find the lower-weight completion of the two-loop ladder.

This leads to a basis of transcendental functions consisting of 15 elements, 
denoted below by $\Q_i(x,\xb)$ with $i=1,\ldots,15$. Organised  
by weight and by parity, even (+) or odd (-) under $x\leftrightarrow\xb$,  
the basis elements are given as follows:\footnote{Exactly the same basis of functions has been 
used in bootstrap computations of one-loop graviton correlators in AdS$_5\times$S$^5$ 
supergravity, from ${\cal N}=4$ SYM theory, see \cite{Aprile:2017bgs,Aprile:2017qoy,Aprile:2019rep}.
}
\begin{align}\label{eq:basis15}
\begin{split}
	\text{weight 4}^-:&\qquad\quad~ \Q_{1,2,3}=\phi^{(2)}(x,\xb)\,,\phi^{(2)}(x',\xb')\,,\phi^{(2)}(1-x,1-\xb)\,,\\
	\text{weight 3}^-:&\qquad\quad~ \hspace{0.23cm}\Q_{4,5}=\log(U)\phi^{(1)}(x,\xb)\,,\log(V)\phi^{(1)}(x,\xb)\,,\\
	\text{weight 3}^+:&\qquad\quad~ \Q_{6,7,8}=\Psi(x,\xb)\,,\Psi(x',\xb')\,,\Psi(1-x,1-\xb)\,,\\
	\text{weight 2}^-:&\qquad\quad~\hspace{0.46cm}\Q_{9}=\phi^{(1)}(x,\xb)\,,\\
	\text{weight 2}^+:&\qquad\hspace{0.07cm}\Q_{10,11,12}=\log^2(U)\,,\log(U)\log(V)\,,\log^2(V)\,,\\
	\text{weight 1}^+:&\qquad\quad\hspace{0.06cm}\Q_{13,14}=\log(U)\,,\log(V)\,,\\
	\text{weight 0}^+:&\qquad\qquad\,\Q_{15}=1\,.
\end{split}
\end{align}
With the above definitions, we consider the following ansatz for the one-loop function $\F(x,\xb)$:
\begin{align}\label{eq:ansatz_G}
	\F^{(2)}(x,\xb) = \frac{1}{V}\sum_{i=1}^{15}\frac{p_i(x,\xb)}{(x-\xb)^{d_\pm}}\,\Q_i(x,\xb)\,,\quad \text{with}~~p_i(x,\xb) = \sum_{n=0}^{d_\pm}\sum_{m=n}^{d_\pm}a^{(i)}_{n,m}(x^n\xb^m+x^m\xb^n)\,.
\end{align}
Let us comment on some relevant features of this ansatz: 
\begin{itemize}
\item The  $\log^2(U)$ predictions \eqref{eq:leading_log_res_S} and \eqref{eq:leading_log_res_A}
imply the values $d_-=7$ and $d_+=6$ 
according to the parity of $\Q_i(x,\xb)$,
so as to ensure that $\F(x,\xb)$ is symmetric under $x\leftrightarrow\xb$. 
\item
There is an explicit factor of $1/V$. This is needed in order to match the $\log^2(U)$ predictions \eqref{eq:leading_log_res_S} and \eqref{eq:leading_log_res_A}, where this denominator is generated by the action of $\dfour$. 
\item The $a^{(i)}_{n,m}$ are unfixed numerical coefficients, and the degree of a given 
numerator should not exceed the degree of the corresponding denominator. The 
ansatz so far contains a total of 468 free parameters.
\end{itemize}
As a technical remark, we take the parameters $a^{(i)}_{n,m}$ to be \textit{algebraic} numbers. The only transcendental number which one may expect to arise, namely $\zeta_3$, is already included in the weight $3^+$ part of our ansatz due to the relation $\Psi(x,\xb)+\Psi(x',\xb')+\Psi(1-x,1-\xb)=12\,\zeta_3$.

\subsection{CFT constraints, crossing and $\emph{n=21}$}\label{subsec:bootstrap}
We will now describe our bootstrap algorithm, which consists of the 
following four constraints:\footnote{These constraints are to be imposed on our ansatz in 
\eqref{eq:ansatz_G}, as a consequence of the fact that the same constraints must 
be obeyed by the  one-loop reduced correlator itself.}
\begin{enumerate}[itemsep=0pt]\renewcommand{\labelenumi}{(\arabic{enumi})}
\item\textbf{Crossing symmetry of the correlator.} The function $\F^{(2)}(x,\xb)$ has to satisfy \eqref{eq:crossing_F}, namely
\begin{align}
	\F^{(2)}(x,\xb)=\frac{1}{U^2}\,\F^{(2)}\left(\frac{1}{x},\frac{1}{\xb}\right).
\end{align}
\item\textbf{Absence of unphysical poles at $x=\bar{x}$.}
For a correlator in Euclidean signature, poles at $x=\xb$ are unphysical and therefore need to be cancelled.
Note that the ansatz in \eqref{eq:ansatz_G} contains up to 7 powers of $(x-\xb)$ in the denominator.
\item\textbf{Matching the leading log$^2U$ predictions.} Starting from ${\cal F}^{(2)}(x,\xb)$, 
we construct the $\log^2(U)$ discontinuity of the reduced correlator ${\cal H}^{(2)}(x,\xb)$ 
and decomposed it in reps via \eqref{eq:irrep_basis}. This decomposition
has to match the OPE predictions derived in Section \ref{subsec:leading_logs}. 
\item\textbf{Absence of long operators with $h$ or $\hb=0$.} Since the reduced correlator 
only contains the exchange of long operators,  
and there are no long double-trace operators with either $h$ or $\hb=0$ 
(see Section \ref{subsec:double-trace_spectrum}), 
the $\log(U)$ projection of the correlator needs to 
vanish for these values of $h$,$\hb$. This comes out for free in the symmetric and 
antisymmetric channels, but not when we cross to the singlet $\I$. 
\end{enumerate}
After implementing our four-step algorithm, we find the following remarkable outcome:
The value of $n$ is fixed to be $n=21$ and all parameters $a_{n,m}^{(i)}$ are determined, 
up to four `tree-level like' ambiguities. Before giving the explicit expression of $\F^{(2)}(x,\xb)$, 
let us break down how the value of $n=21$ emerges.

After imposing $(1)$ crossing, $(2)$ absence of spurious poles, and $(3)$, we find 
no free parameters in the $\log^2(U)$ discontinuity of ${\cal F}^{(2)}(x,\xb)$. This means there 
are no free parameters entering the $\log^2(U)$ discontinuities of $\F^{(2)}(x',\xb')$, as expected, 
since by construction this orientation also contributes to the symmetric and antisymmetric channels, 
but also no free parameters are present in the  $\log^2(U)$  discontinuity of $\F^{(2)}(1-x,1-\xb)$. 
It follows that there are no free parameters in the $\log^2(U)$ discontinuity of the singlet channel 
which is built out of the combination
\begin{align}
{\cal H}^{(2)}_{\I}\Big|_{\log^2(U)} = 
\bigg[ {\cal F}^{(2)}(x,\xb)+\frac{ {\cal F}^{(2)}(x',\xb')}{V^2}+ n\,\F^{(2)}(1-x,1-\xb)\bigg] _{\log^2(U)} \,.
\end{align}
The above outcome provides an opportunity for a consistency check with the OPE predictions 
in the singlet channel given in \eqref{eq:leading_log_res_singlet}.
To this end, we perform a block decomposition and focus on the minimum 
twist prediction at $(h,\hb)=(1,1)$ and the next-to-minimum twist prediction 
at $(h,\hb)=(2,2)$, finding the following result:
\begin{align}\label{wherewenequal21}
	\begin{matrix}
	h=\hb=1:\qquad\qquad & \qquad \displaystyle \frac{1}{2}\times\frac{1}{4}\big({-}2+\frac{n}{3}\big)^2\overset{!}{=}\frac{4+n}{8}\,,\\[.4cm]
	h=\hb=2:\qquad\qquad & \displaystyle \frac{1}{2}\times\frac{1}{100}(5n^2-56 n+200)\overset{!}{=} \frac{200+49n}{200}\,.
	\end{matrix}
\end{align}
By construction, the block decomposition of our one-loop function gives coefficients that 
depend \textit{linearly} on $n$ (the number of tensor multiplets), 
as can be seen from the first line of \eqref{eq:irrep_basis}, in contrast, the $n$-dependence of the OPE predictions 
\eqref{eq:leading_log_res_singlet} is \textit{quadratic}, meaning that the two expressions can not match for generic values 
of $n$! Since no free parameters are involved, we find two non-trivial equations for $n$. 
Remarkably, both of them result in the same constraint on $n$:\footnote{
Both equations in \eqref{wherewenequal21} are  
equivalent to $n(n-21)=0$. The solution $n=0$ should be discarded since $\H^{(2)}_{\I} = \H^{(2)}_{\S}$ as can be seen from \eqref{eq:irrep_basis}. In other words, for $n=0$ the singlet coincides with the symmetric channel for which we already imposed the matching with the leading log prediction in step (3) of our algorithm.
}
\begin{empheq}[box=\fbox]{equation}
\begin{aligned}\label{eq:n=21}
	~n=21~
\end{aligned}
\end{empheq}
Our result \emph{uniquely} selects the theory obtained from compactification of type IIB supergravity on K3.

We would like to emphasize that  the prediction for $(h,\hb)=(1,1)$ only needed input from $\langle s_1 s_1 s_1 s_1 \rangle$. 
However, the prediction for $(h,\hb)=(2,2)$ needed input from $\langle s_1 s_1 s_1 s_1 \rangle$, $\langle s_1 s_1 s_2 s_2 \rangle$  and  $\langle s_1 s_1 \sigma_2 \sigma_2 \rangle$, and
so we used knowledge about the mixed tensor-graviton correlators. 
The fact that $n=21$ is consistent for both values of $(h,\hb)=(1,1)$ and $(2,2)$
is therefore non-trivial, and we conjecture that $n=21$ remains a solution for the tower of predictions at \textit{all} values of $h=\hb$.

Although additional consistency checks on the spectrum would be desirable, it is remarkable that $n=21$ 
arises purely as an algebraic result about the consistency of the OPE underlying the quantum gravity theory that we are bootstrapping from the dual SCFT. Given this result, it is worth discussing why the appearance of type IIB supergravity compactified on K3 is natural, whereas one would not have expected type IIB supergravity on  $T^4$.

From the point of view of the symmetries of the CFT, both the (AdS$_3\times$S$^3)\times$K3 as well as the (AdS$_3\times$S$^3)\times T^4$ compactification are dual to two-dimensional SCFTs with $\mathcal{N}=(4,4)$ supersymmetry. However, the underlying six-dimensional supergravity theories are different (having (2,0) and (2,2) supersymmetry, respectively), resulting in a different spectrum of superconformal primary operators. In particular, the analysis of \cite{deBoer:1998kjm} shows that the 2d SCFT dual to the $T^4$ compactification contains both bosonic and fermionic superprimaries transforming under $SO(5)$, while the SCFT dual to the K3 compactification has only bosonic superprimaries transforming under $SO(21)$. Therefore, in order to study the theory dual to (AdS$_3\times$S$^3)\times T^4$ one has to take into account double-particle operators built from the extra fermionic primaries\footnote{
In particular, in addition to the spectrum of primary operators listed in Table \ref{table:qnumbers}, from each KK-level one finds four fermionic primaries with quantum numbers $(h,j,\hb,\jb)=(\frac{p}{2},\frac{p}{2},\frac{p-1}{2},\frac{p-1}{2})$ and four with $(h,j)\leftrightarrow(\hb,\jb)$ exchanged, each transforming in the four-dimensional spinor representation of $SO(5)$.}
in addition to the ones listed in \eqref{eq:dt_list}. Equivalently, there are additional mixed correlators that one has to consider. As a result, the leading-log predictions are different, meaning that the starting point of our one-loop bootstrap computation is modified. 
This difference can also be understood from the bulk point of view: the one-loop amplitude is sensitive to the field content which runs in the loop, and this differs between the two six-dimensional supergravity theories.

In Section \ref{sec:flat-space} we will provide another piece of  independent evidence for the validity of our one-loop function and the value $n=21$ by considering the flat-space limit
and the four-point amplitudes in 6d (2,0) supergravity recently uncovered in \cite{Huang:2025nyr}.

\subsection{The one-loop correlator in position space}\label{subsec:simplification_dfour}
We find that the explicit result for the one-loop function $\F^{(2)}(x,\xb)$, obtained from the bootstrap 
algorithm described above, can be written in the following form:
\begin{equation}\label{definitionF2explicit}
{\cal F}^{(2)}(x,\xb)=\dfour\,\P(x,\xb) +\frac{U-V+1}{2UV} + \sum_{i=1}^4\beta_i\,B_i(x,\xb)\,.
\end{equation}
The auxiliary function, or `preamplitude', $\P(x,\xb)$ is given by
\begin{align}
	\P(x,\xb)=\frac{1}{U^2}\sum_{i=1}^{15}\widetilde{p}_i(x,\xb)\,\Q_i(x,\xb)\,,
\end{align}
but, remarkably, out of the 15 coefficient functions $\tilde{p}_i$ we find that only 6 are non-zero:
\begin{align}\label{eq:p_coeffs}
\begin{split}
	\widetilde{p}_2&=\frac{(2x\xb-x-\xb)\,U^2}{(x-\xb)^3}\,,\qquad\widetilde{p}_3=\frac{x^3-3x^2\xb-4x^2-3x\xb^2+12x\xb+\xb^3-4\xb^2}{2(x-\xb)^3}\,,\\[3pt]
	\widetilde{p}_5&=-\frac{x+\xb}{4(x-\xb)}\,,\qquad\widetilde{p}_7=\frac{U^2}{(x-\xb)^2}\,,\qquad\widetilde{p}_8=-\frac{x^2-4x\xb+\xb^2}{2(x-\xb)^2}\,,\qquad\widetilde{p}_9=-\frac{2U}{(x-\xb)}\,.
\end{split}
\end{align}
All other $\tilde{p}_i$ vanish.

The rewriting in terms of the preamplitude $\P(x,\xb)$ is motivated by the fact that the $\log^2(U)$ discontinuity in the symmetric and antisymmetric channels is simplified by extracting the differential operator $\dfour$. Due to the relation \eqref{eq:F_as_sum} the same is true for the $\log^2(U)$ contribution to ${\cal F}^{(2)}(x,\xb)$. Our computation then demonstrates a much stronger statement, namely that up to a simple rational term the full one-loop function ${\cal F}^{(2)}(x,\xb)$ can be written as $\dfour$ acting on a substantially simpler preamplitude.\footnote{The preamplitude is not unique as the differential operator $\dfour$ has a kernel. We have made use of this freedom to arrive at the simple form of the coefficient functions $\tilde{p}_i(x,\xb)$ given in \eqref{eq:p_coeffs}.}

The functions $B_i(x,\xb)$ in \eqref{definitionF2explicit} are given by
\begin{align}\label{eq:ambiguities}
	B_i(x,\xb)=\Big\{\dbar{1221},\,\dbar{2222},\,\dbar{3223},\,\dbar{1243}+\dbar{2233}-2\dbar{1232}\Big\}\,.
\end{align}
These are the four ambiguities, whose coefficients are left unfixed by our bootstrap algorithm 
since they solve the constraints (1)-(4) in Section \ref{subsec:bootstrap} for any value of the $\beta_i$.
This is expected because they do not contribute 
to the $\log^2(U)$ discontinuity in any orientation. 
Let us point out the following considerations:
\begin{itemize}
\item The first ambiguity, $B_1=\dbar{1221}$, is the same function that appears at tree-level, i.e.~the $\delta^{i_1i_4}\delta^{i_2i_3}$ 
coefficient of $\H^{(1)}$, see \eqref{eq:H1}. 
This ambiguity, and the free-like term ${\cal G}^{(2)}$ in \eqref{generalG2},  contribute to the protected sector of the correlator.
In total, the relevant holomorphic contribution to the protected sector is found from the leading term in the small $\xb$ expansion of the 
$s$-channel (the $\delta^{i_1i_2}\delta^{i_3i_4}$ component of the full correlator $G^{(2)i_1i_2i_3i_4}$ in \eqref{eq:largeNexp}), which is given by
\begin{equation}
	G^{(2)}_s = \G^{(2)}_s + \frac{(y-x)}{y}\bigg[\frac{(1-\beta_1)\log(1-x)}{x}-\frac{(x-2)\log^2(1-x)}{x^2}+O(\bar{x})\bigg],
\end{equation}
where $\G^{(2)}_s$ is a constant and the prefactor  ${(y-x)}/{y}$ is the holomorphic part of ${(x-y)(\bar{x}-\bar{y})}/{y\bar{y}}$ in \eqref{generalG2}.
The coefficient $\beta_1$ appears in the $\log(1-x)$ contribution, 
while the $\log^2(1-x)$ term is fully fixed by the double-particle bootstrap. 
The value of $\G^{(2)}_s$  and $\beta_1$ can be inferred by reasoning on the exchange of the half-BPS 
multiplet $V^+_1$, which contains the holomorphic stress tensor as a descendant.
%
The term proportional to $\log^2(1-x)$ provides new CFT data for the exchange of protected double-$V^+_{1}$ multiplets and their derivatives 
(thus containing holomorphic double-stress tensors with derivatives as descendants).
We expect that one can achieve a non-trivial validation of our result  
by extending the analysis of the identity super-Virasoro block performed 
in \cite{Giusto:2018ovt} to order $1/N^2$, but we leave the precise statement to future work.

\item The second ambiguity, $B_2=\dbar{2222}$, is a contact-term ambiguity which plays the role of a finite one-loop counter term. We expect that $\beta_2$ can be fixed by embedding the one-loop amplitude into the full string theory.\footnote{A similar contact-term ambiguity is present in the 
one-loop supergraviton amplitude in AdS$_5\times$S$^5$ \cite{Aprile:2017bgs}, and its coefficient has been fixed by considering the integrated correlator computed from supersymmetric localisation \cite{Chester:2019pvm}.}

\item The last two ambiguities, $B_3$ and $B_4$, are on a different footing. They have maximal 
denominator power $(x-\xb)^7$, and as such they contribute to the flat-space limit. 
\end{itemize}
The statement that $\dbar{2222}$ corresponds to a contact term is more 
transparent in Mellin space, and we will come back to this in Section \ref{sec:flat-space}. As it turns out, 
upon considering the flat-space amplitude constructed in \cite{Huang:2025nyr}, the flat space limit provides 
two additional constraints on the ambiguities, 
so that we restrict to two the number of ambiguities of the one-loop correlator $\F^{(2)}(x,\xb)$.\footnote{See 
also Appendix \ref{app:bpl}, where we the consider the analogous formulation of the flat-space limit 
in position space, known as the bulk-point limit.}
In particular, we shall see that 
\begin{align}
	\beta_4=0\,,
\end{align}
and that the value of $\beta_3$ can be related to a specific contribution in the flat-space amplitude constructed in \cite{Huang:2025nyr}. 
We will give more details in the next section. 

To summarize, the genuine ambiguities of the double-particle bootstrap are $\beta_2$ and $\beta_3$ and we expect them to be fixed by
the one-loop amplitude in the full string theory.

\subsection{Extracting new CFT data} 
With the one-loop correlator  at hand, we can now proceed as to extract some of the new 
CFT data encoded in it. We will focus on the one-loop anomalous 
dimensions $\gamma^{(2)}_{\a}$ of the lowest dimension double-trace operators with $h=1$ or $\hb=1$, since there is no degeneracy in this case.

The anomalous dimension $\gamma^{(2)}_{\a}$ is found from the $\log(U)$ contribution to $\Ht^{(2)}$, which reads
\begin{align}\label{eq:H2_logu_OPE}
	\Ht^{(2)}_{\a}\Big|_{\log(U)}=\sum_{h,\hb}\bigg[ 
	\big[A_{\a}^{(1)}\gamma^{(1)}_{\a}\big]_{h,\hb}+
	\big[A_{\a}^{(0)}\gamma_{\a}^{(2)}\big]_{h,\hb}+ 
	\big[A^{(0)}_{\a}(\gamma_{\a}^{(1)})^2\big]_{h,\hb}\,{\partial}_{h,\hb}^{\text{no-log}}\bigg]B_{h,\hb}(x,\xb)\,,
\end{align}
where the derivative ${\partial}_{h,\hb}^{\text{no-log}}\equiv\partial+\bar{\partial}\equiv\partial_h+\partial_{\hb}$ 
acts on the long blocks $B_{h,\hb}(x,\xb)$ and one is instructed to remove all logarithmic 
terms generated by the action of derivatives as these are already accounted for in 
\eqref{eq:logu_stratification} by the terms involving explicit $\log(U)$ factors. 

In order to isolate the middle term in \eqref{eq:H2_logu_OPE}, i.e.~the one 
containing the anomalous dimensions, we first need to subtract the other contributions, as we do below.

The combination $[A^{(0)}_{\a}(\gamma^{(1)})^2_{\a}]_{h,\hb}$ is already known from 
our analysis in Section \ref{subsec:double-trace_spectrum},  and can thus be easily generated. 
On the other hand, the term $[A_{\a}^{(1)}\gamma^{(1)}_{\a}]_{h,\hb}$ needs to be computed.
Since there is no degeneracy we can obtain this quantity directly from tree level correlator, recalling that 
\begin{equation}
\Ht^{(1)}_\a(x,\xb)\Big\vert_{\log^0(U)}=\sum_{h,\hb}\bigg[\big[A_{\a}^{(1)}\big]_{h,\hb}+
\big[A^{(0)}_{\a}(\gamma_{\a}^{(1)})\big]_{h,\hb}{\partial}_{h,\hb}^{\text{no-log}}\bigg]B_{h,\hb}(x,\xb)\,.
\end{equation}
Extracting $[ A_{\a}^{(1)}]_{h,\hb}$ from \eqref{eq:L1}, we find that the usual derivative relation holds, i.e.
\begin{align}\label{derrelatiopn}
	\big[ A_{\a}^{(1)}\big]_{h,\hb} = (\partial_h+\partial_{\hb})\,\big[A^{(0)}_{\a}(\gamma_{\a}^{(1)})\big]_{h,\hb}\,,
\end{align}
where $\big[A^{(0)}_{\a}(\gamma_{\a}^{(1)})\big]_{h,\hb}$ was given in \eqref{eq:A0g1_11pp}, upon setting $p=1$. 
This derivative relation is satisfied for all three flavour channels and for all values of $(h,\hb)$ 
(in particular also below the double-trace threshold $h=\hb=1$). We emphasize that in order to establish 
\eqref{derrelatiopn} it is crucial to identify that $\Ht^{(1)}$ is given in \eqref{eq:L1}, by including 
both $\H^{(1)}$ and the free-like contributions. Only after doing so one finds that the derivative relation holds.

We now have all the necessary ingredients to solve for $\gamma^{(2)}_{\a}$. 
For the unique double-particle operators at  $h=1$ or $\hb=1$, in
the symmetric and antisymmetric channels, we find\footnote{This is also the case for $h=\hb=1$ in the singlet, see Table \ref{tab:degeneracies}. However, it turns out that also the contact-term ambiguity $B_2$ contributes for this value of $(h,\hb)$, preventing us from extracting a definite value for $\gamma^{(2)}$.} 
\begin{align}\label{gamma2sym}
\gamma^{(2)}_{\mathbb{S}} = -\frac{24}{(\ell-1)(\ell+4)}-2\beta_1-\frac{2}{3}\beta_2\,\delta_{\ell,0}\,,\qquad (\tau=2\,,~\ell\text{ even})\,,
\end{align}
and we recall that $\beta_1$ is the coefficient of the tree-level ambiguity, while $\beta_2$ is 
the coefficient of the contact-term ambiguity which contributes only at spin 0.
Note that $\gamma^{(2)}$ has a pole at spin $\ell=1$, warning us that analyticity 
in spin breaks down below spin 1, thus at spin $0$, where knowing the value of $\beta_2$ is crucial.
The same phenomenon arises also in ${\cal N}=4$ SYM theory, see \cite{Aprile:2017bgs,Alday:2017vkk,Chester:2019pvm}.

A similar calculation as the one above in the antisymmetric channel yields
\begin{align}\label{gamma2asym}
	\gamma^{(2)}_{\mathbb{A}} = -\frac{16}{\ell(\ell+3)}-2\beta_1\,,\qquad (\tau=2\,,~\ell\text{ odd})\,.
\end{align}
We note that while the tree-level anomalous dimensions \eqref{eq:anom_dims_S,A} are 
given by the same formula for both channels, the one-loop corrections in \eqref{gamma2sym} and \eqref{gamma2asym} turn out to be different.

\section{The one-loop amplitude in Mellin space}\label{sec_Mellin_amp}
Mellin space is a natural language to write holographic correlators \cite{Fitzpatrick:2011ia}. 
The key advantage  is the fact that the Mellin transform of the reduced correlator in position 
space defines  a Mellin amplitude whose analytic structure closely parallels that of flat-space 
scattering amplitudes. This property is manifested by the flat-space limit \cite{Penedones:2010ue}.

For the correlator at hand, whose reduced correlator does not depend on the R-symmetry cross-ratios, 
we define the Mellin transform of $\H^{i_1 i_2 i_3 i_4}(x,\xb)$ through the integral\footnote{For correlators 
with non-trivial R-symmetry structure  the Mellin space formalism can be extended 
to incorporate an additional discrete Mellin transform acting on the 
R-symmetry cross-ratios. This formalism is called  AdS$\times$S Mellin 
space \cite{Aprile:2020luw}, see also \cite{Aprile:2021mvq} for the specific 
case of AdS$_3\times$S$^3$. }
\begin{equation}\label{eq:Mellin_H}
	\H^{i_1 i_2 i_3 i_4}(x,\xb) = \oint ds\,dt\, U^s V^t\,\Gamma(-s)^2\Gamma(-t)^2\Gamma(-u)^2\,\mathcal{M}^{i_1 i_2 i_3 i_4}(s,t)\,,
\end{equation}
where $u$ is defined as  $s+t+u=-2$ and 
\begin{equation}\label{defMgeneral}
	\mathcal{M}^{i_1i_2i_3i_4}(s,t) 
	= \delta^{i_1i_2}\delta^{i_3i_4}\,\mathcal{M}^{}(t,s) + \delta^{i_1i_4}\delta^{i_2i_3}\,\mathcal{M}^{}(s,t) + \delta^{i_1i_3}\delta^{i_2i_4}\,\mathcal{M}^{}(s,u)\,.
\end{equation}
The function ${\cal M}(s,t)$ satisfies the symmetry ${\cal M}(s,t)={\cal M}(u,t)$. 
This formula parallels the one given in \eqref{eq:H2_F}.  From there we read that 
the Mellin transform of ${\cal F}(x,\xb)$ in \eqref{eq:H2_F}
is ${\cal M}(s,t)$ in \eqref{defMgeneral}.

In analogy to the correlator in position space, the Mellin amplitude $\mathcal{M}^{i_1i_2i_3i_4}(s,t)$ admits a large-$N$ expansion. 
The tree-level contribution  in \eqref{eq:G1} in Mellin space is simply given by
\begin{equation}\label{eq:M1111}
	\mathcal{M}^{(1),i_1 i_2 i_3 i_4}(s,t) = -\frac{\delta^{i_1 i_2}\delta^{i_3 i_4}}{s+1} - \frac{\delta^{i_1 i_4}\delta^{i_2 i_3}}{t+1} - \frac{\delta^{i_1 i_3}\delta^{i_2 i_4}}{u+1}\,.
\end{equation}
Already at tree level, the simplicity of the Mellin amplitude is evident. As shown in \cite{Rastelli:2019gtj,Giusto:2019pxc,Giusto:2020neo}, 
this amplitude is the simplest generalisation of the tree-level flat-space amplitude.

Below we construct the Mellin transform of the one-loop reduced correlator,
focusing on the Mellin transform of
${\cal F}^{(2)}(x,\xb)$, thus ${\cal M}^{(2)}(s,t)$.
Of course the full amplitude in \eqref{eq:Mellin_H} is recovered by using crossing symmetry.

As for the position space result described in the previous section, $\mathcal{M}^{(2)}(s,t)$ 
is given by a sum over transcendental functions in the variables $s,t$ that multiply rational functions. 
There is however a major simplification to start with: In Mellin space there are only three basis elements.
Apart from unity, the other two non-trivial basis elements are ${\cal B}(s,t)$ and ${\cal C}(s,t)$, where,
\begin{align}\label{eq:Mellin_kernels}
\begin{split}
	{\cal B}(s,t)&=\frac{1}{2}\left( \psi^{(1)}(-s)+\psi^{(1)}(-t)-\big[\psi(-s)-\psi(-t)\big]^2-\pi^2 \right),\\
	{\cal C}(s,t)&=\psi(-s)-\psi(-t)\,. 
\end{split}
\end{align}
The first object, ${\cal B}(s,t)$ in \eqref{eq:Mellin_kernels}, is simply the Mellin amplitude of 
the two-loop ladder 
\begin{align}\label{eq:Mellin_basis}
	\frac{\phi^{(2)}(x',\xb')}{x-\bar{x}}&=-\oint_{-i\infty}^{+i\infty}\!\!\!\!\!ds\,dt\,\, U^s V^t\,\Gamma(-s)^2\Gamma(-t)^2\Gamma(s+t+1)^2\times {\cal B}(s,t)\,.
\end{align}
This is the only orientation that has double poles at $s=0,1,2,\ldots $ and $t=0,1,2,\ldots\,$, and therefore contributes to 
$\log^2(U)\log^2(V)$. Then, the Mellin amplitude 
${\cal C}(s,t)$ is obtained from the two-loop ladder by using \eqref{eq:Mellin_basis} and 
taking derivative as in \eqref{weight3derivphi2}. One can see that the result gives a 
combination of terms, one of which involves ${\cal C}(s,t)$. 

Let us note that the transcendental counting for ${\cal M}^{(2)}(s,t)$ works differently
compared to position space. For example, by counting the degree of the polygammas 
in ${\cal B}(s,t)$ and ${\cal C}(s,t)$ it is apparent that they have degree two and one, 
respectively. A rational Mellin amplitude counts as zero degree.  
This is because the string of $\Gamma$ functions provide double-poles in $s$ and $t$ 
and therefore manifest double-logs to start with. Also, one should note that   the three 
seeds correspond to antisymmetric transcendental functions and, analogously to what 
happens with D-function, when a seed is multiplied by a rational function in $s,t$, 
it will correspond in position space to a function with non-trivial rational prefactors 
and a non-trivial lower weight completion. 

The Mellin space representation gives a stratification of the correlator in the form 
\begin{equation}
\sum_{0\leq a,b,c,d\leq 2}\log^a(U)\log^b(V)(\pi^2)^c (\zeta_3)^d f_{abcd}(U,V)\,,
\end{equation}
where $f_{abcd}(U,V)$ have a Laurent expansion in small $U$ and small $V$ with  
coefficients depending on $s$ and $t$ evaluated at integers, through the residue 
operation. This same stratification can be readily obtained from the position space 
correlator and matched against the integration performed in Mellin space, assuming 
an ansatz of the form
\begin{equation}
{\cal M}^{(2)}(s,t) = R_2(s,t)  {\cal B}(s,t)+ R_1(s,t) {\cal C}(s,t) + R_0(s,t) + (s\leftrightarrow u)\,,
\end{equation}
where $R_i(s,t)$ are rational functions to be determined. 

Following the above procedure, we shall Mellinise the function ${\cal F}^{(2)}(x,\xb)$ 
in \eqref{definitionF2explicit} with the following redefinition of ambiguities: 
\begin{equation}
\beta_1=\alpha_1-11\,,\qquad \beta_2= \alpha_2+8\,,\qquad \beta_{3}=\alpha_{3}\,,\qquad \beta_{4}=\alpha_{4}\,.
\end{equation}
As a result we obtain
\begin{equation}\label{ampliMellin}
{\cal M}^{(2)}(s,t)= \widetilde{\cal M}^{\twoladder}(s,t)+ \widetilde{\cal M}^{\twoladder}(u,t) + \sum_{i=1}^{4} \alpha_i\, {\cal M}_{B_i}\,,
\end{equation}
where $u=-s-t-2$, the amplitude $\widetilde{\cal M}^{\twoladder}(s,t)$ stands for
\begin{equation}\label{ampliMellinn}
\widetilde{\cal M}^{\twoladder}(s,t)= -\left[\frac{2s(1+s)}{1+u}+\frac{4s^2}{2+u}\right]{\cal B}(s,t) + \left[ \frac{2(t-s)}{2+u}+\frac{2(s-u)}{1+t}\right]{\cal C}(s,t) + \left[ \frac{4}{2+u}  + \frac{4}{t+1}\right],
\end{equation}
and the ${\cal M}_{B_i}$ denote the Mellin amplitudes of the ambiguities:
\begin{align}\label{Mellin_amb}
	\mathcal{M}_{B_1}=-\frac{1}{t+1}\,,\qquad \mathcal{M}_{B_2}=1\,,\qquad\mathcal{M}_{B_3}=\,{-}t\,,\qquad\mathcal{M}_{B_4}=\frac{su}{t+1}\,.
\end{align}
We will shortly explain why we have introduced the notation $\widetilde{\cal M}^{\twoladder}(s,t)$.

Let us first comment on some important features of the result in \eqref{ampliMellin}.
As pointed out already, ${\cal B}(s,t)$ is the only one to contribute to the $\log^2(U)\log^2(V)$ 
coefficient function in position space. Its rational amplitude in \eqref{ampliMellinn} 
vanishes at $s=0$, therefore the $\log^2(U)\log^2(V)$ coefficient function in the small 
$U$ expansion will start at twist four, one layer above the minimum,  which corresponds 
to twist-two, i.e.~$h=1$ or $\bar{h}=1$. This is a reflection of the twist-two peculiarity
that we observed in Section \ref{subsec:leading_logs}. Also, we would like to emphasize 
that the simple poles that appear at tree level, see \eqref{eq:M1111}, corresponding to the 
values of $s,t,u=-1$, also appear in ${\cal M}^{(2)}(s,t)$. However,  in ${\cal M}^{\twoladder}(s,t)$ 
we find an additional simple pole at $u=-2$ and by crossing, we will find addition simple poles 
at $s,t=-2$. These poles are below unitarity and so the total residue must be vanishing. 
This is indeed the case, and the cancellation is non-trivial since it involves all transcendental functions.

Lastly, the ${\cal M}_{B_i}$ in \eqref{ampliMellin} are the Mellin amplitudes of the four 
ambiguities $B_i(x,\xb)$ given in \eqref{eq:ambiguities} in their position space representation, quoted above in \eqref{Mellin_amb}. Let us notice that $\mathcal{M}_{B_1}$ is nothing but the tree-level correlator, now written in Mellin space. Next, the functions 
$\mathcal{M}_{B_2}$ and ${\cal M}_{B_3}$ have a neat interpretation as coming from an effective action containing higher-derivative terms: $\mathcal{M}_{B_2}$ corresponds to a contact term of the schematic form $\phi^4$, 
while ${\cal M}_{B_3}$ corresponds to a two-derivative term of the form $(\partial\phi)^2\phi^2$.
Finally, we see that ${\cal M}_{B_3}$ and ${\cal M}_{B_4}$ scale linearly  in the limit of large Mellin variables, differently from ${\cal M}_{B_1}$ and ${\cal M}_{B_2}$.  The same 
linear behaviour is observed for the coefficient of ${\cal B}(s,t)$. 
Thus, as anticipated at the end of Section \ref{subsec:simplification_dfour}, we expect to constrain the coefficients $\alpha_3$ and $\alpha_4$ by considering the flat-space limit. This will be discussed later on in Section \ref{sec:flat-space}.

\subsection{Differential representation}
In preparation for the next section, we will now write ${\cal M}^{\twoladder}(s,t)$ in \eqref{ampliMellinn} manifesting that this function descends entirely from the two-loop ladder, thus explaining the 
notation $\twoladder$ previously introduced.  This is the so called `differential representation' 
that was introduced in \cite{Huang:2023ppy,Huang:2024dck,Huang:2024rxr} in the study of 
graviton amplitudes in AdS$_5\times$S$^{5}$ and gluon amplitudes 
in AdS$_5\times$S$^{3}$. It turns out to be conceptually very useful also in our case.

Given ${\cal M}^{\twoladder}(s,t)$, that we repeat below for convenience,
\begin{equation}\label{ampliMellinnn}
\widetilde{\cal M}^{\twoladder}(s,t)= -\left[\frac{2s(1+s)}{1+u}+\frac{4s^2}{2+u}\right]{\cal B}(s,t) + \left[ \frac{2(t-s)}{2+u}+\frac{2(s-u)}{1+t}\right]{\cal C}(s,t) + \left[ \frac{4}{2+u}  + \frac{4}{t+1}\right],
\end{equation}
we can rewrite it in the following form, 
\begin{equation}\label{diff_rep_M}
{\cal M}^{\twoladder}(s,t) = \sum_{i=1}^{7}  {\cal V}^{}_i(s,t) {M}_i^{}(s,t) \,,
\end{equation}
upon introducing the vectors,
\begin{equation}
\!\!\!\!\!\!\!\begin{array}{c}
\begin{tikzpicture}
\draw (0,0) node[scale=.85] {$
\begin{array}{rccccccc}
{\cal V}_i^{}=\Big\{&\!\! {\cal B}(s,t) ,&\!\!\!\!\!\! {\cal B}(s+1,t),  & {\cal B}(s,t+1), & {\cal B}(s-1,t), & {\cal B}(s,t-1),&\!\!\!\! {\cal B}(s-1,t-1), &\!\!\!\!{\cal B}(s+1,t+1)\ \ \Big\},\\[0.7cm]
{M}_i^{}=\Bigg\{&\!\!\!\! \displaystyle - \frac{2s(1+s)}{u+1}, &\!\! \displaystyle\frac{ (4s-2)(s+1) }{u+2}, &  \displaystyle -\frac{(4t+2)s+2}{u+2}, &  \displaystyle -\frac{4s^2}{u+2}, &  \displaystyle\frac{4s(t-s)}{u+2},  & \displaystyle\!\!\!\!  \frac{4s^2}{u+2}, & \displaystyle \!\!\frac{4(1-s)(1+s)}{u+2}\Bigg\}.
\end{array}$};
\end{tikzpicture}
\end{array}
\end{equation}
The various components of ${\cal V}^{}_i(s,t)$ are double boxes with particular 
shifts in their arguments. For example, the second one is defined as ${\cal V}^{}_{i=2}(s,t)\equiv {\cal B}(s+1,t)$. Note that one can write 
explicitly the result of the shift, finding
\begin{equation}\label{exampleB1t}
{\cal B}(s+1,t)= {\cal B}(s,t)-\frac{1}{s+1}{\cal C}(s,t)\,.
\end{equation}
More generally, the action of a shift simply causes a cascade onto the lower 
transcendental basis, with particular rational coefficients depending on the shift 
The table of all shifts is given below 
\begin{equation}\label{rationalprefactorEllis}
\begin{array}{c}
\begin{tikzpicture}
\draw (0,0) node[scale=.90] {$
\begin{array}{c|c|c|c|c|c|c|c}
 		& { \cal B}(s,t)  & {\cal B}(s+1,t) &{\cal B}(s,t+1) & {\cal B}(s-1,t) & {\cal B}(s,t-1) & {\cal B}(s-1,t-1) & {\cal B}(s+1,t+1) \\[2ex]
\hline 
\rule{0pt}{.8cm}{\cal B}(s,t) & 1 & 1 & 1 & 1 & 1 & 1 & 1 \\[2ex]
\hline 
\rule{0pt}{.8cm}{\cal C}(s,t) & 0 & -\frac{1}{s+1} & \frac{1}{t+1} & \frac{1}{s} & -\frac{1}{t}  & -\frac{(s-t)}{st} & \frac{(s-t)}{(s+1)(t+1)} \\[2ex]
\hline
\rule{0pt}{.8cm} 1& 0&  0 & 0 & -\frac{1}{s^2} & -\frac{1}{t^2}  & -\frac{(s^2-st+t^2)}{s^2t^2} & \frac{1}{(s+1)(t+1)}
\end{array}$};
\end{tikzpicture}
\end{array}
\end{equation}
The column corresponding to each ${\cal V}^{}_i$ on top gives its decomposition in the basis 
$\{ {\cal B}(s,t),{\cal C}(s,t),1\}$. 

By looking at the table in \eqref{rationalprefactorEllis}, and comparing with \eqref{ampliMellinnn},
it is immediately apparent that $(1+u)$ and $(2+u)$ are the only denominators that we need in 
order to match the rational coefficients that multiplies ${\cal B}(s,t)$ in \eqref{ampliMellinn}. 
Moreover we observe that $(1+u)$ must be the `central node' accompanying ${\cal V}^{}_{i=5}$ 
in the first column, i.e.~${\cal B}(s,t)$ with no shifts, and the reason is that $(1+u)$ does not 
appear at lower transcendental weight.  Finally, the coefficient in the amplitude with denominator 
$(1+t)$ must be generated enterily by the shifts. Remarkably, there is a solution to these 
constrains,  and the result is precisely \eqref{diff_rep_M}.

Overall, we conclude that the Mellin space representation takes a very compact form.

\subsection{Matching the flat-space amplitude}\label{sec:flat-space}
In this section we supplement our symmetry and OPE constraints, used in the 
previous section to bootstrap the one-loop function ${\cal M}^{(2)}(s,t)$ in \eqref{ampliMellin}, 
with constraints from  the flat-space limit \cite{Penedones:2010ue}.
First we shall explain how to implement the flat-space limit on the one-loop AdS correlator, 
which we will then compare with the result of the four-tensor amplitude in 6d (2,0) supergravity recently computed in \cite{Huang:2025nyr}. For a complementary formulation of the flat-space limit in position space, see Appendix \ref{app:bpl}.

There are several reasons for looking into the flat-space limit. To start with, the flat-space limit provides 
a non-trivial consistency check of our bootstrap result for the one-loop correlator, that so far only 
used input from the CFT.  Then, a detailed analysis of this matching will constrain the two 
ambiguities $\alpha_3$ and $\alpha_4$. Finally, we will provide another perspective on 
the emergence of $n=21$.

In the following, let ${\bs}$, ${\bt}$ and ${\bu}$ denote the 6d {Mandelstam invariants} 
of the scalar fields at the bottom of the tensor multiplets in the 6d (2,0) supergravity, thus
\begin{equation}
\bs=-(p_1+p_2)^2,\qquad \bt=-(p_1+p_4)^2,\qquad \bu=-(p_1+p_3)^2,
\end{equation}
obeying $\bs+\bt+\bu=0$. 
The flat-space amplitude of four tensor multiplets then reads 
\begin{equation}\label{oneloopamp}
    \mathcal{A}^{i_1i_2i_3i_4}(\bs,\bt) =  
    \delta^{i_1i_2}\delta^{i_3 i_4}\mathcal{A}_{}(\bt,\bs)  
    + \delta^{i_1i_4}\delta^{i_2i_3} \mathcal{A}_{}(\bs,\bt) 
    +\delta^{i_1i_3}\delta^{i_2i_4}\mathcal{A}_{}(\bs,\bu)\,,
\end{equation}
where the function $\mathcal{A}_{}(\bs,\bt)$ has the symmetry $\mathcal{A}_{}(\bs,\bt)
=\mathcal{A}_{}(\bu,\bt)$. It will be useful to note that these are the same conventions 
as for ${\cal M}^{i_1i_2i_3i_4}(s,t)$ in \eqref{defMgeneral}.  

The flat-space amplitude admits an expansion in the six-dimensional Newton's constant, 
$G_N$, whose first two terms are
\begin{equation}\label{eq:flat_expansion}
\mathcal{A}^{i_1i_2i_3i_4}(\bs,\bt)= (8\pi G_N)  \mathcal{A}^{(1)i_1i_2i_3i_4}(\bs,\bt) +\frac{(8\pi G_N)^2}{(4\pi)^3}\mathcal{A}^{(2)i_1i_2i_3i_4}(\bs,\bt) + \ldots \,.
\end{equation}
The tree-level term $\mathcal{A}^{(1)}_{}(\bs,\bt)$ is well known and given by \cite{Lin:2015dsa,Heydeman:2018dje}
\begin{align}\label{eq:flat_tree}
\mathcal{A}^{(1)i_1i_2i_3i_4}(\bs,\bt)= - \frac{  \delta^{i_1i_2}\delta^{i_3 i_4} }{\bf s} - \frac{  \delta^{i_1i_4}\delta^{i_2 i_3} }{\bf t} - \frac{  \delta^{i_1i_3}\delta^{i_2 i_4} }{\bf u}\,.
\end{align}
In particular, ${\cal A}^{(1)}(\bs, \bt)=-\frac{1}{\bt}$ in \eqref{oneloopamp}. 
The one-loop contribution $ \mathcal{A}^{(2)}_{}(\bs,\bt)$, and its crossing images, 
have been recently calculated in \cite{Huang:2025nyr} by combining constraints 
from the two-particle unitarity cut and crossing symmetry. In this section we will use 
the result quoted in \cite{Huang:2025nyr}, specified to $n=21$, and then consider 
the $n\neq 21$ result in the next section. In our notation, the amplitude of \cite{Huang:2025nyr} 
reads\footnote{\label{foot:scheme_dependence}The last term $-\frac{2}{3}\bt$, being cut- and singularity-free, is scheme-dependent 
and has been computed by using the double-copy construction in dimensional regularisation.}
\begin{align}\label{eq:Fs}
	\mathcal{A}^{(2)}_{}(\bs,\bt)\Big|_{n=21}=-\bs^2 {\cal B}^{\oneladder}_{\text{6d}}(\bs,\bt) 
	-\bu^2 {\cal B}^{\oneladder}_{\text{6d}}(\bu,\bt)
	-\frac{2}{3}\bt\,,
\end{align}
where the function ${\cal B}^{\oneladder}_{\text{6d}}(\bs,\bt)$ is the 6d scalar box 
diagram. Assuming the Euclidean region $\bs,\bt<0$,  the 6d scalar box  diagram 
evaluates to \cite{Henn:2014qga}
\begin{equation}\label{eq:box_6d}
	{\cal B}^{\oneladder}_{\text{6d}}(\bs,\bt)=-\frac{1}{2}\frac{1}{\bu}\Big(\log^2\Big(\frac{-\bs}{-\bt}\Big)+\pi^2 \Big)\,.
\end{equation}
(Note that we have not simplified the signs for reasons that will become clear in \eqref{rel_db_flat}).

An important observation that we will use later on is the following: The flat space amplitude 
cannot have a non-zero residue in the Mandelstam variables apart from those of the graviton pole at tree-level.
This implies that the residue at ${\bf t}=0$, or ${\bf s}=0$, or ${\bf u}=0$, of the one-loop contribution must be vanishing. 
In fact, one can check that in the physical region (${\bf s}<0$ and ${\bf t}>0$) there are no such poles in ${\cal A}^{(2)}({\bf s},{\bf t})$.\footnote{We 
recall that the physical region is ${\bf s}<0$ and ${\bf t}>0$, where the logarithm in ${\cal B}(s,t)$ is continued as $\log +i\pi$.}

The flat-space limit is straightforward in Mellin space, since it corresponds to taking the limit of large Mellin variables  of {the} Mellin amplitude, and match it with the formula given 
in \cite[Eq.~(9)]{Penedones:2010ue}. Compared to this reference, there are two additional things 
to do: First, the dimensionality of the CFT should be specified to $d=2$.  Second, we have to replace 
$\Sigma_{\text{there}}\rightarrow \frac{p_1+p_2+p_3+p_4}{2} + 2$ in order to account for the fact that 
we study the flat-space limit of the reduced correlator. Finally,  upon identifying the Mellin variables 
in the Mack's gamma functions, we arrive at the result, 
\begin{equation}\label{Penedones_2d}
\lim_{s,t\gg 1}{\cal M}_{p_1p_2p_3p_4}(s,t) = {\cal N}_{\Sigma}\,\frac{L}{\Gamma[1+\Sigma]}\, 
\int_0^{\infty}\!\!d\beta\, e^{-\beta} \beta^{\Sigma} {\cal A}_{}\left(\frac{4\beta}{L^2} s, \frac{4\beta}{L^2} t\right),
\end{equation}
where ${\cal M}_{p_1p_2p_3p_4}(s,t)$ is the Mellin amplitude, $\Sigma=\frac{p_1+p_2+p_3+p_4}{2}$ and ${\cal N}_\Sigma$ is a normalisation 
that we fix so as to normalise to unity the result of the integral over $\beta$ at tree-level, see \eqref{eq:flat_tree}. 
This gives ${\cal N}_\Sigma= \frac{4\Sigma}{(8\pi G_N) L^3}$.

We shall now consider \eqref{Penedones_2d} at one-loop. We notice that ${\cal A}^{(2)}(\bs,\bt)$ 
is homogenous with respect to a rescaling of the Mandelstam variables. In particular, the rescaling 
by $\beta$ in \eqref{Penedones_2d} cancels out in the transcendental funtions, and only the 
rational prefactors carry information about $\beta$, indeed through homogeneous functions of 
some degree (essentially because the amplitude describes massless scattering). Then, the $\beta$ 
integral will evaluate to a number. Finally, to relate RHS and LHS we have to use the AdS/CFT 
dictionary and write $G_N$ in terms of the CFT parameter $N$. Note that we have to consider 
the six-dimensional $G_N$ which is related to $N$ through the relation,
 \begin{equation}\label{adscft}
 G_N=\frac{\pi^2 L^4}{2 N}\,.
 \end{equation}
We can check \eqref{adscft} by considering the AdS$_3$ effective action, where one would find 
$\text{Vol(S}^3\text{)}/(16\pi G_N)=N/(4\pi L^2)$, in agreement with the conventions of \cite{Taylor:2007hs}.

In the above framework, we will compute
\begin{equation}\label{limitpenedones1oneloop}
\lim_{s,t\gg 1}{\cal M}^{(2)}_{1111}(s,t) = {\cal N}_{\Sigma}\,\frac{L}{\Gamma[1+\Sigma]} 
\int_0^{\infty}\!\!d\beta\, e^{-\beta} \beta^{\Sigma} {\cal A}^{(2)}\left(\frac{4\beta}{L^2} s, \frac{4\beta}{L^2} t\right).
\end{equation}
Dropping the subscript, since it is clear that we will focus on $p_i=1$, the Mellin amplitude on the LHS of the previous equation is the one in \eqref{ampliMellin}, namely
\begin{equation}
{\cal M}_{}^{(2)}(s,t)= \widetilde{\cal M}^{\twoladder}(s,t)+ \widetilde{\cal M}^{\twoladder}(u,t) + \sum_{i=1}^{4} \alpha_i {\cal M}_{B_i}\,,
\end{equation}
with $u=-s-t-2$.
Let us first comment on the limit  for the seed functions in $\widetilde{\cal M}^{\twoladder}(s,t)$. 
Consider the (kernel of the) double box $\phi^{(2)}$ in \eqref{eq:Mellin_kernels}, which we 
rename as ${\cal B}^{\twoladder}(s,t)$. Upon taking the limit of large Mellin variables we find the relation
\begin{equation}\label{rel_db_flat}
\lim_{\Lambda\rightarrow \infty} {\cal B}^{\twoladder}(\Lambda \bs,\Lambda \bt)= -\frac{1}{2}\bigg( \big( \log(-\bs) -\log(-\bt)\big)^{\!2}+\pi^2 \bigg)=  \bu\, {\cal B}^{\oneladder}_{\text{6d}}(\bs,\bt)\,.
\end{equation}
Note that the result is homogenous of degree zero, and that $\log(\Lambda)$ terms cancel 
out. Now, even though the double-box function is a weight four function in position space, 
we have to consider that in Mellin space there is a reduction due to the Gamma function, 
which aposteriori explains why the relation \eqref{rel_db_flat} holds. In Mellin space we also 
find the result, 
\begin{equation}
\lim_{\Lambda\rightarrow \infty} {\cal C}(\Lambda \bs,\Lambda \bt)= \log(-\bs)-\log(-\bt)\,.
\end{equation}
From the point of view of the differential representation \eqref{diff_rep_M}, the limit of ${\cal C}(\bs,\bt)$ 
arises from those shifts of the Mellin variables which act as first order derivatives when $s,t$ are large. 

Considering the expression of $\widetilde{\cal M}^{\twoladder}(s,t)$ in \eqref{ampliMellinnn}, it is 
simple to see that only the ${\cal B}(s,t)$ term contributes at leading order, which is $O(\Lambda)$.  
In particular, its coefficient function  becomes  $-6s^2/u$ in the limit of large Mellin variables.
Then, the two ambiguities ${\cal M}_{B_3}$ and ${\cal M}_{B_4}$ give a contribution at the same order, $O(\Lambda)$, while ${\cal M}_{B_1}$ and ${\cal M}_{B_2}$ are subleading. 
From the explicit computation of formula \eqref{limitpenedones1oneloop} we finally obtain 
\begin{equation}\label{matching_flat}
\lim_{\Lambda\rightarrow \infty} \frac{1}{\Lambda}{\cal M}^{(2)}_{1111}(\Lambda \bs,\Lambda \bt)=  \bigg[-{6\bs^2} {\cal B}^{\oneladder}_{\text{6d}}(\bs,\bt) + (\bs\leftrightarrow \bu) \bigg] 
-\alpha_3\bt+\alpha_4 \frac{\bs\bu}{\bt}
=6\times {\cal A}^{(2)}(\bs,\bt)\Big|_{n=21}\,,
\end{equation}
where the integral over $\beta$ is responsible for the factor of $6$ on the RHS, coming from the Pochhammer $(\Sigma)_2=6$, with $\Sigma=\frac{1+1+1+1}{2}$.

Remarkably, the 6d one-loop box part of the equality in \eqref{matching_flat} matches on the nose, and we are left with a constraint on the values of the two ambiguities $\alpha_3$ and $\alpha_4$. For $\alpha_4$, we immediately obtain
\begin{equation}\label{fixingalpha3alpha4}
\alpha_4=0\,.
\end{equation}
This is a consequence of the statement that in the one-loop flat space 
amplitude there cannot be a non-zero residue in the Mandelstam variable ${\bf t}=0$, 
(or in the other two orientations, ${\bf s}=0$ or ${\bf u}=0$, respectively) and as we explained around \eqref{eq:box_6d}
the 6d box has  already vanishing residue.

The fact that $\alpha_4=0$ shows that the genuine ambiguities of the double-particle bootstrap 
are indeed polynomials in the Mellin variables. The coefficients of ${\cal M}_{B_2}$ and ${\cal M}_{B_3}$ can then be interpreted as capturing higher-derivative corrections in a gravitational EFT, where $\M_{B_2}$ corresponds to a quartic contact term interaction
and $\M_{B_3}$ corresponds to a two-derivative term.

Finally, we can match the coefficient of the linear term in $\bt$ in formula \eqref{matching_flat}, yielding $\alpha_3=4$.
However, in the calculation of the flat space amplitude \eqref{eq:Fs} in \cite{Huang:2025nyr}, the linear term in $\bt$ is scheme-dependent. 
As explained there, the particular value $-\frac{2}{3}$ has been obtained in dimensional regularisation. It would be interesting to resolve this scheme-dependence in the full string theory already in flat space, but this task is beyond the scope of our work.

\section{Lessons from UV finiteness in flat space and $n\neq21$}\label{sec:one-loop_extended}
In this section we come back to the problem of $n\neq 21$. First we would like to recall how we 
arrived at the constraint on the value of $n$. This  originated from combining crossing symmetry 
and the OPE predictions in the singlet of $SO(n)$, namely
\begin{align}\label{equation_crossing_sec6}
{\cal H}^{(2)}_{\I}\Big|_{\log^2(U)} = 
\bigg[ {\cal H}^{(2)}_{t}+ {\cal H}^{(2)}_{u}+ n {\cal H}^{(2)}_{s}\bigg] _{\log^2(U)} = 
\frac{1}{2} \sum_{h,\hb} \big[A^{(0)}_{\I}(\gamma^{(1)}_{\I})^2\big]_{h,\hb} B_{h,\bar{h}}\,.
\end{align}
On the one hand, the RHS is quadratic in $n$ when $h=\hb$. On the other hand,
crossing symmetry implies that there is only one independent function on the LHS.
Hence the value of $n$ cannot be arbitrary if such a function is $n$-independent .
It is clear from our reasoning how to relax this constraint: We have to consider 
a one-loop function with explicit $n$ dependence, so that 
both LHS and RHS are polynomials of the same degree in $n$. 

Let us proceed our analysis in Mellin space, taking advantage of the results from the 
previous section. Following from above, the most general parametrisation of the one-loop 
correlator is thus
\begin{equation}\label{genstruct1111}
{\cal M}^{(2)}(s,t)= \Bigg[ \widetilde{\cal M}^{\twoladder}(s,t)+ \widetilde{\cal M}^{\twoladder}(u,t)+ \sum_{i=1}^{4} \alpha_i\, {\cal M}_{B_i}\Bigg] + (n-21) {\cal R}(s,t)\,,
\end{equation}
where the contribution in parenthesis is precisely the one computed in previous sections, 
see \eqref{ampliMellin}-\eqref{ampliMellinn}, for which we know that $n=21$, while 
the last function, ${\cal R}(s,t)={\cal R}(u,t)$, is the new remainder.

We would like to rephrase the constraint on $n=21$ as the absence of a specific 
function, namely ${\cal R}(s,t)$, subject to its own bootstrap constraints.  
In fact, ${\cal R}(s,t)$ cannot be arbitrary. First, it must not contribute to the $\log^2(U)$ 
discontinuity of ${\cal M}^{\twoladder}(s,t)$, which is $n$-independent
and captured already by our previous result. Therefore ${\cal R}(s,t)$ 
cannot have simple poles in the $s$-plane. Second, the combination of functions that 
enter the singlet channel must be consistent with the $\log^2(U)$ predictions 
explained in \eqref{refine_singletpred1}-\eqref{refine_singletpred2}, 
this time for all values of $n$. More explicitly, by writing the singlet channel as
\begin{equation}\label{Randthesinglet}
{\cal M}^{(2)}_{\I}= \bigg[ \widetilde{\cal M}^{\twoladder}(s,t) + \widetilde{\cal M}^{\twoladder}(s,u)  
+n\,\widetilde{\cal M}^{\twoladder}(t,s) +\ldots \bigg] + (n-21)\bigg[ {\cal R}(s,t)+{\cal R}(s,u) + n {\cal R}(t,s) \bigg],
\end{equation}
(omitting the ${\cal M}_{B_i}$ in the parenthesis)
we must find that the amplitude in \eqref{Randthesinglet} computes a $\log^2(U)$ 
discontinuity that is consistent with
\begin{align}
	\label{refine_singletpred11}
	\big[A^{(0)}_{\I}(\gamma^{(1)}_{\I})^2\big]_{h= \bar{h}}  = &\
	 \textrm{polynomial in \emph{n} of degree}\,\leq 2 \,,\\[.2cm]
	\label{refine_singletpred22}
	\big[A^{(0)}_{\I}(\gamma^{(1)}_{\I})^2\big]_{h\neq \bar{h}}  = &\
	\textrm{polynomial in \emph{n} of degree}\,\leq 1\,.
\end{align}
We note that \eqref{refine_singletpred11} is consistent with \eqref{Randthesinglet} only if the 
last term, $(n-21)n{\cal R}(t,s)$, contributes to the $\log^2(U)$ discontinuity. In this way 
we can at least match the power of $n$ on both sides of \eqref{equation_crossing_sec6}, i.e. a polynomial in $n$ of degree $\leq 2$.
As a result, we learn that ${\cal R}(s,t)$ is a single-variable function of $t$ with infinitely many simple poles.\footnote{Proof: The combination ${\cal R}(s,t)+{\cal R}(s,u)$ does not contribute 
to the $\log^2(U)$ discontinuity because of the first constraint.  Therefore 
\eqref{refine_singletpred11}-\eqref{refine_singletpred22} is compatible with \eqref{Randthesinglet} 
only if ${\cal R}(t,s)$ contributes to the $\log^2(U)$ discontinuity at $h=\hb$. Then,  ${\cal R}(t,s)$ 
must be a single-variable function of $s$ with infinitely many simple poles in $s$.}

It is useful at this point to note the the flat-space limit \cite{Penedones:2010ue} provides a direct translation 
between our CFT reasoning and the flat-space amplitude. In particular, it must be the case that the flat-space 
amplitude has the same structure as in \eqref{genstruct1111}. This expectation can be immediately validated 
by looking at the result of \cite{Huang:2025nyr}, from which we find 
\begin{equation}\label{eq:A2}
\mathcal{A}^{(2)}_{}(\bs,\bt)\ = \ 
	\bigg[ -\bs^2 {\cal B}^{\oneladder}_{\text{6d}}(s,t)(\bs,\bt) 
	+(\bs\leftrightarrow \bu)
	-\frac{2}{3}\bt\, \bigg]- (n-21) \frac{\bt}{12} \log\Big({-}\frac{\bt}{\mu^2}\Big)\,, 
\end{equation}
where $\mu$ is a renormalisation scale. Quite interestingly, the flat-space result manifests 
the fact that $n=21$ implies UV finiteness of the amplitude and vice versa. 

In sum, recalling that the poles of the $\Gamma$ functions that define the Mellin amplitude 
${\cal M}(s,t)$ are in correspondence with the quantum numbers of the double-particle operators, 
and that in our conventions these $\Gamma$ functions are $\Gamma[-s]^2\Gamma[-t]^2\Gamma[s+t+2]^2$,  
we conclude that
\begin{itemize}  
\item ${\cal R}(s,t)$ is constrained to be a single-variable function of $t$ with infinitely many simple 
poles at the location of double-particle operators, namely $t=0,1,2,\ldots$.
\item ${\cal R}(s,t)$ is constrained by the flat-space limit 
\begin{equation}
\lim_{\Lambda\rightarrow \infty} \frac{1}{\Lambda}{\cal R}^{}_{}(\Lambda \bs,\Lambda \bt)= - \frac{\bt}{2} \log\Big({-}\frac{\bt}{\mu^2}\Big)\,.
\end{equation}
The fact that the flat-space limit depends on the scaling parameter is 
analogous to the claim of \cite{Huang:2025nyr} that the flat-space amplitude is not UV 
finite when $n\neq21$. 
\end{itemize}
The above constraints restrict the space of possibilities to the following ansatz,
\begin{equation}\label{MellinforRst}
{\cal R}(s,t) = \left( a_1 t + a_2  + \frac{a_3}{t+1} \right) \big( \psi(-t) +\gamma_E\big)\,,
\end{equation}
where $\gamma_E$ is the Euler-Mascheroni constant, and $a_1,a_2,a_3$ 
are numerical coefficients that have to be determined.
The presence of $a_3$ seems unnatural at first, but it can not be excluded since above
Mellin amplitude in \eqref{MellinforRst} has vanishing residue at $t=-1$, as expected for 
a pole below the two-particle threshold.

The flat-space limit determines the value of $a_1$ and then we can use the 
OPE predictions at $h=\hb=1,2$ to find $a_2,a_3$.
By doing so we find that the Mellin amplitude $\mathcal{R}(s,t)$ is simply given by
\begin{align}\label{remainder_mellin}
	\mathcal{R}(s,t)=-\frac{1}{2}\left(t+\frac{2}{3}\right)\,\big(\psi(-t)+\gamma_E\big)\,,
\end{align}
and in particular $a_3$ turns out to vanish.

At this point it is interesting to ask about the position space result for ${\cal R}(s,t)$, 
since it must lie outside the space of SVHPL functions that we used in Section \ref{sec:one-loop}.
The missing element is $f^{(3)}(x,\xb)$: the \emph{unique} weight-three, single-valued, 
fully antisymmetric function that involves the additional letter $x-\xb$. Due to the appearance 
of this new letter, the function lies outside the space of harmonic polylogarithms,
but precisely this feature is the one that provides the desired scale-dependent logarithmic term
when taking the bulk-point limit (see the discussion around equation \eqref{eq:bpl_fs} in Appendix \ref{app:bpl}). 
The same function was discussed in \cite{Drummond:2019hel} as a crucial building block of one-loop string corrections to graviton scattering on AdS$_5\times$S$^5$, but here we need this new ingredient already in supergravity.

In parallel with our previous analysis in Mellin space, we introduce the function 
\begin{align}\label{frack_fs}
\begin{split}
\frac{{\mathfrak{f}}_s(x, \xb)}{x-\xb} &=\oint ds\,dt\, U^s V^t\,\Gamma(-s)^2\Gamma(-t)^2\Gamma(s+t+1)^2\times \big( \psi(-s)+ \gamma_{E} \big) \,.
\end{split}
\end{align}
Then, the precise combination of weight-three functions that nicely maps to \eqref{frack_fs} 
is a combination of $f^{(3)}$ and a weight-three SVHPL function \cite{Heslop:2023gzr}:\footnote{The definitions given in  \cite{Heslop:2023gzr} differ by an overall factor from those given in this paper, specifically  $f^{(3),\text{there}}=2f^{(3),\text{here}}$ and $\mathfrak{f}_s^{\text{there}} = \frac{2}{x-\xb}\mathfrak{f}_s^{\text{here}}$.}
\begin{align}\label{eq:fs_def}
{\mathfrak{f}}_s(x, \xb)= \frac{1}{3}  \Big( f^{(3)}(x,\xb)+ \frac{1}{2} \log(U)\phi^{(1)}(x,\xb) +  \frac{1}{2}  \log \Big(\frac{U}{V}\Big)\phi^{(1)}(x,\xb) \Big) \,.
\end{align}
This function is fully characterised by its total derivative:
\begin{align}
\begin{split}
 d \, {\mathfrak{f}}_s(x,\xb) &= \frac{1}{2}\log(U) \log\left(\frac{U}{V}\right) d \log\left(\frac{1-x}{1-\xb}\right)
	-\frac{1}{2} \phi^{(1)}(x,\xb) \, \left( d\log(U){+} \ d\log \left(\frac{U}{V}\right) \right)\\
	&\quad -2\phi^{(1)}(x,\xb)\,d\log\left(\frac{x-\xb}{x \xb}\right).
\end{split}
\end{align}
Note the invariance under the $x\to x'$ transformation, corresponding to $1\leftrightarrow2$ exchange symmetry. 
The other two orientations are $ {\mathfrak{f}}_t(x, \xb) = - {\mathfrak{f}}_s(1-x, 1-\xb) $ and ${\mathfrak{f}}_u(x, \xb) = - {\mathfrak{f}}_s(1/x,1/ \xb)$, whose Mellin amplitudes are given by $\psi(-t)+\gamma_{E}$ and $ \psi(-u)+\gamma_{E}$, respectively.

We now reconsider the bootstrap analysis of ${\cal R}(x,\xb)$ from the point of view of position space, 
and by doing so we will corroborate our findings in Mellin space. 
The basis of functions, including the lower weight completion, is built out of the following ten elements, 
\begin{align}\label{eq:basis10}
\begin{split}
	\text{weight 3}^-:&\qquad\quad{\mathfrak{f}}_s(x, \xb)\,,{\mathfrak{f}}_t(x, \xb)\,,{\mathfrak{f}}_u(x, \xb)\,,\\
	\text{weight 2}^+:&\qquad\quad\log^2(U)\,,\log(U)\log(V)\,,\log^2(V)\,,\\
	\text{weight 2}^-:&\qquad\quad\phi^{(1)}(x,\xb)\,,\\
	\text{weight 1}^+:&\qquad\quad\log(U)\,,\log(V)\,,\\
	\text{weight 0}^+:&\qquad\quad1\,.
\end{split}
\end{align}
Calling $\widetilde{Q}_i$ the elements of this basis, 
we make an ansatz based on \eqref{eq:basis10} with a form 
similar to the one described in \eqref{eq:ansatz_G}, 
\begin{align}
	{\cal R}(x,\xb) = \sum_{i=1}^{10}\frac{\tilde{p}_i(x,\xb)}{(x-\xb)^{d_\pm}}\,\widetilde{\Q}_i(x,\xb)\,,\quad \text{with}~~\tilde{p}_i(x,\xb) = \sum_{n=0}^{d_\pm}\sum_{m=n}^{d_\pm}\tilde{a}^{(i)}_{n,m}(x^n\xb^m+x^m\xb^n)\,.
\end{align}
Note that here we omitted the factor of $1/V$, as this was required to match the structure of the leading 
log predictions which here are already taken care of by the main function ${\cal F}^{(2)}(x,\xb)$ in \eqref{definitionF2explicit}.

We then proceed by imposing (1) crossing symmetry, (2) absence of unphysical poles, and (3) \textit{absence} of log$^2(U)$ 
contributions to the $t$- and $u$-channel -- for the same reason that the main function already matches the leading logs 
in the symmetric and antisymmetric channels. In addition, we need to enforce that $s$-channel contribution to the leading 
log contributes only to spin $\ell=0$. This leads to the result expected from Mellin space, where differently from the 
calculation in Section \ref{subsec:bootstrap} the $\log^2(U)$ part in the singlet channel still contains 3 
unfixed parameters. These correspond precisely to the parameters $a_1,a_2,a_3$ in \eqref{MellinforRst}.

The function $\mathcal{R}(x,\xb)$  can be given very compactly in terms the following function:
\begin{equation}\label{eq:Rxxb}
\mathcal{R}(x,\xb) = -\frac{1}{6}\big(U \partial_U+V \partial_V+1 \big)^2\big(3V \partial_V+ 2\big)\,\bigg[\frac{\mathfrak{f}_t(x, \xb)}{x-\xb}\bigg]\,,
\end{equation}
whose Mellin transform can be readily verified to coincide with \eqref{remainder_mellin}.
This completes the bootstrap of the remainder function when $n\neq21$.

\section{Conclusions and outlook}\label{sec:discussion}
In this work we used the double-particle bootstrap in AdS to explicitly construct the 
one-loop correction  
to the four-point correlator $\langle s^{i_1}_1s^{i_2}_1s^{i_3}_1s^{i_4}_1\rangle$, 
dual to the scattering of tensor multiplets in AdS$_3\times$S$^3$ supergravity. As 
we explained, our approach is based on leveraging data from the consistency of the 
OPE in the dual field theory. In particular, we used disconnected and tree-level 
correlators to make predictions for the maximal logarithmic discontinuities at all 
loop orders, and then we used these predictions to set up a bootstrap problem 
for the reduced one-loop correlator, $\H^{(2)i_1i_2i_3i_4}(x,\xb)$. As a result we 
obtained a unique function, up to a handful of tree-like ambiguities. Remarkably, given our one-loop function, we found that crossing symmetry and the 
consistency of the OPE in the singlet channel of $SO(n)$
determines the value of $n$ to be $n = 21$. 

In terms of the explicit one-loop result, we presented our function $\H^{(2)i_1i_2i_3i_4}(x,\xb)$ 
both in position space and Mellin space. In position space, we found very useful to 
extract a differential Casimir operator $\dfour$, which allowed us to write the correlator 
in few lines. We then considered the Mellin amplitude and found that it can be written as a 
differential representation, using only the double-box $\mathcal{B}(s,t)$ with 
shifted arguments. This result is even simpler: it fits one line. Then, from Mellin space, 
we built our way to the  flat-space limit finding perfect agreement with a recent computation 
of \cite{Huang:2025nyr}, thus providing another non-trivial consistency check of our 
one-loop correlator.

Ideally we would like to test the consistency of the OPE, in relation to $n=21$, for all exchanged
double-particle operators. Besides the known tree-level correlators  \cite{Rastelli:2019gtj,Giusto:2020neo,
Behan:2024srh,Aprile:2025kfk}, involving the tensor multiplets $s^I_p$ and the 
gravitons $\sigma_q$, these tests would require knowledge of tree-level correlators 
involving the spin-1 primaries $V_p^{\pm}$, namely the correlators $\langle s_1^{i_1}s_1^{i_1}V_p^{\pm}V_p^{\pm}\rangle$ 
which are currently unknown. A solution to this problem would pave the way to systematise 
the bootstrap approach to four-point correlators with arbitrary KK-modes in AdS$_3\times$S$^3$,  
in various directions: one-loop correlators with both tensors and gravitons, 
mixed correlators with single and multiparticle external operators, along the lines of \cite{Giusto:2019pxc,Giusto:2020neo,Ceplak:2021wzz,Aprile:2025hlt, Aprile:2026uxe}, 
and two-point functions in the presence of a defect, along the lines of \cite{Chen:2024orp,Chen:2025yxg}.
 We hope to come back to this point in the future.\footnote{The unmixing 
of double-particle operators is crucial also for computing string corrections \cite{Chester:2024wnb,Jiang:2025oar}.}

In this work we have taken a first concrete step in the direction of unmixing of the singlet of $SO(n)$. First, we counted 
the degeneracy of the double-particle operators  for all quantum numbers. Second, we solved the 
mixing of operators for the first non-trivial system in which both tensors and gravitons 
participate. Notably, this calculation revealed a surprise: the anomalous 
dimensions become rational for $n=21$ and moreover one of them vanishes.

We emphasise that so far two apparently {independent} calculations lead  
to $n=21$:\vspace{-0.2cm}
\begin{itemize}\setlength\itemsep{-2pt}
\item[$1)$] at tree level for $h=\hb=2$: in the unmixing of the anomalous dimensions 
of long double-trace operators in the singlet channel, 
\item[$2)$] at one loop for $h=\hb=1$ and $h=\hb=2$: in the one-loop bootstrap calculation as a 
consequence of crossing symmetry together with the leading log OPE prediction.
\end{itemize}\vspace{-0.3cm}
We find it intriguing that in both circumstances something remarkable happens when 
$n=21$ and we would like to comment on some interesting open questions that are related to this.

We mentioned above that the spectrum of double-particle operators in the singlet channel with  
$h=\hb=2$ is special when $n=21$. In particular, we found that one of the three anomalous 
dimensions vanishes at tree level. This implies that even though there are three 
operators which mix, the CFT data that we extract from the tree-level correlators sees only two of them. 
It  is unlikely that supersymmetry alone can explain this result, but we can speculate  
about two features of our setup that might be connected to this. Firstly, there is the tree-level 
hidden 6d symmetry \cite{Rastelli:2019gtj,Giusto:2020neo}. In practise we know that correlators 
$\langle {\cal O}_{p_1}{\cal O}_{p_2} {\cal O}_{p_3} {\cal O}_{p_4}\rangle$ with ${\cal O}_p\in\{s_p,\sigma_p\}$
admit generating functions from master correlators in 6d. For tensor-tensor correlators the 
master integral is conformal and there is a relation between the block decomposition 
in AdS$_3$ and the block decomposition in 6d. This symmetry is also present for the tensor-graviton 
and graviton-graviton correlators but a similar relation between their corresponding block decompositions 
has not been explored. What  we can do is to connect with the analysis done in the flat-space limit. In 
this case we observe that the  partial-wave decomposition of the graviton cut to the one-loop flat-space 
amplitude \cite{Huang:2025nyr} has vanishing coefficient for spin $\ell_{\rm 6d}=0$. 
This might imply a version of the phenomenon called rank-constraints studied in \cite{Aprile:2020mus} 
and thus it might be related to the vanishing anomalous dimension that we see in AdS$_3$. 
If this relation is correct, one might expect to find vanishing anomalous dimensions for higher 
values of $h=\hb$ which correspond to  the aforementioned $\ell_{\rm 6d}=0$.
Secondly, the theory with $n=21$ can be embedded into string theory, so perhaps a greater symmetry enhancement  
might be responsible for the observed vanishing.
For instance, symmetry enhancement arises sometimes in string compactifications, and in particular it is known 
to take place for K3. More generally, an enhancement of the superconformal algebra through the addition 
of spectral flow generators is expected for worldsheet sigma-models with target space that are Ricci flat 
and Kähler, such as K3, see e.g. \cite{Odake:1988bh,Eguchi:1988vra}.\footnote{We thank Nikolay Bobev 
for bringing this to our attention.}

Another interesting extension of our computation would be to consider the next correction to the 
$\langle s_1^{i_1}s_1^{i_2}s_1^{i_3}s_1^{i_4}\rangle$ correlator at two-loop order. 
For scattering of gravitons on AdS$_5\times$S$^5$ \cite{Huang:2021xws,Drummond:2022dxw} 
and gluons on AdS$_5\times$S$^3$ \cite{Huang:2023oxf},  this problem has been successfully 
tackled by using a differential Casimir operator in order to simplify the form of the ansatz, 
which now contains transcendental functions up to weight six.  
In our setup, the analogous of this differential operator is our $\dfour$.  
It appears then that the same approach is viable in AdS$_3\times$S$^3$  and we hope to make 
progress in this direction in the near future.

Also, it would be important to get access to more bootstrap constraints for the D1-D5 
system. One such constraint could arise from knowledge about the superVirasoro
 blocks at all orders in the large-$c$ expansion. Another constraint could arise from developing 
 integrated correlators technology analogously to what has been done in ${\cal N}=4$ SYM 
 during the last few years \cite{Binder:2019jwn,Chester:2025kvw}, 
see \cite{Dorigoni:2022iem} for a recent review. Solving the $G_N$-expansion of the flat-space amplitude 
could be another recourse.\footnote{A dream would be to reformulate the problem in terms of nonlinear 
integral equations, see \cite{Gumus:2026mhb} for recent progress in this direction.}

\section*{Acknowledgments}
We are grateful to Connor Behan, Nikolay Bobev, Stefano Giusto, Rodrigo Pitombo, Rodolfo Russo, Congkao Wen for useful discussions at various stages of this work, and to Nikolay Bobev, Yu-tin Huang, Henrik Johansson, and Congkao Wen for valuable comments on the draft.  We also would like to thank an anonymous referee for their insightful comments and suggestions which helped improve the quality of the paper.
MS would like to thank the ITF at KU Leuven, Universidad Complutense, and Nordita for their kind hospitality during the final stages of this work.

FA is supported by
RYC2021-031627-I funded by MCIN/AEI/10.13039/501100011033 and by the NextGeneration EU/PRTR. 
FA also acknowledges support from the
``HeI" staff exchange program \href{https://cordis.europa.eu/project/id/101182937}{DOI 10.3030/101182937} 
financed by the MSCA.  The work of HP is supported in part by the FWO projects G003523N, G094523N, and G0E2723N, 
and the KU Leuven C1 project C16/25/01. The research of MS is funded by the European Union (ERC Synergy Grant MaScAmp 101167287). 
Views and opinions expressed are however those of the author(s) only and do not necessarily reflect those of the European 
Union or the European Research Council Executive Agency. Neither the European Union nor the granting authority 
can be held responsible for them.

\appendix
\section{Details on the unmixing}\label{APPA}

\subsection{Long blocks for generic R-symmetry representations}\label{app:blocks}
In this appendix we provide the necessary details on the long blocks decomposition that is relevant in the case of correlators with pairwise equal external single-particle operators. 
In general these have a non-trivial dependence on $(y,\yb)$,  See Appendix \ref{app:KK_correlators} for examples of such correlators. 

For correlators wtih pairwise equal external single-particle operators, we shall decompose ${\cal L}$ into an auxiliary bosonic decomposition that reads as follows, 
\begin{align}\label{eq:block_decomposition_general}
	\Ht_{\mathbf{a}}(x,\xb, y,\yb) = \sum_{h,\bar{h},j,\bar{j}} A_{\mathbf{a},h,\bar{h},j,\bar{j}} \,L_{h,\bar{h},j,\bar{j}}(x,\xb, y,\yb) \,,
\end{align}
where the relevant building blocks take the form of a product of the bosonic block ${B}_{h,\bar{h}}(x,\xb)$ from \eqref{eq:long_block} and the spherical harmonics of the R-symmetry group $SU(2)_R\times SU(2)_L$:
\begin{equation}
L_{h,\bar{h},j,\bar{j}} ={B}_{h,\bar{h}}(x,\xb) \times Y_{j,\bar{j}} (y,\yb)\,,
\end{equation}
with $Y_{j,\bar{j}} (y,\yb)$ given by
\begin{align}
 {Y}_{j,\bar{j}}(y,\yb) &= \frac{1}{2}\Big( g_{-j-1} (y)g_{-\bar{j}-1}(\bar{y})+  g_{-j-1} (\yb)g_{-\bar{j}-1}({y}) \Big) \,,
\end{align}
and we recall the definition of $g_h (x)=x^{h} {}_2 F_1 (h+1,h+1,2h+2;x)$.

\subsection{Summary of tree-level correlators of higher KK-modes}\label{app:KK_correlators}
We collect the tree-level Mellin amplitudes that have been used to compute one-loop leading logs and anomalous dimensions in the main text. 
For the one-loop leading logs, we needed four-tensor and mixed tensor-graviton correlators in the charge configuration $p_1=p_2=1$, $p_3=p_4=p$. In addition to these, the computation of the $h=\bar{h}=2$ anomalous dimensions also requires tensor and graviton correlators in the configurations $p_1=p_2=p_3=p_4=2$.

In the following we shall adopt the conventions of \cite{Aprile:2025kfk}, with the two-point functions of $s_p^{i}$ and $\sigma_p$ normalised so that:
\begin{align}\label{twopt_norm}
\begin{split}
\langle s_p^{i_1}(z_1,w_1) s_p^{i_2}(z_2,w_2) \rangle  & = \frac{\delta^{i_1 i_2}}{p} \frac{|w_{12}|^{2p}}{|z_{12}|^{2p} }\,, \qquad~~\, p\geq 1\,, \\
 \langle  \sigma_p(z_1,w_1) \sigma_p(z_2,w_2) \rangle  & = \frac{p^2-1}{p} \frac{|w_{12}|^{2p}}{|z_{12}|^{2p} }\,, \qquad p\geq 2\,.
\end{split}
\end{align}

\noindent {\bf Four-tensor correlator.}  The tree-level four-tensor correlators $\langle s_1^{i_1}s_1^{i_2}s_p^{i_3}s_p^{i_4}\rangle$ are given by
\begin{equation}
{\cal H}_{11pp}^{i_1 i_2 i_3 i_4}= \frac{1}{\Gamma(p)} \oint ds\,dt\, U^s V^t\,\Gamma[-s] \Gamma[-s+1-p]  \Gamma[-t]^2\Gamma[-u]^2 \mathcal{M}_{11pp}^{i_1 i_2 i_3 i_4}\,,
\end{equation}
with $s+t+u=-p-1$ and the Mellin amplitude reads
\begin{equation}
\mathcal{M}_{11pp}^{i_1 i_2 i_3 i_4} = -\frac{\delta^{i_1 i_2}\delta^{i_3 i_4}}{s+p} - \frac{\delta^{i_1 i_4}\delta^{i_2 i_3}}{t+1} - \frac{\delta^{i_1 i_3}\delta^{i_2 i_4}}{u+1}  \,.
\end{equation}
For $p=1$, this agrees with the $\langle s_1^{i_1}s_1^{i_2}s_1^{i_3}s_1^{i_4}\rangle$ correlator given in the main text, c.f. \eqref{eq:Mellin_H}-\eqref{eq:M1111}.

In the charge configuration $p_i=2$, i.e. for the correlator $\langle s_2^{i_1}s_2^{i_2}s_2^{i_3}s_2^{i_4}\rangle$, we have
\begin{equation}
{\cal H}_{2222}^{i_1 i_2 i_3 i_4}= \frac{1}{\Gamma(p)} \oint ds\,dt\, U^s V^t\,\Gamma[-s]^2  \Gamma[-t]^2\Gamma[-u]^2 \mathcal{M}_{2222}^{i_1 i_2 i_3 i_4}\,,
\end{equation}
with $s+t+u=-3$ and Mellin amplitude given by
\begin{align}
{\cal M}_{2222}^{i_1 i_2 i_3 i_4} =& - \delta^{i_1 i_2}\delta^{i_3 i_4}\Big( \frac{1+ \tilde V}{s+1}+\frac{\tilde U}{s+2}\Big)-\delta^{i_1 i_4}\delta^{i_2 i_3}\Big( \frac{1+\tilde U}{t+1}+\frac{\tilde V}{t+2}\Big)-\delta^{i_1 i_3}\delta^{i_2 i_4}\Big( \frac{\tilde U+\tilde V}{u+1}+\frac{1}{u+2}\Big)\,.
\end{align}

\noindent {\bf Mixed tensor-graviton correlator. }
The mixed tensor-graviton correlators $\langle s_1^{i_1}s_1^{i_2}\sigma_p\sigma_p\rangle$ read
\begin{equation}
{\cal H}_{11pp}^{i_1 i_2}= \frac{1}{\Gamma(p)} \oint ds\,dt\, U^s V^t\,\Gamma[-s]\Gamma[-s+1-p] \Gamma[-t]^2\Gamma[-u]^2\, \mathscr{M}_{1s}^+\ \delta^{i_1i_2}\,, \quad  
 \mathscr{M}_{1s}^+ = -\frac{(p^2-1)}{s+p} \,,
\end{equation}
with the on-shell constraint $s+t+u=-p-1$.

For the charge configuration $p_i=2$, the correlator $\langle s_2^{i_1}s_2^{i_2}\sigma_2\sigma_2\rangle$ is slightly more involved as the R-symmetry structure opens up:
\begin{equation}
{\cal H}_{2222}^{i_1 i_2}= \frac{1}{\Gamma(p)} \oint ds\,dt\, U^s V^t\,\Gamma[-s]^2  \Gamma[-t]^2\Gamma[-u]^2 \mathscr{M}_{2222}^{}\ \delta^{i_1i_2}\,,
\end{equation}
where $s+t+u=-3$ and the Mellin amplitude reads
\begin{align}
\begin{split}
\mathscr{M}_{2222}=&\bigg[\Big(\frac{1}{s+1}  + \frac{2}{(s+1)(t+1)}\Big) + \Big( {-}\frac{2}{s+1} -\frac{3}{s+2}  + \frac{2}{(t+1)(u+1)} \Big)\,\tilde U\\[5pt]
&+\Big( \frac{1}{s+1}  + \frac{2}{(s+1)(u+1)} \Big)\,\tilde V \bigg]\,.
\end{split}
\end{align}

\noindent {\bf Four-graviton correlator. } Finally, the four-graviton correlator $\langle \sigma_2\sigma_2\sigma_2\sigma_2\rangle$ reads:
\begin{equation}
{\cal H}_{2222}= \frac{1}{\Gamma(p)} \oint ds\,dt\, U^s V^t\,\Gamma[-s]^2  \Gamma[-t]^2\Gamma[-u]^2 \mathcal{M}_{2222}\,,
\end{equation}
with
\begin{align}
\begin{split}
{\cal M}_{2222}=& -\frac{1}{s+1}  - \frac{1}{t+1}
-\frac{4}{u+1} -\frac{9}{u+2} + \frac{12}{(s+1)(t+1)}
\\[5pt] 
& + \Big({-}\frac{1}{t+1} -\frac{1}{u+1}  -\frac{4}{s+1} - \frac{9}{s+2}+ \frac{12}{(t+1)(u+1)}
  \Big)\,\tilde U  \\[5pt]
& + \Big( {-}\frac{1}{s+1} - \frac{1}{u+1}-\frac{4}{t+1}  - \frac{9}{t+2} + \frac{12}{(s+1)(u+1)}\Big)\,\tilde V\,,
\end{split}
\end{align}
where again $s+t+u=-3$.

\subsection{Calculation of anomalous dimensions \eqref{eq:anom_dim_I1} and \eqref{eq:anom_dim_I2}}\label{app:unmixing}
Here we provide the details of the calculation of tree-level anomalous dimensions in the singlet channel. To address the mixing of operators, we need information from a matrix of correlators of size $d_\I\times d_\I$, where $d_\I$ is the number of degenerate double-trace operators given in \eqref{eq:degeneracy_I}. 

For $(h,\hb)=(1,1)$, there is only one operator ($d_\I=1$) and there is no mixing to resolve. We only need the $\langle s_1^{i_1}s_1^{i_2}s_1^{i_3}s_1^{i_4}\rangle$ correlator at disconnected and tree-level order. We have
\begin{align}
	A^{(0)}_\I = \frac{1}{4}\,, \qquad A^{(0)}_\I\gamma^{(1)}_\I = \frac{n-6}{12}\,,\qquad\qquad (h=\hb=1)
\end{align}
from which we readily obtain \eqref{eq:anom_dim_I1}.

For the case $(h,\hb)=(2,2)$, the degeneracy of operators is threefold and we need to consider the (symmetric) matrix of correlators
\begin{align}
\begin{pmatrix}
	\langle s_1^{i_1}s_1^{i_2}s_1^{i_3}s_1^{i_4}\rangle & \langle s_1^{i_1}s_1^{i_2}s_2^{i_3}s_2^{i_4}\rangle & \langle s_1^{i_1}s_1^{i_2}\sigma_2\sigma_2\rangle \\
	\# & \langle s_2^{i_1}s_2^{i_2}s_2^{i_3}s_2^{i_4}\rangle & \langle s_2^{i_1}s_2^{i_2}\sigma_2\sigma_2\rangle \\
	\# & \# & \langle\sigma_2\sigma_2\sigma_2\sigma_2\rangle
\end{pmatrix},
\end{align}
where the correlators with flavour indices need to be projected onto the singlet channel. In order to extract the OPE coefficients from the long part of the correlators involving higher KK-levels, we used the long superconformal blocks discussed in Appendix \ref{app:blocks}.

At disconnected level, we find the diagonal matrix
\begin{align}
\mathbf{A}^{(0)}_\I=\begin{pmatrix}
	\frac{1}{36} & 0 & 0 \\
	0 & \frac{1}{36} & 0 \\
	0 & 0 & \frac{1}{4}
\end{pmatrix},\qquad\qquad (h=\hb=2)
\end{align}
Here, we have used the fact that the disconnected free-theory coefficients for the four-graviton correlator are the same as the one for four-tensor correlator except for a different overall normalisation inherited from the two-point function normalisation \eqref{twopt_norm}.

For the $\log(U)$ part at tree-level we have 
\begin{align}
\mathbf{M}_\I=\begin{pmatrix}
	\frac{n-10}{60} & \frac{5-n}{30} & -\frac{1}{10}\sqrt{n}\\
	\frac{5-n}{30} & \frac{1}{60}(4n-25) & -\frac{1}{20} \sqrt{n} \\
	-\frac{1}{10}\sqrt{n} & -\frac{1}{20}\sqrt{n} & -\frac{63}{20}
\end{pmatrix},\qquad\qquad (h=\hb=2)
\end{align}
where the  factor of $\sqrt{n}$ in the mixed tensor-graviton correlator is needed to ensure the correct normalisation of the singlet channel projector, recall the relation $P_{\I}^{i_1i_2i_3i_4} =\frac{1}{n}\delta^{i_1 i_2}\delta^{i_3 i_4}$.
The anomalous dimensions can then be obtained as the {eigenvalues} of the matrix product
\begin{align}
\gamma^{(1)}_{\I,k}={\tt Eigenvalues}\Big[\mathbf{M}_\I\cdot{\mathbf{A}_\I^{(0)}}^{-1}\Big]\,,
\end{align}
which leads to the result quoted in \eqref{eq:anom_dim_I2}.

\section{Matching via the bulk-point limit}\label{app:bpl}
Here we revisit the flat-space limit from the position space point of view, known as the bulk-point limit. 
The idea is to consider sufficiently localised wave-packets in AdS to focus on a point in the bulk, such that 
curvature effects become negligible. The bulk fields then scatter as if they were in flat-space, and in this limit 
one therefore recovers the corresponding flat-space scattering process. As such, the bulk-point limit is another 
manifestation of the flat-space limit in Mellin space discussed in Section \ref{sec:flat-space}.

The above-mentioned property of AdS amplitudes manifests itself in the dual holographic correlators 
as a singularity at $x=\xb$ \cite{Gary:2009ae}, i.e. schematically one has
\begin{align}
	\H(x,\xb)~\xrightarrow{~\text{bpl}~}~\frac{\mathcal{K}(x)}{(x-\xb)^k}\,,
\end{align}
for some positive integer $k$ and a function $\mathcal{K}(x)$ which is related to the amplitude in flat-space. Note that this bulk-point singularity can be accessed only in the Lorentzian regime, and we thus first need to analytically continue the Euclidean correlator to Lorentzian signature as otherwise one would find a vanishing result in the limit $\xb\to x$. At the level of the conformal cross ratios $(x,\xb)$, this procedure is implemented by continuing $x$ counter-clock wise around 0 and $\xb$ around 1. Then, the bulk-point singularity is exposed by taking the limit $\xb\to x$.

In order to proceed it is useful to apply the analytic continuation and the bulk-point limit described above to the basis of 15 transcendental functions listed in \eqref{eq:basis15}. Explicitly, one has 
\begin{align}\label{eq:bpl_basis15}
\begin{split}
	\phi^{(2)}(x,\xb)~&\xrightarrow{~\text{bpl}~}~-2\pi^2\log(x)\big[\log(x)+2\pi i\big]\,,\\
	\phi^{(2)}(x',\xb')~&\xrightarrow{~\text{bpl}~}~2\pi^2\big[\big(\log(1-x)-\log(x)\big)^2+\pi^2\big]\,,\\
	\phi^{(2)}(1-x,1-\xb)~&\xrightarrow{~\text{bpl}~}~2\pi^2\log(1-x)\big[\log(1-x)+2\pi i\big]\,,\\
	\log(U)\,\phi^{(1)}(x,\xb)~&\xrightarrow{~\text{bpl}~}~8\pi^2\big[\log(x)+\pi i\big]\,,\\
	\log(V)\,\phi^{(1)}(x,\xb)~&\xrightarrow{~\text{bpl}~}~8\pi^2\big[\log(1-x)+\pi i\big]\,,\\
	\phi^{(1)}(x,\xb)~&\xrightarrow{~\text{bpl}~}~4\pi^2\,,\\
\end{split}
\end{align}
with all other basis elements giving a vanishing result.

We now have all the ingredients to compute the bulk-point limit of the correlator $\H^{i_1i_2i_3i_4}(x,\xb)$ and compare to the corresponding flat-space amplitude $\mathcal{A}^{i_1i_2i_3i_4}(\bs,\bt)$. To this end, we introduce a dimensionless variable $z$ defined as
\begin{align}\label{eq:z}
	z\equiv\frac{1+\cos\theta}{2}=1+\frac{\bt}{\bs}\,,
\end{align}
where $\theta$ is the scattering angle in the center of mass frame. Moreover, we note that after having implemented the bulk-point limit of the correlator, the cross-ratio $x$ needs to be identified with the variable $z$ as $x\equiv1/z$.\footnote{
To establish this, one can compare how the two variables $z$ and $x$ transform under crossing. For instance, under $1\leftrightarrow2$ exchange, from the definition \eqref{eq:z} we have that $z\mapsto1-z$ whereas the cross-ratio $x$ from \eqref{eq:cross-ratios} transforms as $x\mapsto x/(x-1)$. Reconciling these two different transformation properties leads to the identification $x\equiv1/z$, and one can verify that all other crossing transformations are consistent with this assignment.
}

Let us first illustrate this matching at tree level. Writing the tree-level flat-space amplitude from \eqref{eq:flat_tree} in terms of the variable $z$ introduced above, we find
\begin{align}
	\mathcal{A}^{(1),i_1i_2i_3i_4}(z) = -\frac{8\pi G_N}{\bs}\bigg(\delta^{i_1i_2}\delta^{i_3i_4}+\frac{\delta^{i_1i_4}\delta^{i_2i_3}}{z-1}-\frac{\delta^{i_1i_3}\delta^{i_2i_4}}{z}\bigg),
\end{align}
where we have included the 6d gravitational coupling constant $8\pi G_N$. This is matched by the bulk-point limit of the tree-level correlator given in \eqref{eq:H1} as
\begin{align}
\begin{split}
	\frac{1}{N}\,\H^{(1),i_1i_2i_3i_4}(x,\xb)~&\xrightarrow{~\text{bpl}~}~-\frac{8\pi^2}{N}\bigg((1-x)\,\delta^{i_1i_2}\delta^{i_3i_4}+x\,\delta^{i_1i_4}\delta^{i_2i_3}-x(1-x)\,\delta^{i_1i_3}\delta^{i_2i_4}\bigg)\\
	&\qquad=\frac{2(1-x)}{\pi}\times\bs\,\mathcal{A}^{(1),i_1i_2i_3i_4}(z=1/x)\,,
\end{split}
\end{align}
where we have used the relation \eqref{adscft} to write $N$ in terms of the 6d gravitational constant $G_N$.

Proceeding to one-loop order, we choose to perform the analogous matching only on the $t$-channel contribution of the full amplitude, i.e. the coefficient of $\delta^{i_1i_4}\delta^{i_2i_3}$. By crossing symmetry, the other orientations are guaranteed to match as well. We start again by writing the flat-space amplitude in terms of $z$ defined in \eqref{eq:z}. Let us first consider the 6d scalar box diagram given in \eqref{eq:box_6d}, which enters the one-loop amplitude $\mathcal{A}^{(2)}(\bs,\bt)$ as a basic building block. When expressed in terms of the variable $z$, and after appropriate analytic continuation to have manifestly real logarithms for $z\in[0,1)$, we have
\begin{align}
	\B^{\oneladder}_{\text{6d}}(z) = \frac{1}{2z\,\bs}\log(1-z)\big[\log(1-z)+2\pi i\big].
\end{align}
For the full $t$-channel contribution from \eqref{eq:Fs} we then have
\begin{align}
	\mathcal{A}^{(2)}(z) = -\frac{(8\pi G_N)^2\,\bs}{(4\pi)^3}~\bigg[\frac{\log(1-z)\big[\log(1-z)+2\pi i\big]-z^3\big[\big(\log(z)-\log(1-z)\big)^2+\pi^2\big]}{2z}+\frac{2}{3}(z-1)\bigg]\,,
\end{align}
where we also included its prefactor appearing in \eqref{eq:flat_expansion}. This expression is matched by the bulk-point limit of our result \eqref{definitionF2explicit} for the one-loop correlator $\F(x,\xb)$. Explicitly, we find
\begin{align}\label{eq:bpl_match}
	\frac{1}{N^2}\,\F(x,\xb)\Big|_{\beta_3=4,\beta_4=0}~&\xrightarrow{~\text{bpl}~}~\frac{2^8\,45 x^4(1-x)^3}{\pi}\times\frac{1}{\bs}\,\,\mathcal{A}^{(2)}(z=1/x)\Big|_{n=21}\,,
\end{align}
where as indicated the matching enforces $n=21$ in the flat-space amplitude and moreover determines the coefficients of the ambiguities $B_3$ and $B_4$ to be
\begin{align}
	\beta_3=4\,,\quad\beta_4=0\,.
\end{align}
We thus arrive at the same conclusions as in the Mellin space formulation of the flat-space limit described in Section \ref{sec:flat-space}, from an independent calculation formulated entirely in position space.

Lastly, let us consider bulk-point limit of the one-loop correlator for general values of $n$. As discussed in Section \ref{sec:one-loop_extended}, in order to match the flat-space amplitude \eqref{eq:A2} we needed to extend the space of basis functions by including the function $f^{(3)}(x,\xb)$ which contains the additional letter $x-\xb$. In the bulk-point limit, the presence of the letter $x-\xb$ leads to an enhanced (logarithmic) divergence which can be identified with the scale-dependent logarithm in the flat-space amplitude.\footnote{Explicitly, one finds that the bulk-point limit of the function $f^{(3)}(x,\xb)$ results in
\begin{align}
	f^{(3)}(x,\xb)~\xrightarrow{~\text{bpl}~}~-4\pi^2\big[\log(x)+\log(1-x)+3\log(-s/\mu^2)+2\pi i\big]\,,
\end{align}
see Section 3.2 of \cite{Drummond:2022dxw} for more details on how the scale-dependent term arises.} For the function $\mathfrak{f}_s(x, \xb)$ defined in \eqref{eq:fs_def}, we then find that it contributes to the bulk-point limit as
\begin{align}\label{eq:bpl_fs}
	\mathfrak{f}_s(x,\xb)~\xrightarrow{~\text{bpl}~}~4\pi^2\log\Big({-}\frac{\bs}{\mu^2}\Big)\,,
\end{align}
for some choice of scale $\mu^2$. Similar relations hold for $\mathfrak{f}_t(x,\xb)$ and $\mathfrak{f}_u(x,\xb)$ with $\bs$ replaced by $\bt$ and $\bu$, respectively. Note that here we already made use of the identification $x\equiv1/z$ to map to the correct flat-space variables \eqref{eq:z}.

With this, the bulk-point limit of the remainder $\mathcal{R}(x,\xb)$ given in \eqref{eq:Rxxb} correctly reproduces the scale-dependent logarithmic term in the flat-space amplitude, i.e. we have
\begin{align}
	\frac{1}{N^2}\,\mathcal{R}(x,\xb)~\xrightarrow{~\text{bpl}~}~\frac{2^8\,45 x^4(1-x)^3}{\pi}\times\frac{1}{\bs}\,\mathcal{A}^{(2)}_{\log}(z=1/x)\,,
\end{align}
notably with the same overall factor as in \eqref{eq:bpl_match}.
Here $\mathcal{A}^{(2)}_{\log}$ denotes the logarithmic term in \eqref{eq:A2},
\begin{align}
	\mathcal{A}^{(2)}_{\log}(z)= -\frac{(8\pi G_N)^2}{(4\pi)^3}\frac{\bt}{12}\log\left(-\frac{\bt}{\mu^2}\right) = -\frac{(8\pi G_N)^2}{(4\pi)^3}\frac{\bs(z-1)}{12}\log\left(-\frac{\bs(z-1)}{\mu^2}\right),
\end{align}
and we have again included the $G_N$-dependent prefactor from \eqref{eq:flat_expansion}.

\bibliography{references.bib}
\bibliographystyle{JHEP}
\end{document}